\documentclass[twocolumn]{aastex631}

\received{October 30, 2022}
\revised{May 9, 2023}
\accepted{May 28, 2023}
\submitjournal{ApJS}

\shorttitle{FacetClumps for identifying molecular clumps}
\shortauthors{Jiang et al.}
\graphicspath{{./}{figures/}}
\usepackage{mathrsfs}
\usepackage{amsmath}
\usepackage{appendix}
\usepackage{array}
\usepackage{url}
\usepackage{hyperref}
\usepackage{soul} 
\usepackage{color, xcolor} 
\usepackage{makecell}
\usepackage{booktabs}
\usepackage{threeparttable}

\shortauthors{Jiang et al.}

\begin{document}

\title{FacetClumps: A Facet-based Molecular Clump Detection Algorithm}

\author[0000-0002-3549-5029]{Yu Jiang}
\affiliation{Center for Astronomy and Space Sciences, China Three Gorges University,
	\\8 University Road, 443002 Yichang, China}
\affiliation{College of Science, China Three Gorges University, \\
	8 University Road, Yichang, China}
\affiliation{Purple Mountain Observatory, Chinese Academy of Sciences, \\
	10 Yuanhua Road, 210023 Nanjing, China}

\author[0000-0003-0849-0692]{Zhiwei Chen}
\affiliation{Purple Mountain Observatory, Chinese Academy of Sciences, \\
	10 Yuanhua Road, 210023 Nanjing, China}

\author{Sheng Zheng}
\affiliation{Center for Astronomy and Space Sciences, China Three Gorges University,
	\\8 University Road, 443002 Yichang, China}
\affiliation{College of Science, China Three Gorges University, \\
	8 University Road, Yichang, China}

\author{Zhibo Jiang}
\affiliation{Purple Mountain Observatory, Chinese Academy of Sciences, \\
	10 Yuanhua Road, 210023 Nanjing, China}

\author{Yao Huang}
\affiliation{Center for Astronomy and Space Sciences, China Three Gorges University,
	\\8 University Road, 443002 Yichang, China}

\author{Shuguang Zeng}
\affiliation{Center for Astronomy and Space Sciences, China Three Gorges University,
	\\8 University Road, 443002 Yichang, China}

\author{Xiangyun Zeng}
\affiliation{Center for Astronomy and Space Sciences, China Three Gorges University,
	\\8 University Road, 443002 Yichang, China}

\author{Xiaoyu Luo}
\affiliation{Center for Astronomy and Space Sciences, China Three Gorges University,
	\\8 University Road, 443002 Yichang, China}

\correspondingauthor{Zhiwei Chen}
\email{zwchen@pmo.ac.cn}
\correspondingauthor{Sheng Zheng}
\email{zsh@ctgu.edu.cn}

\begin{abstract}
	A comprehensive understanding of molecular clumps is essential for investigating star formation. We present an algorithm for molecular clump detection, called FacetClumps. This algorithm uses a morphological approach to extract signal regions from the original data. The Gaussian Facet model is employed to fit the signal regions, which enhances the resistance to noise and the stability of the algorithm in diverse overlapping areas. The introduction of the extremum determination theorem of multivariate functions offers theoretical guidance for automatically locating clump centers. To guarantee that each clump is continuous, the signal regions are segmented into local regions based on gradient, and then the local regions are clustered into the clump centers based on connectivity and minimum distance to identify the regional information of each clump. Experiments conducted with both simulated and synthetic data demonstrate that FacetClumps exhibits great recall and precision rates, small location error and flux loss, a high consistency between the region of detected clump and that of simulated clump, and is generally stable in various environments. Notably, the recall rate of FacetClumps in the synthetic data, which comprises $^{13}CO$ ($J = 1-0$) emission line of the MWISP within $11.7^{\circ} \leq l \leq 13.4^{\circ}$, $0.22^{\circ} \leq b \leq 1.05^{\circ}$ and 5 km s$^{-1}$ $\leq v \leq$ 35 km s$^{-1}$ and simulated clumps, reaches 90.2\%. Additionally, FacetClumps demonstrates satisfactory performance when applied to observational data. 
\end{abstract} 


\section{Introduction}
Molecular clouds contain a significant proportion of gas and dust, and are the birthplace of many prominent young objects. Star formation processes take place on the scale of giant molecular clouds or even smaller. \citep{StarFormation,StarFormationInClusters}. Great efforts have been invested in characterizing the feature of molecular gas \citep[e.g.][]{Lifetime,AnalysisofClumps}, deriving the stellar initial mass function \citep[e.g.][]{MassFunction1,MassFunction2,MassFunction3}, and observing the local star formation rate \citep[e.g.][]{StarFormationRate0, StarFormationRate2}. Carbon monoxide surveys are a crucial way in unveiling the mysteries of stellar formation, stellar evolution, and galactic structure \citep[e.g.][]{CO1,CO2,CO3,CO4,M161,MWISP}. Giant molecular clouds often exhibit substructures, such as filaments, clumps, and cores \citep{MC}, while faint sources can be easily obscured by noise. A critical challenge in many research projects is how to accurately segment giant molecular clouds and detect faint targets in the data from carbon monoxide surveys. 

Some of the available clump-finding algorithms include GaussClumps \citep{GaussClumps}, ClumpFind \citep{ClumpFind1}, ReinHold \citep{CUPID2}, FellWalker \citep{CUPID2}, Local Density Clustering (LDC) \citep{LDC}, and ConBased\citep{ConBased}. The location of peaks is a crucial aspect of GaussClumps, ClumpFind, ReinHold, and LDC. GaussClumps starts from the brightest peak in the data cube to fit the ellipsoid clump, subtracts the fitting clump, and then performs the fitting from the brightest peak in the residuals. ClumpFind contours the data array at many different levels, with a peak being an isolated contour, then works from the highest contour levels to a specified minimum contour level. ReinHold and LDC both rank the data in descending order, with those at the top and above a minimum intensity being considered as potential peaks; For ReinHold, a peak is deemed significant if the pixels spanned by the peak along any one dimension are greater than a specified minimum number; For LDC, a peak is deemed significant if no pixels with greater intensity exist within a specified neighborhood. FellWalker and ConBased do not rely on alternative peaks but instead take into account the relationships between the nearest peaks. FellWalker ascends the line of greatest gradient until a peak is reached, then jumps to the pixel with the highest value in an extended neighborhood to identify clumps, and merges adjacent clumps if their peak-dependent dip is less than a specified value. ConBased divides signals into small regions and merges them from the regions with the smallest volume, using a merging rule based on connectivity, peak distance and intensity differences, and volume. 

Molecular clumps are irregular in shape and are characterized by faint gas enveloping a denser central source. To address the two main objectives of molecular clump detection, namely identifying the location and region of molecular clump, we propose a novel algorithm called FacetClumps. The location of peaks, particularly those of faint clumps, is readily impacted by noise. To identify the location of the denser central source and reduce the reliance on the peak in the detection process, FacetClumps utilizes morphology \citep[e.g.][]{Morphology0, Morphology1, Morphology2, ConBased} to extract signal regions from the original data, and incorporates the Gaussian Facet model \citep{HaralickFacet, HaralickFacet2, Operators, GaussianFacet} and extremum theory of multivariate function to locate clump centers in the signal regions. A single molecular clump is relatively smooth and continuous; however, FellWalker may detect multiple distinct components as a single clump, while LDC obtains the connectivity by selecting the subpart with the largest volume from the potential clumps and discarding the smaller, discontinuous subparts. To improve the accuracy of regional segmentation, FacetClumps utilizes a gradient-based method \citep{CUPID2} to segment the signal regions into local regions, and then applies a connectivity-based minimum distance clustering method to cluster the local regions to the clump centers. To improve the adaptability, the parameters of FacetClumps are automatically adjusted according to different local situations, and are optimized for detecting faint and overlapping clump.

We illustrate the processes and details of FacetClumps by combining text and schematic diagrams in Section 2. In Section 3, we determine suitable values for FacetClumps parameters in simulated clumps, and compare its performance with that of other algorithms in simulated and synthetic clumps. We then apply FacetClumps to observational data, and present the results in Section 3. In Appendix A, we conduct experiments on larger synthetic data with different signal densities. In Appendix B, we analyze the performance of FacetClumps in different resampled synthetic data. Finally, we summarize our work in Section 4. 

\begin{figure*}
	\centering
	\centerline{\includegraphics[width=6.6in]{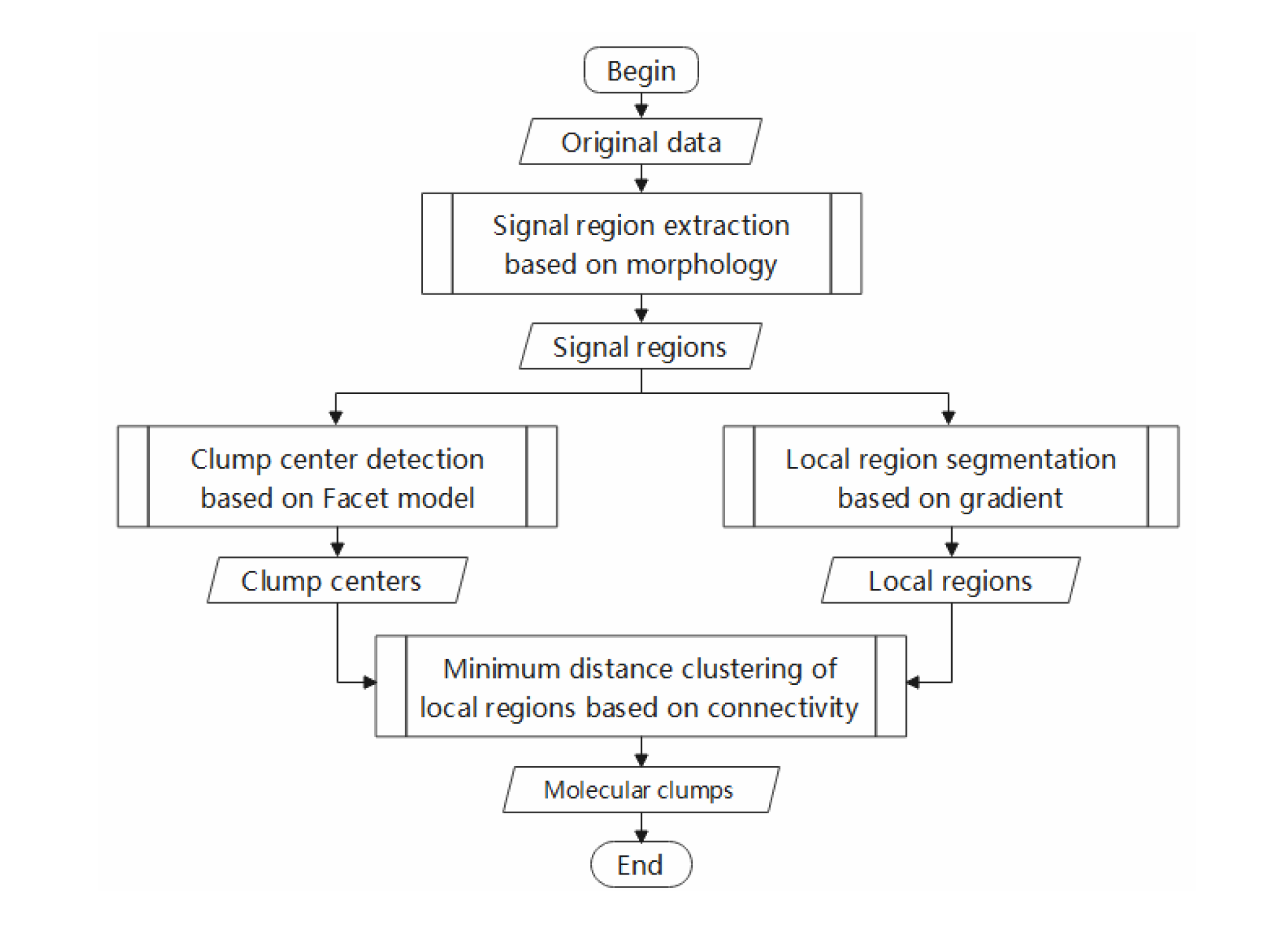}}
	\caption{Flow chart. FacetClumps consists of four sub-processes: (1) signal region extraction based on morphology, (2) clump center detection based on the Facet model, (3) local region segmentation based on the gradient, and (4) minimum distance clustering of local regions based on connectivity.}
	\label{Fig1}
\end{figure*}

\section{The FacetClumps algorithm}
FacetClumps primarily consists of four sub-processes. The first sub-process is signal region extraction based on morphology, which includes threshold segmentation, opening, dilation, and connected domain labeling of the original data, followed by combining the connected domain with the original data to acquire the signal regions. The second sub-process is clump center detection based on the Facet model. The Gaussian Facet model operators are convolved with the signal data, resulting in a fitting surface and corresponding fitting coefficients. The first and second derivatives of the fitting surface are then calculated from these coefficients. The extremum determination theorem is employed to identify potential maxima by utilizing the first derivatives and the eigenvalues of the Hessian matrix constructed from the second derivatives. To account for the interference of noise, the maximum regions which may contain maxima are extracted by adaptively adjusting the thresholds of the first derivatives and the eigenvalues. The centroid of a maximum region is taken as the clump center. The third sub-process is local region segmentation based on local gradients. In the smallest neighborhood of voxeles, each signal region is segmented into local regions by ascending along the highest gradient of intensity \citep{ConBased}. The fourth sub-process is minimum distance clustering of local regions based on connectivity. Matching clump centers with local regions, each matched region is regarded as the target region of a clump. The connected-nearest local regions of a target region are then merged into the target region, forming clumps. This merging process continues until all local regions are clustered.

Finally, the statistics of each clump in pixel and WCS coordinate systems are collected in two tables, and the regional information of each clump is recorded in a mask. The flow of FacetClumps is shown in Figure \ref{Fig1}. FacetClumps can be applied to two-dimensional (position-position space, hereafter PP) and three-dimensional (position-position-velocity space, hereafter PPV) observational data, but we will mainly introduce the PPV flow. We have shared the code on Github under a permissive MIT license\footnote{\href{https://github.com/JiangYuTS/FacetClumps}{https://github.com/JiangYuTS/FacetClumps}}, made it publicly available as a Python package called FacetClumps\footnote{\href{https://pypi.org/project/FacetClumps/}{https://pypi.org/project/FacetClumps/}}, and deposited the latest version to Zenodo \citep{zenodov4}. We warmly welcome community contributions to its optimization. 

\begin{figure*}
	\centering
	\vspace{0cm}
	\begin{minipage}[t]{0.24\textwidth}
		\centering
		\centerline{\includegraphics[width=1.8in]{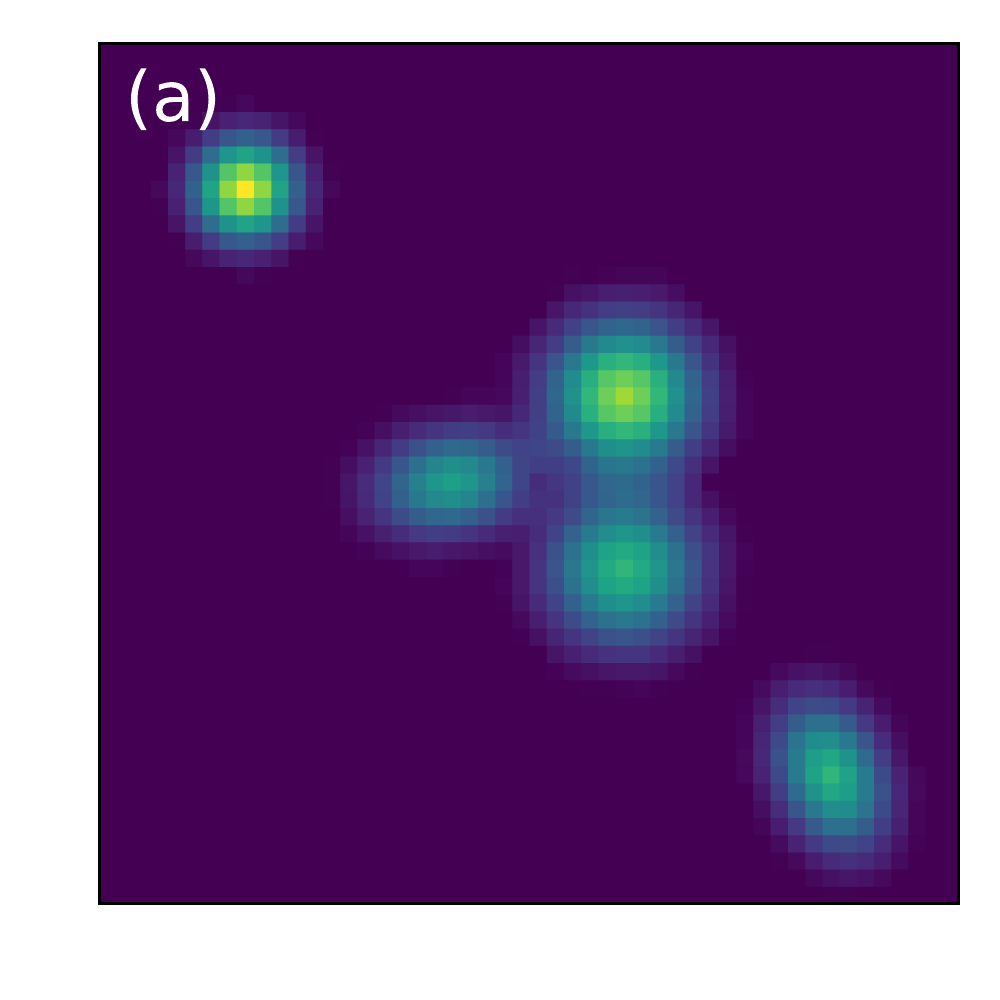}}
\end{minipage}\begin{minipage}[t]{0.24\textwidth}
		\centering
		\centerline{\includegraphics[width=1.8in]{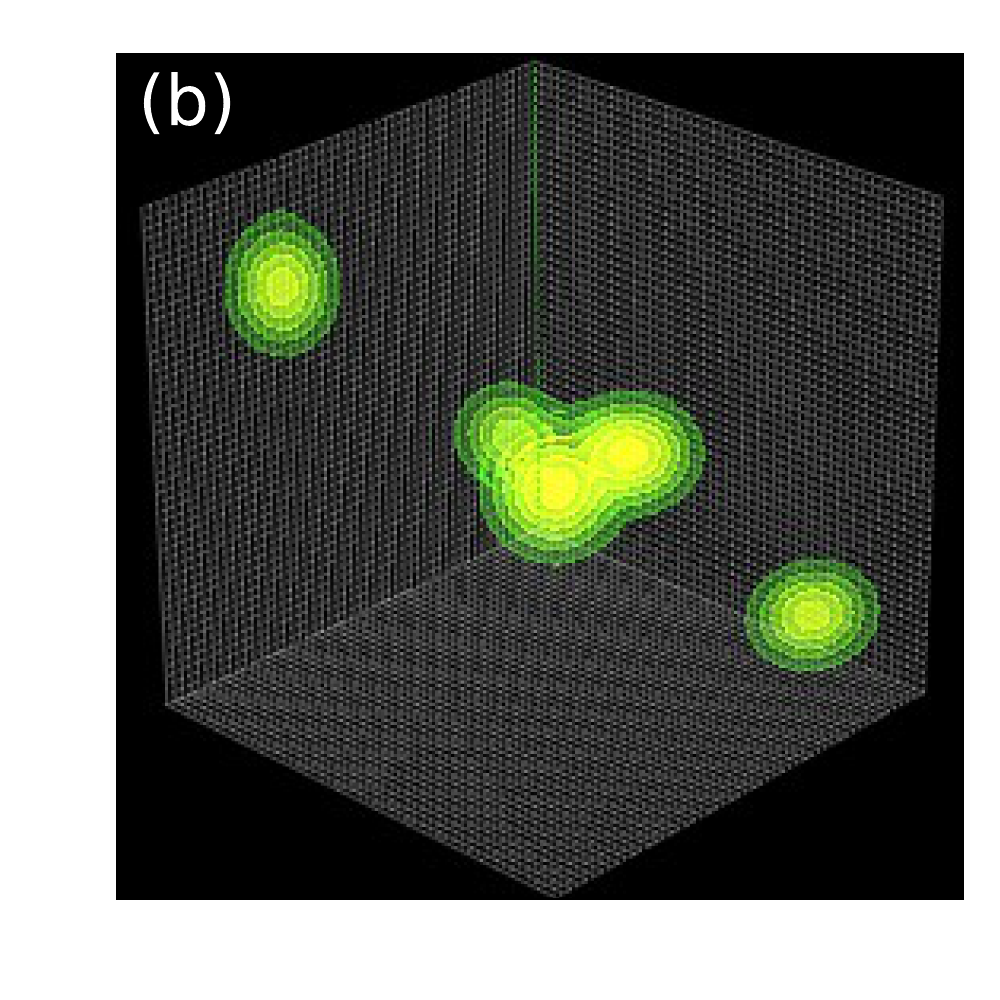}}
\end{minipage}\begin{minipage}[t]{0.24\textwidth}
		\centering
		\centerline{\includegraphics[width=1.8in]{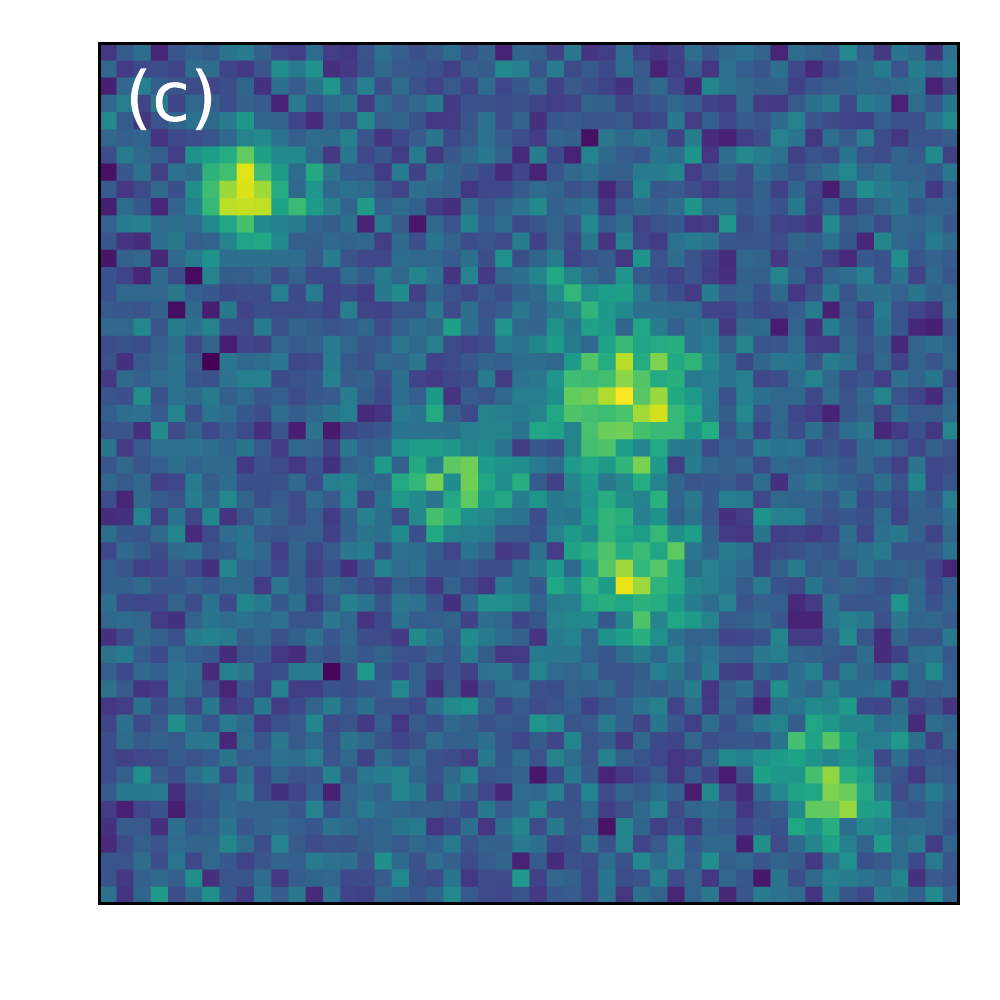}}
\end{minipage}\begin{minipage}[t]{0.24\textwidth}
		\centering
		\centerline{\includegraphics[width=1.8in]{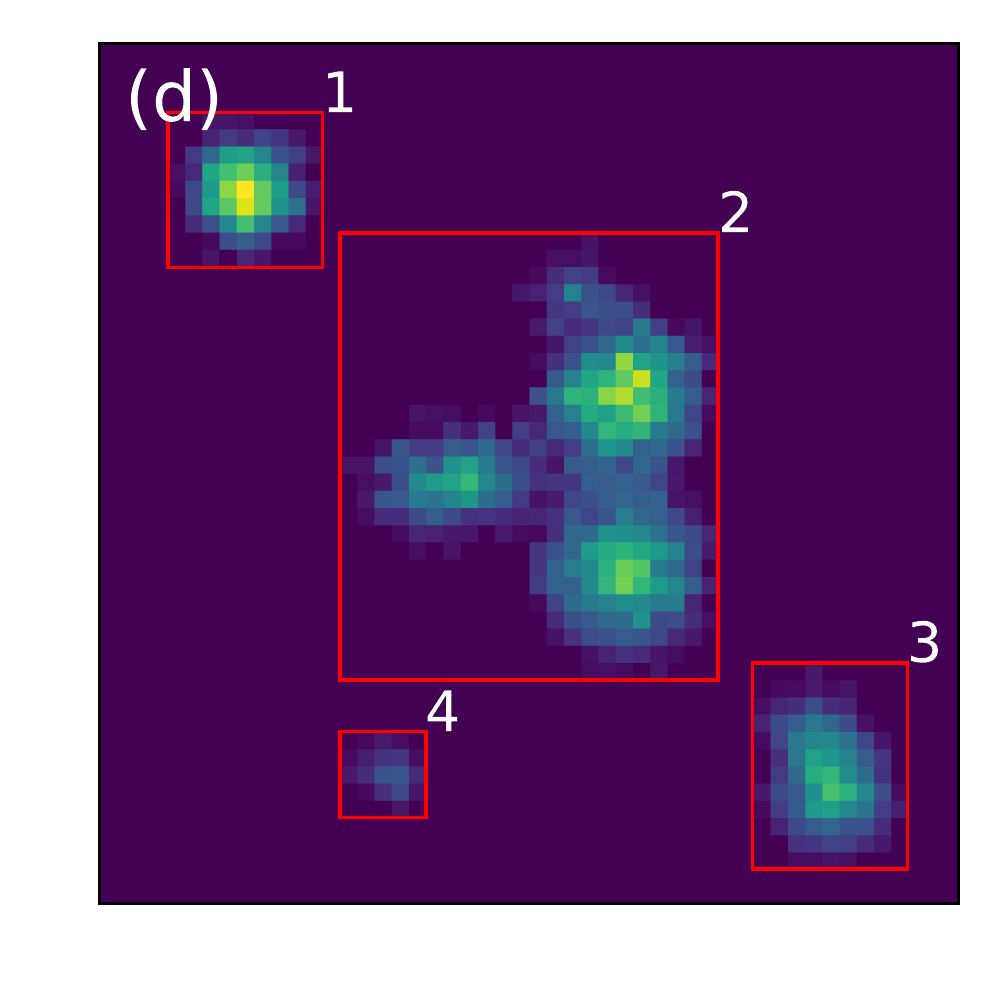}}
\end{minipage}

	\caption{(a) The original noise-free data; (b) The original noise-free data with a PPV view; (c) Noisy data; (d) The signal regions,  with a red box being the smallest enclosing rectangle of a region.}
	\label{Fig_Morphology}
\end{figure*} 

\subsection{Signal region extraction based on morphology}\label{Sec2.1}
To enhance the efficiency and robustness of FacetClumps, the original data is preprocessed based on morphology. As illustrated in Figure \ref{Fig_Morphology}(a), there are five clumps, three of which are overlapping. Figure \ref{Fig_Morphology}(b) shows the PPV view of Figure \ref{Fig_Morphology}(a), and Figure \ref{Fig_Morphology}(c) displays the clumps after introducing noise. 

Firstly, a free parameter $Threshold$ (typically $2\times RMS$, where $RMS$ is the noise RMS of the data) is selected to binarize the original data. Voxel values greater than the threshold are labeled as binarization mask. Secondly, a morphological opening operation \citep{opening} is performed on the binarization mask to obtain an opening mask. The opening operation separates the boundaries of the regions from the noise, which eliminates noise points and has a negligible impact on the size of the regions. Thirdly, to reduce the loss of flux and enhance the detection of faint clumps, a morphological dilation is performed on the opening mask, and the signals below the noise level and above the threshold of the dilated voxels are removed to obtain a dilation mask. The kernels of the opening and dilation operators are ball operators of radius one. Fourthly, a connected domain operator is applied to the dilation mask to obtain a connected domain mask. 

Connectivity is a relationship between neighborhoods. In PPV, there are three distinct types of connectivity \citep{ClumpFind1}: $Type \uppercase\expandafter{\romannumeral1}$, $Type \uppercase\expandafter{\romannumeral2}$, and $Type \uppercase\expandafter{\romannumeral3}$. Each type requires that the maximum number of orthogonal hops of a voxel considered to be a neighbor are 1, 2, and 3, respectively \citep{CType1,CType2}. A schematic diagram of this is shown in Figure 1 of \cite{ClumpFind1}.

The connected domain masks with intensity are indicative of signal regions, where the connectivity type is $Type \uppercase\expandafter{\romannumeral3}$, as depicted in Figure \ref{Fig_Morphology}(d). It is evident that the signal region in box 2 consists of three overlapping clumps, and the signal region in box 4 is composed of noise. 

\subsection{Clump center detection based on Facet model}\label{Sec2.2}

\subsubsection{Basic principle of PPV Gaussian Facet model}\label{Sec2.2.2}
Facet model utilizes the concept of sub-sections to obtain the most accurate analytic function in a regular region centered around a specific pixel, using the least-squares fitting method.  \cite{Wu2DTo3D} extended the two-dimensional directional derivative edge detector to three-dimensional. \cite{WeijerLinear} investigated the Gaussian weighted Facet model from the linear scale-space category. Unlike the traditional Haralick Facet model, the Gaussian Facet model employs the weighted least-squares fitting method to obtain the operators of the model. The Gaussian function serves as the weight, indicating the varying importance of different sampling points in determining the final operators. The weight is higher when a sampling point is closer to the center of the region. \cite{GaussianFacet} derived the three-dimensional edge detection operator based on the Gaussian Facet model and applied it to three-dimensional subvoxel surface detection, achieving good performances. 

For simple images, the intensity function can be approximated by a piecewise constant or piecewise bivariate linear function. For complex images, higher-order polynomials should be chosen. In this paper, we use a polynomial of ternary cubic with integer coefficients to establish the PPV Gaussian Facet model. The original molecular clump surface $I(x,y,z)$ can be approximated by a linear combination of a set of bases $g_i(i=1,2,\ldots,20)$, and the approximation function $f(x,y,z)$ is defined as: 

\begin{equation}\label{fxyz}
\begin{gathered}
f(x,y,z)=\Phi \boldsymbol{a}\\
\boldsymbol{a}=(a_1,a_2,\ldots,a_{20})^\mathrm{T}\\
\Phi=(g_1,g_2,\ldots,g_{20})=(1,x,y,z,x^2,y^2,z^2,xy,xz,yz,\\
xyz,xy^2,xz^2,x^2y,yz^2,x^2z,y^2z,x^3,y^3,z^3)\\
\end{gathered}
\end{equation}

The least squares fitting minimizes the difference $\varepsilon$ of the molecular clump surface $I(x,y,z)$ and approximation function $f(x,y,z)$:

\begin{equation}\label{epsilon}
\it{\varepsilon=\iiint_{\Omega}(I(x,y,z)-f(x,y,z))^2W(x,y,z)dxdydz}
\end{equation}

\noindent where $W(x,y,z)$ is the window function defining the locality of the model fitting. The Gaussian Facet model employs the Gaussian window function:

\begin{equation}\label{Wxyz}
\it{W(x,y,z)=\frac{1}{2\pi s^2}exp(-\frac{x^2+y^2+z^2}{2s^2})}
\end{equation}

\noindent where $s$ represents the window radius. The window scale $SWindow$ is a free parameter, and $s=\lfloor SWindow/2\rfloor$. The optimal parameter vector $\boldsymbol{a}$ is obtained by projecting the function $f$ onto the subspace spanned by the basis functions in $\Phi$. The inner product in this function space is expressed as: 

\begin{equation}\label{pq}
\begin{split}
p*q&\equiv\langle p,q\rangle W\\
&=\iiint_{\Omega}p(x,y,z)q(x,y,z)W(x,y,z)dxdydz\\
\end{split}
\end{equation}

\noindent where $p,q$ are arbitrary functions of three variables. So the difference $\varepsilon$ in (\ref{epsilon}) is rewrited as (\ref{epsilon_T}):

\begin{equation}\label{epsilon_T}
\begin{aligned}
\varepsilon&=\langle I-\Phi\boldsymbol{a},I-\Phi\boldsymbol{a}\rangle W\\
&=(\langle I,I\rangle-2\langle I,\Phi\boldsymbol{a}\rangle+\langle \Phi\boldsymbol{a},\Phi\boldsymbol{a}\rangle)W\\
&=I^\mathrm{T}*I-2\boldsymbol{a}^\mathrm{T}\Phi^\mathrm{T}*I+\boldsymbol{a}^\mathrm{T}\Phi^\mathrm{T}*\Phi\boldsymbol{a}
\end{aligned}
\end{equation}

Taking the derivative of (\ref{epsilon_T}) for $\boldsymbol{a}$, setting $\frac{\partial \varepsilon}{\partial \boldsymbol{a}}=0$, and solving for $\boldsymbol{a}$, we obtain:

\begin{equation}\label{coefficient}
\it{\boldsymbol{a}=(\Phi^\mathrm{T}\Phi)^{-1}\Phi^\mathrm{T}*I=K*I}
\end{equation}

The convolution operators $K$ can be obtained by substituting $\Phi, W$ into (\ref{coefficient}) and utilizing the weighted inner product of (\ref{pq}). The convolution of $K$ with the original surface yields the coefficient $\boldsymbol{a}$ of its fitting surface. The continuous form of the inner product not only provides higher accuracy than the discrete form, but also incorporates the window scale into the definitive $\boldsymbol{a}$, which is advantageous for multi-scale analysis. 

\begin{figure}
	\centering
	\includegraphics[width=2.8in]{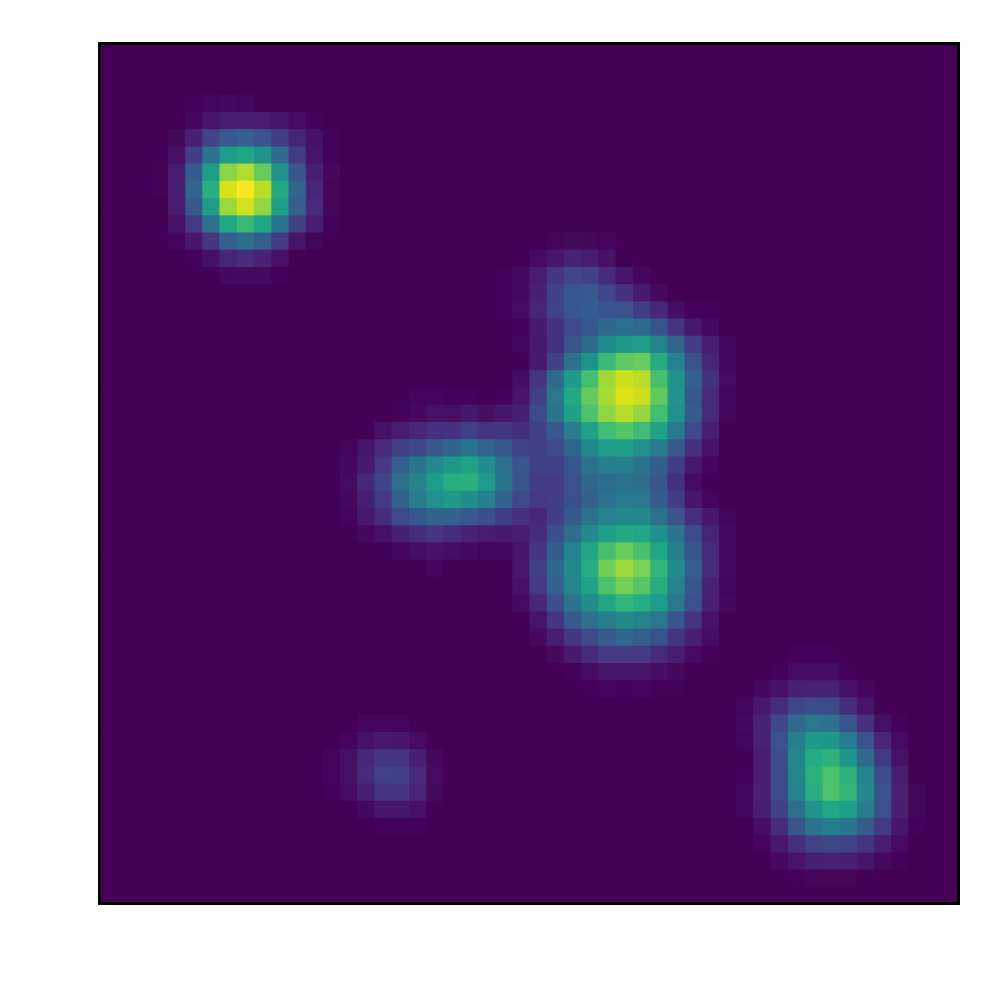}
	\caption{The integral diagram of the fitting surface of the singal regions shown in Figure \ref{Fig_Morphology}(d).}
	\label{Fig_Hook_Face}
\end{figure}

\begin{figure*}
	\centering
	\includegraphics[width=5.6in]{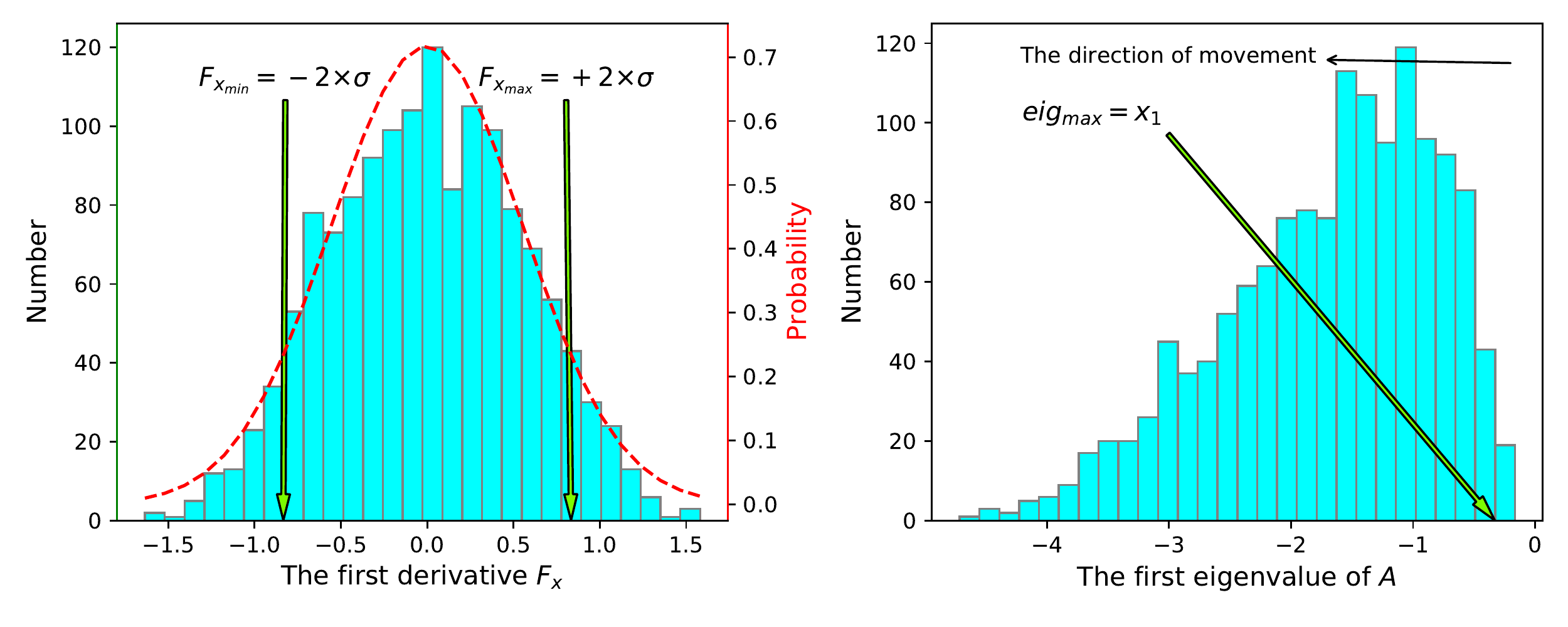}
	\caption{Distribution of the first derivative $F_x$ and the first eigenvalue of the region numbered 2 in Figure \ref{Fig_Morphology}(d). The distribution of $F_x$ is fitted with the Gaussian function. The thresholds of the first derivative $F_x$ are the double negative and positive standard deviations ($-2\times\sigma$ and $+2\times\sigma$) of $F_x$. Coordinates with $F_x$ value between the thresholds are used to judge the maximum regions. Similarly, the thresholds of $F_y$ and $F_z$ are determined in the same manner as $F_x$. The initial threshold of the first eigenvalue is the x-coordinate of the first bin on the right, and the x-coordinate is shifted one bin to the left after each recursion. Eigenvalues that are less than the threshold are taken into consideration for the judgment. Similarly, the thresholds of the second and third eigenvalues are determined in the same manner as the first eigenvalue.}
	\label{Fig_Grad_Eig}
\end{figure*}

\begin{figure*}
	\centering
	\vspace{0cm}
	\begin{minipage}[t]{0.24\textwidth}
		\centering
		\centerline{\includegraphics[width=1.8in]{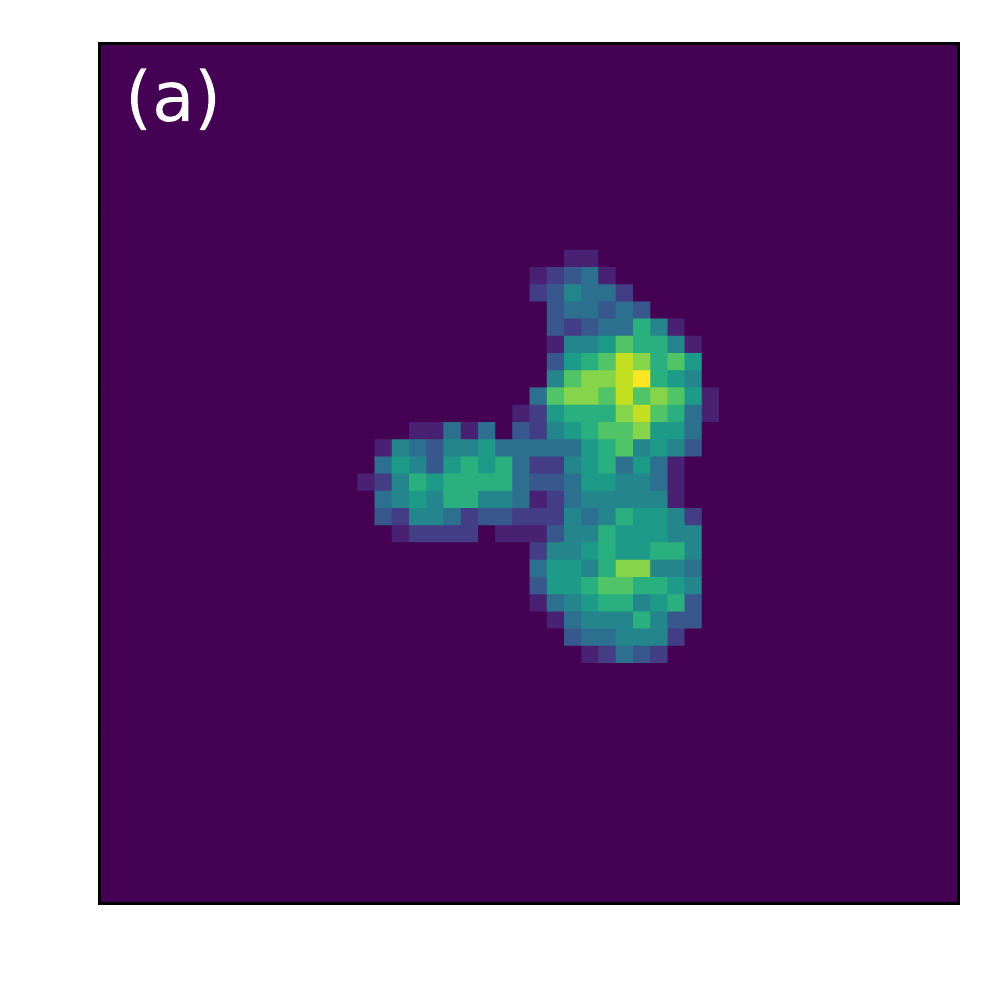}}
\end{minipage}\begin{minipage}[t]{0.24\textwidth}
		\centering
		\centerline{\includegraphics[width=1.8in]{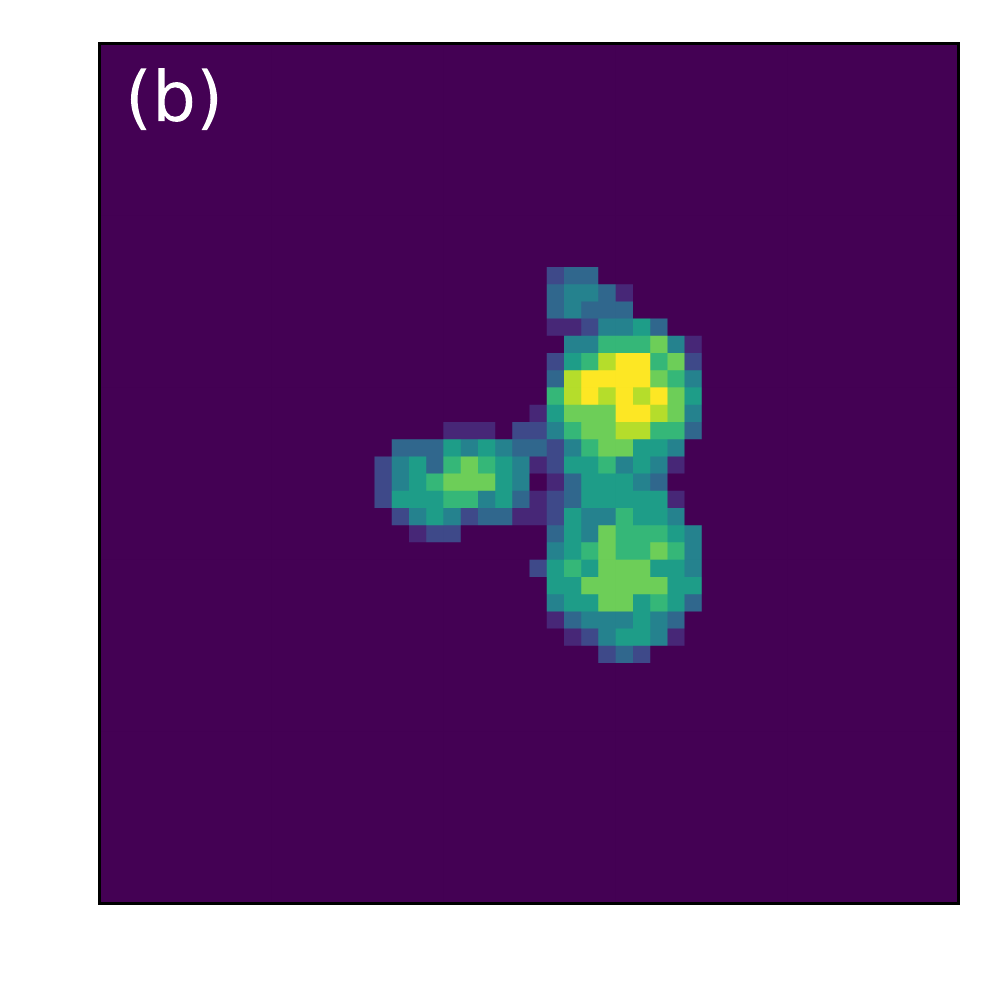}}
\end{minipage}\begin{minipage}[t]{0.24\textwidth}
		\centering
		\centerline{\includegraphics[width=1.8in]{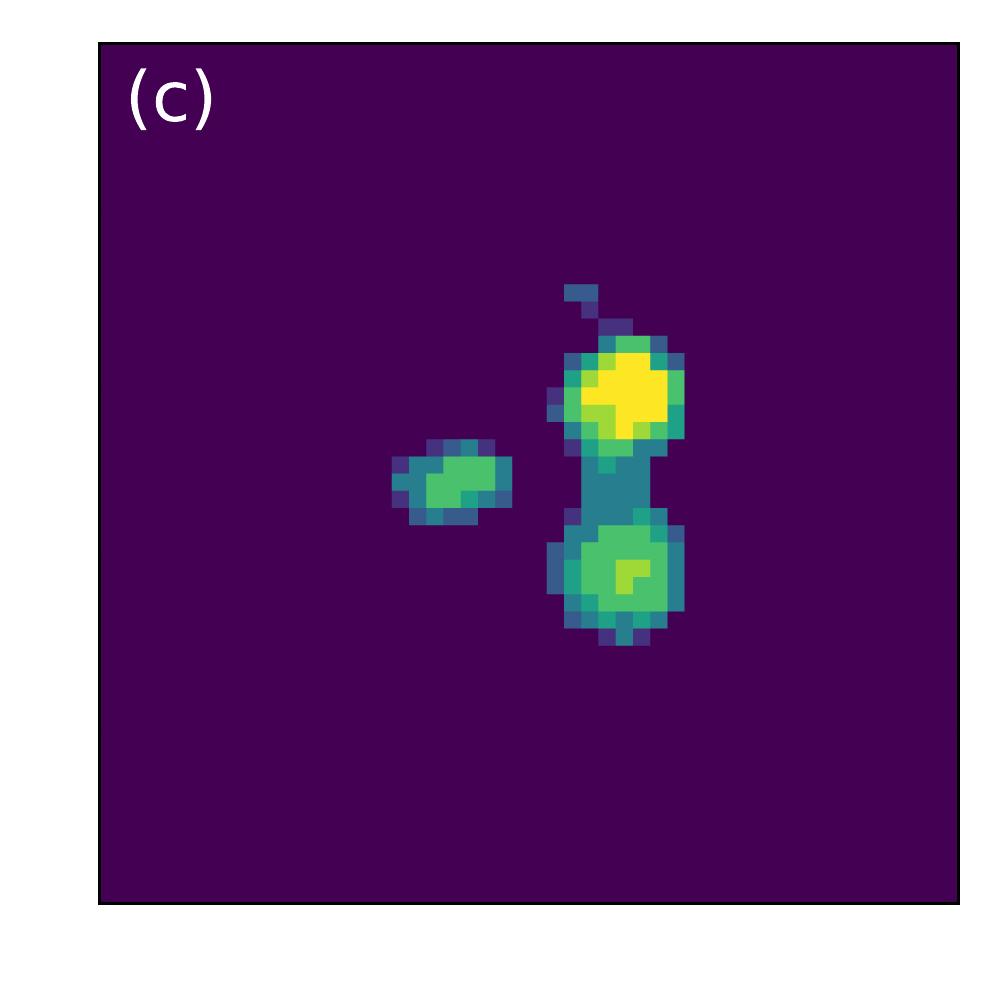}}
\end{minipage}\begin{minipage}[t]{0.24\textwidth}
		\centering
		\centerline{\includegraphics[width=1.8in]{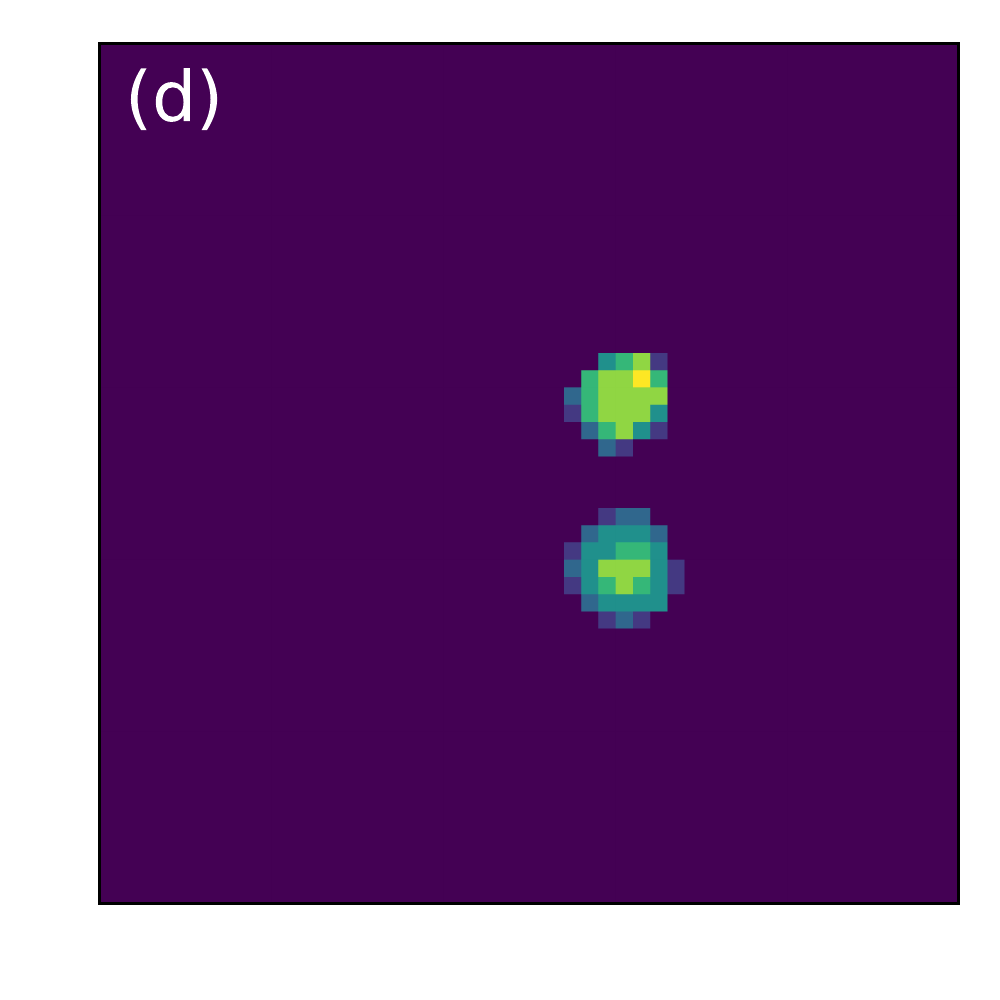}}
\end{minipage}\caption{Some of the recursion masks of the region numbered 2 in Figure \ref{Fig_Morphology}(d). (a) The sub-maximum region after the first recursion. (b) The sub-maximum region after the second recursion. (c) The sub-maximum regions separating one of the clumps and the maximum region of the noise cluster. (d) The sub-maximum regions separating the last two clumps.}
	\label{Fig_Recursion_Lable}
\end{figure*}

\begin{figure*}
	\centering
	\vspace{0cm}
	\begin{minipage}[t]{0.4\textwidth}
		\centering
		\centerline{\includegraphics[width=2.5in]{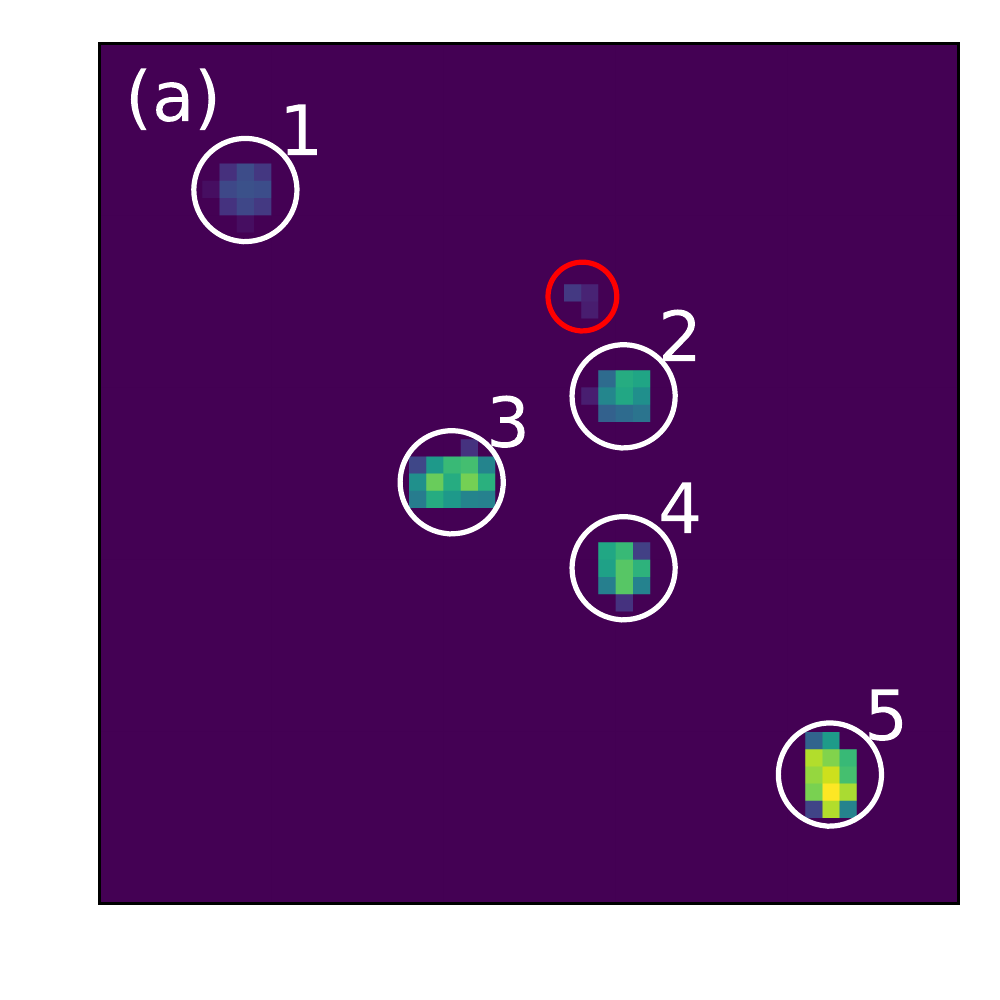}}
\end{minipage}\begin{minipage}[t]{0.4\textwidth}
		\centering
		\centerline{\includegraphics[width=2.5in]{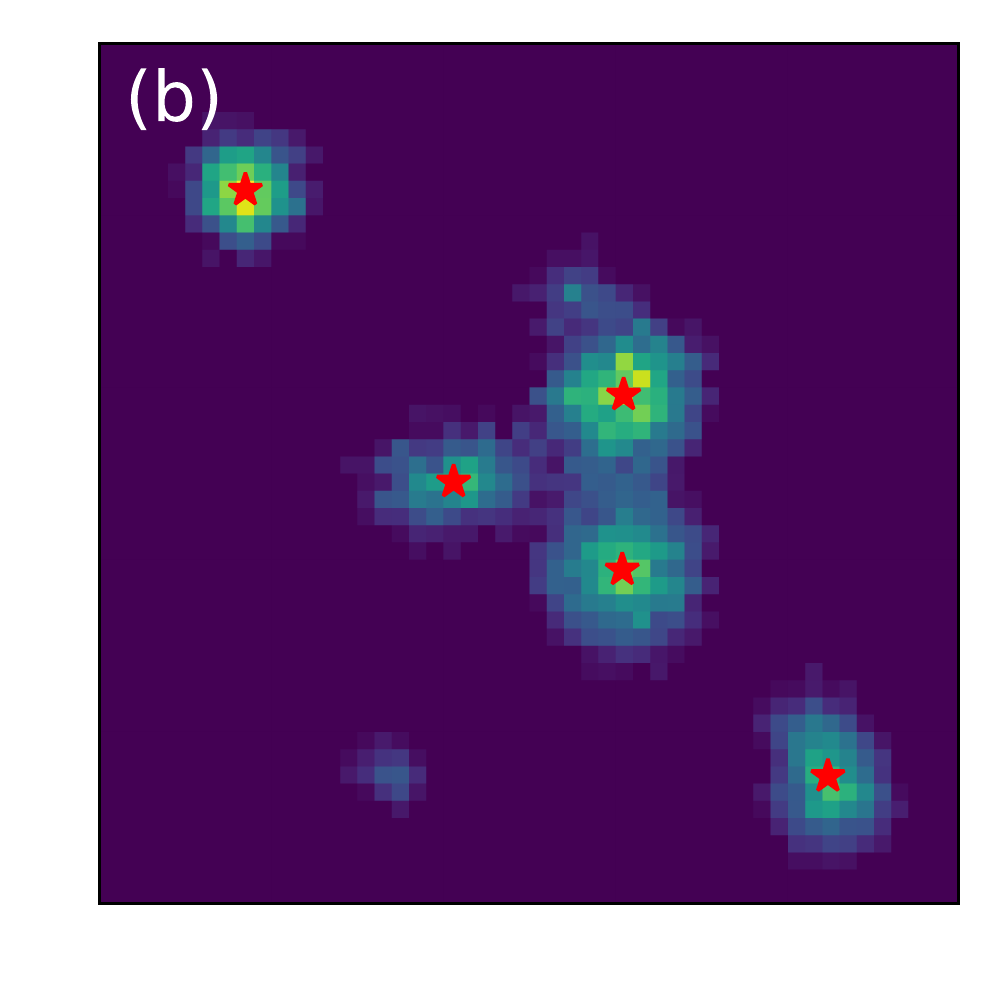}}
\end{minipage}\caption{(a) The integrated intensity of the final maximum regions; the white circles represent the valid maximum regions. (b) The clump centers detected by FacetClumps; the background is the signal regions and the red asterisks denote the locations of the clump centers.}
	\label{Fig_Hook_Centers}
\end{figure*}

\subsubsection{Combine multivariate function maximum determination theorem and Facet model}\label{Sec2.2.3}
Let a multivariate real function $f(x,y,z)$ have the continuous second derivative in the neighborhood of point $P_0(x_0,y_0,z_0)$. If $\frac{\partial f}{\partial \mathbf{x}}|_{(x_0,y_0,z_0)}=0$, and $H$ constructed by the second derivative of $f$ is a negative definite matrix, $f(x_0,y_0,z_0)$ is a maximum. The necessary and sufficient condition for a matrix to be negative definite is that all its eigenvalues are less than zero.

Convolve the data cube of the signal regions in Section \ref{Sec2.1} with the operators $K$ in Section \ref{Sec2.2.2} to derive the first and second derivatives of the fitting surface. The fitting surface is shown in Figure
 \ref{Fig_Hook_Face}. The first derivatives $(F_x,F_y,F_z)$ of $f(x,y,z)_{(x,y,z)=(0,0,0)}$ are obtained by (\ref{partial1}):

\begin{equation}\label{partial1}
\begin{aligned}
F_x&=\frac{\partial f}{\partial x}&=a_2,F_y&=\frac{\partial f}{\partial y}&=a_3,F_z&=\frac{\partial f}{\partial z}&=a_4
\end{aligned}
\end{equation}

Formula (\ref{partial2}) reveals the second derivatives of the central point of neighborhood, which are used to construct the Hessian matrix, as demonstrated in (\ref{Hessian}). 

\begin{equation}\label{partial2}
\begin{aligned}
\frac{\partial^2f}{\partial x^2}&=2a_5,\frac{\partial^2f}{\partial y^2}=2a_6,\frac{\partial^2f}{\partial z^2}=2a_7,\\
\frac{\partial^2f}{\partial xy}&=a_8,\frac{\partial^2f}{\partial xz}=a_9,\frac{\partial^2f}{\partial yz}=a_{10}\\
\end{aligned}
\end{equation}

\begin{equation}\label{Hessian}
\it{H=\left(
	\begin{array}{ccc}
	2a_5 &a_8& a_9 \\
	a_8&2a_6&a_{10} \\
	a_9 &a_{10}&2a_7\\
	\end{array}
	\right)}
\end{equation}

According to the multivariate function maximum determination theorem and considering the interference of noise, any one of the first derivatives around the center of a clump is near zero, and the eigenvalues of the Hessian matrix around the center of a clump are less than zero. As a result, during clump center detection, we do not simply search for maximum values of the fitting surface directly, but instead extract a certain range of maximum regions.

\subsubsection{Locate the clump centers}\label{Sec2.2.4}
We define operations OP1, OP2, OP3, and OP4 to obtain adaptive thresholds for the first derivatives and eigenvalues, and use them to search for the maximum regions in each signal region. Figure \ref{Fig_Grad_Eig} depicts the distribution of the first derivatives $F_x$ and the first eigenvalues of the signal region which is composed of three overlapping clumps as shown in box 2 of Figure \ref{Fig_Morphology}(d). 

Operation OP1: Determine the preselected maximum regions using the first derivatives. A Gaussian function is used to fit the distribution of the first derivative, and its mean is around zero. The standard deviation of the first derivative values is computed for each region, and the thresholds are negative two standard deviations and positive two standard deviations. Voxels whose values fall within these thresholds are selected as candidates for the preselected maximum region.

Operation OP2: Determine the preselected maximum regions using the eigenvalue. To better accommodate the distribution characteristics of different regions, the eigenvalues are binned according to the number of bins ($Bins$), which is calculated by (\ref{FormulaKBins}). The threshold is the x-coordinate of the $i$th bin on the right side of the distribution, as shown in Figure \ref{Fig_Grad_Eig}. (The initial value of $i$ is minus one, and after each recursion, which will be described in operation OP4, the value decreases by one.) Voxels whose values are less than the threshold are selected as the candidates for the preselected maximum region. 

\begin{equation}\label{FormulaKBins}
\it{Bins_i=\lfloor KBins\times \ln{Vsr_{i}} \rfloor}
\end{equation}

\noindent where, $KBins$ is a free parameter, $Vsr_i$ is the volume of the $i$th signal region, $i=1,2,\ldots,n1$, with $n1$ being the number of signal regions.

Operation OP3: Determine the connected maximum regions using the preselected maximum regions. Both the criteria for the first derivatives and eigenvalues must be simultaneously satisfied. Voxels in the preselected sets that meet the criteria are marked as one. The connected domain operator is then applied to the marked data to identify connected domains, and each connected domain is considered to be a maximum region. To improve the ability to detect faint and overlapping clumps, the connectivity type used is $Type \uppercase\expandafter{\romannumeral1}$.

Operation OP4: Determine sub-maximum regions using recursion. Figure \ref{Fig_Recursion_Lable}(a) illustrates that a maximum region extracted from the crowded signal region may still contain multiple clumps. To separate these overlapping clumps, recursion operations OP1, OP2, and OP3 are applied to each maximum region if its area in the spatial direction exceeds the parameter $SRecursionLB$ or its length in the velocity channels exceeds the parameter $SRecursionV$ to extract sub-maximum regions. During the recursion process, the $Bins$ of each sub-region is the same as its signal region, the thresholds of the first derivatives and eigenvalues for each subregion are updated as described in operations OP1 and OP2, and the connected sub-maximum regions are determined as described in operation OP3. It can be seen from Figure \ref{Fig_Recursion_Lable}, the overlapping clumps are successfully separated after recursions. 

The resulting sub-maximum regions that have undergone a recursion are displayed in Figure \ref{Fig_Hook_Centers}(a). Some small sub-regions are commonly caused by noise, as indicated by the red circle in Figure \ref{Fig_Hook_Centers}(a). To determine a valid maximum region, the formula in (\ref{FormulaVmm}) can be used.

\begin{equation}\label{FormulaVmm}
\begin{aligned}
\it{Region_j=\begin{cases}
	True,&\text{if $Vmr_j\geq \log{Vsr_i}$}\\
	False,&\text{else}\\
	\end{cases}}
\end{aligned}
\end{equation}

\noindent where, $Region_j$ represents the $j$th maximum sub-region, $Vmr_j$ is the volume of $Region_j$, $j=1,2,\ldots,n2$, with $n2$ being the number of maximum sub-regions, $Vsr_i$ is the volume of the $i$th signal region, $i=1,2,\ldots,n1$, with $n1$ being the number of signal regions. 

After performing the operations mentioned above, all valid maximum regions are obtained. These regions are combined with the fitting surface to extract intensity information. The centroid of each valid maximum region is calculated using equation (\ref{Cen}). This centroid represents the location of a denser central source and is recorded as a clump center. The resulting clump centers are shown in Figure \ref{Fig_Hook_Centers}(b) and are found to be consistent with the simulated central coordinates. 

\begin{equation}\label{Cen}
\it{\boldsymbol{Cen}=\frac{\sum_{k=1}^{n3}f[\boldsymbol{u_k}]\cdot \boldsymbol{u_k}}{\sum_{k=1}^{n3}f[\boldsymbol{u_k}]}}
\end{equation}

\noindent where, $\boldsymbol{Cen}$ represents the centroid of a valid maximum region, $f$ is the fitting surface, $\boldsymbol{u_k}$ is the coordinate of k, $\boldsymbol{u_k}=\{x, y, z\}=\boldsymbol{U}[k]$, and $\boldsymbol{U}$ is the coordinate set of a valid maximum region, $n3$ is the number of the coordinates. 

\begin{figure}
	\centering
	\centerline{\includegraphics[width=2.8in]{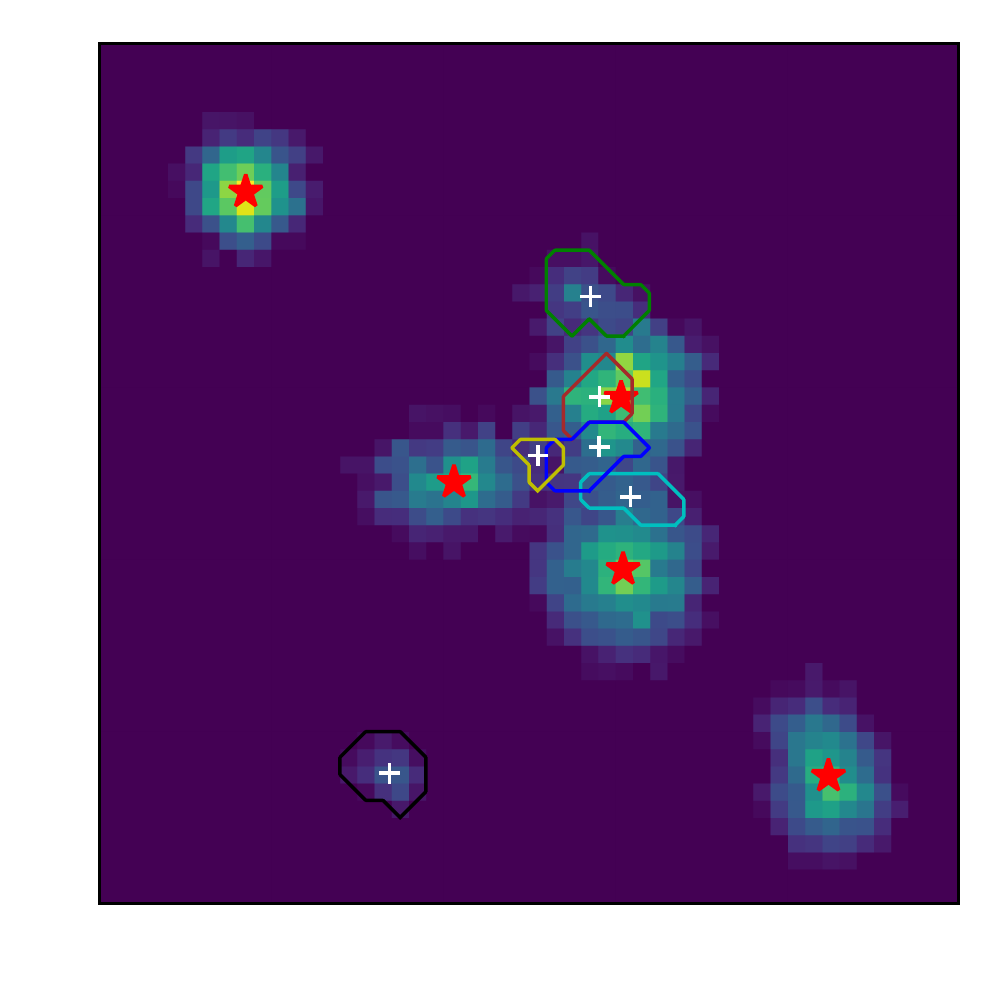}}
	\caption{The PP view of some of the local regions and local centers. Each outline delineates the boundary of a local region, the white plus signs denote the local centers, and the red asterisks denote the clump centers.}
	\label{Fig_Local_Regions}
\end{figure}

\subsection{Local region segmentation based on gradient}\label{Sec2.3}
The approach of segmenting the signal region based on local gradients has been widely applied and has been shown to be effective \citep{CUPID2, ConBased}. FacetClumps utilizes a similar approach to identify local regions and then merges them together. It differs from the FellWalker in that, when the path reaches a local maximum, FacetClumps does not continue to search for voxels with higher intensity in a larger neighborhood. It differs from the ConBased in that FacetClumps uses the centroid of a local region as the reference position for subsequent clustering.

Starting at any signal region, a random voxel is selected as the center of a box whose size is $3\times3\times3$ voxels, then calculate the gradients of intensity between it and each neighbor in the box. Move to the neighbor with the highest gradient, which is regarded as the center for the next movement. The way of finding the next central location is described as (\ref{local region}). The moving progress is repeated until the highest gradient is less than zero. 

\begin{equation}\label{local region}
\begin{aligned}
\it{\boldsymbol{u_{n}}=\begin{cases}
	\boldsymbol{u_0},&\text{if $max(I[\boldsymbol{u_i}]-I[\boldsymbol{u_0}])<0$} \\
	\boldsymbol{U}[i_{max(I[\boldsymbol{u_i}]-I[\boldsymbol{u_0}])}],&\text{else}
	\end{cases}}
\end{aligned}
\end{equation}

\noindent where, $\boldsymbol{u_n}$ represents the next central coordinate. $max()$ stands for the maximal value of its arguments. $I$ is intensity map, $\boldsymbol{U}$ is the set of neighbor coordinates in a box. $\boldsymbol{u_0}$ is the central location of $\boldsymbol{U}$, $\boldsymbol{u_i}\in \boldsymbol{U},i=1,2,\ldots,26$. $I[\boldsymbol{u_i}]$ is the intensity of $\boldsymbol{u_i}$, and $I[\boldsymbol{u_0}]$ is the intensity of $\boldsymbol{u_0}$. $\boldsymbol{U}[i_{max(I[\boldsymbol{u_i}]-I[\boldsymbol{u_0}])}]$ is the neighbor coordinate corresponding to the maximum intensity. 

All voxels that have been traversed the movement process are recorded as a path, and the end of each path is recorded as a local maximum. Once a local maximum is reached, new local maxima are sought from the unvisited voxels until all voxels in all signal regions have been searched. The voxels on the paths leading to the same local maximum form a local region, and the connectivity type of each local region meets $Type \uppercase\expandafter{\romannumeral3}$. The centroid of each local region is calculated and recorded as a local center. By applying this approach, it is possible to obtain all local centers and regions for each signal region. Examples of partial local centers and local regions for signal regions 2 and 4 in Figure \ref{Fig_Morphology}(d) are shown in Figure \ref{Fig_Local_Regions}.

\subsection{Minimum distance clustering of local regions based on connectivity}\label{Sec2.4}
The following process relies on the signal regions, clump centers, local centers, and local regions, the purpose of which, is to let local regions be clustered to their corresponding clump centers with precision. A local region is deemed to match a clump center if the coordinates of the clump center fall within the region. In PPV space, the distance between two voxels is measured by (\ref{PPV Dist}).
\begin{equation}\label{PPV Dist}
\it{Dist=\sqrt{\frac{\Delta x^2+\Delta y^2}{FwhmBeam^2}+\frac{\Delta z^2}{VeloRes^2}}}
\end{equation}
\noindent where, $Dist$ represents the distance between $(x_1,y_1,z_1)$ and $(x_2,y_2,z_2)$, $FwhmBeam$ is the full width at half maximum (FWHM) of the instrument beam, and $VeloRes$ is the velocity resolution of the instrument. 

Firstly, the clump centers match with signal regions to filter out local regions that cannot be clustered to any one clump center. If the match fails, the signal region and its local regions and local centers will be removed. This commonly occurs when the signal region is composed of noise, as shown by the black contour in Figure \ref{Fig_Local_Regions}. 

Secondly, clump centers match with local regions to obtain filtered clump centers and target regions. If a match is successful, the matched local region is re-recorded as the target region of the matched clump center. The matched local region and its local center will be removed from the record obtained in Section \ref{Sec2.3}. If a match fails, this means that a local region may correspond to more than one clump center, due to the difference in connectivity types between the maximum region shown in Figure \ref{Fig_Hook_Centers}(a) and the local region shown in Figure \ref{Fig_Local_Regions}. The local region matched by the failed clump center has already been matched with another clump center and become a target region. In this scenario, to choose a more appropriate clump center for the target region, one of the failed clump center and the matched clump center that is closer to the local center of the target region will be retained, and the other one will be removed. After all the matches have been made, the remaining clump centers are recorded as filtered clump centers, each of which is associated with a target region. 

\begin{figure}
	\centering
	\centerline{\includegraphics[width=7.8in]{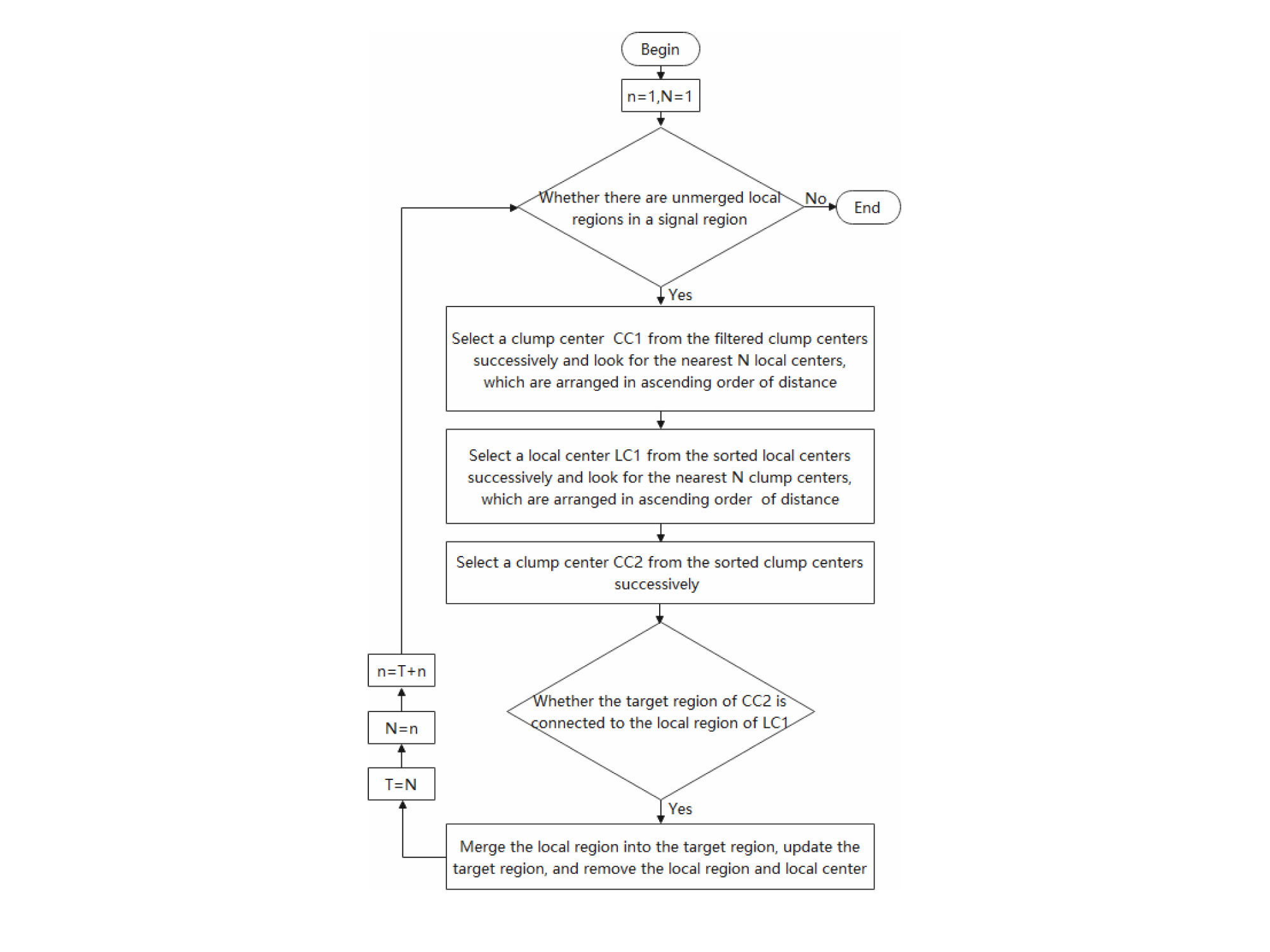}}
	\caption{Diagram of minimum distance clustering of local regions based on connectivity. $N$ is a continuously updated member of the Fibonacci sequence, and $n$ and $T$ are temporary variables. Connectivity is the necessary condition for merging. In the case of connectivity, a filtered local region is merged into a target region whose clump center is closest to its local center.}
	\label{Fig_Three_Loop}
\end{figure}

\begin{figure}
	\centering
	\centerline{\includegraphics[width=2.8in]{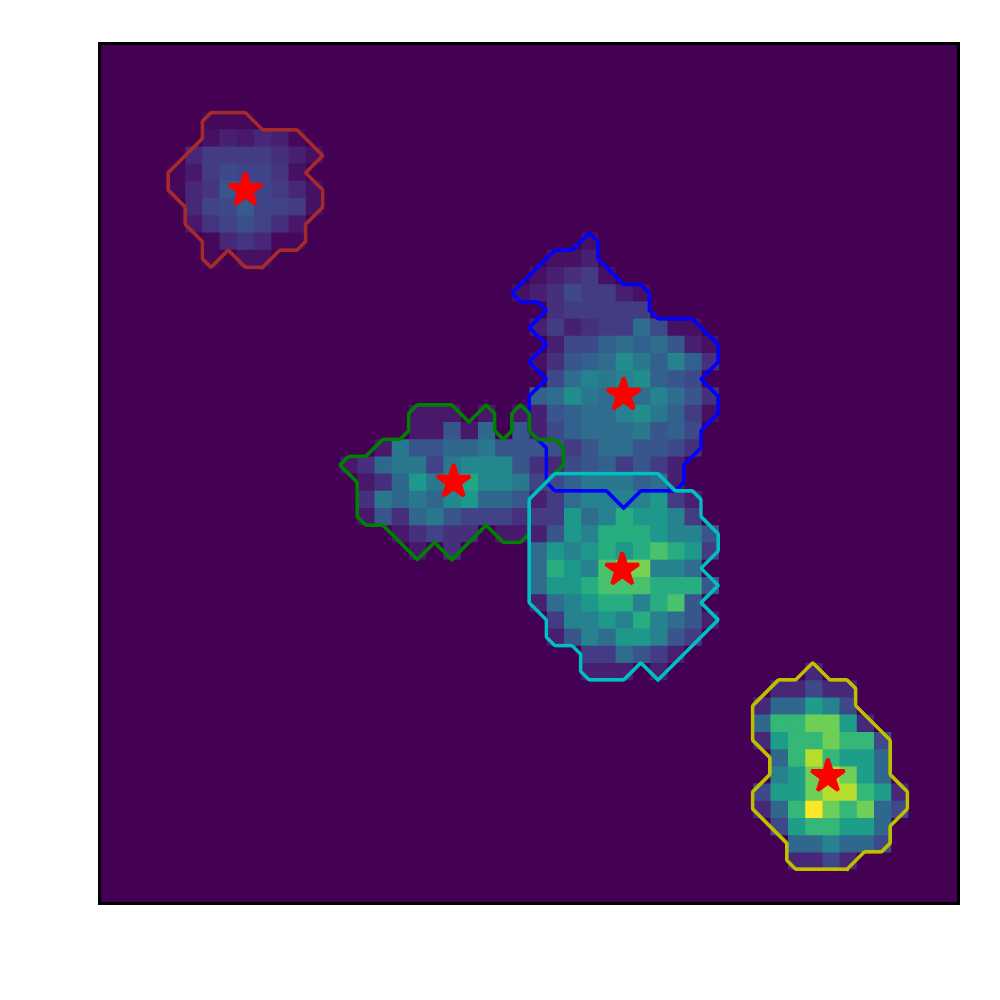}}
	\caption{The PP view of the integral mask of the detected clumps. Each outline delineates the boundary of a clump, and the red asterisks denote the clump centers.}
	\label{Fig_Result_2D}
\end{figure}

\begin{figure}
	\centering
	\centerline{\includegraphics[width=3.4in]{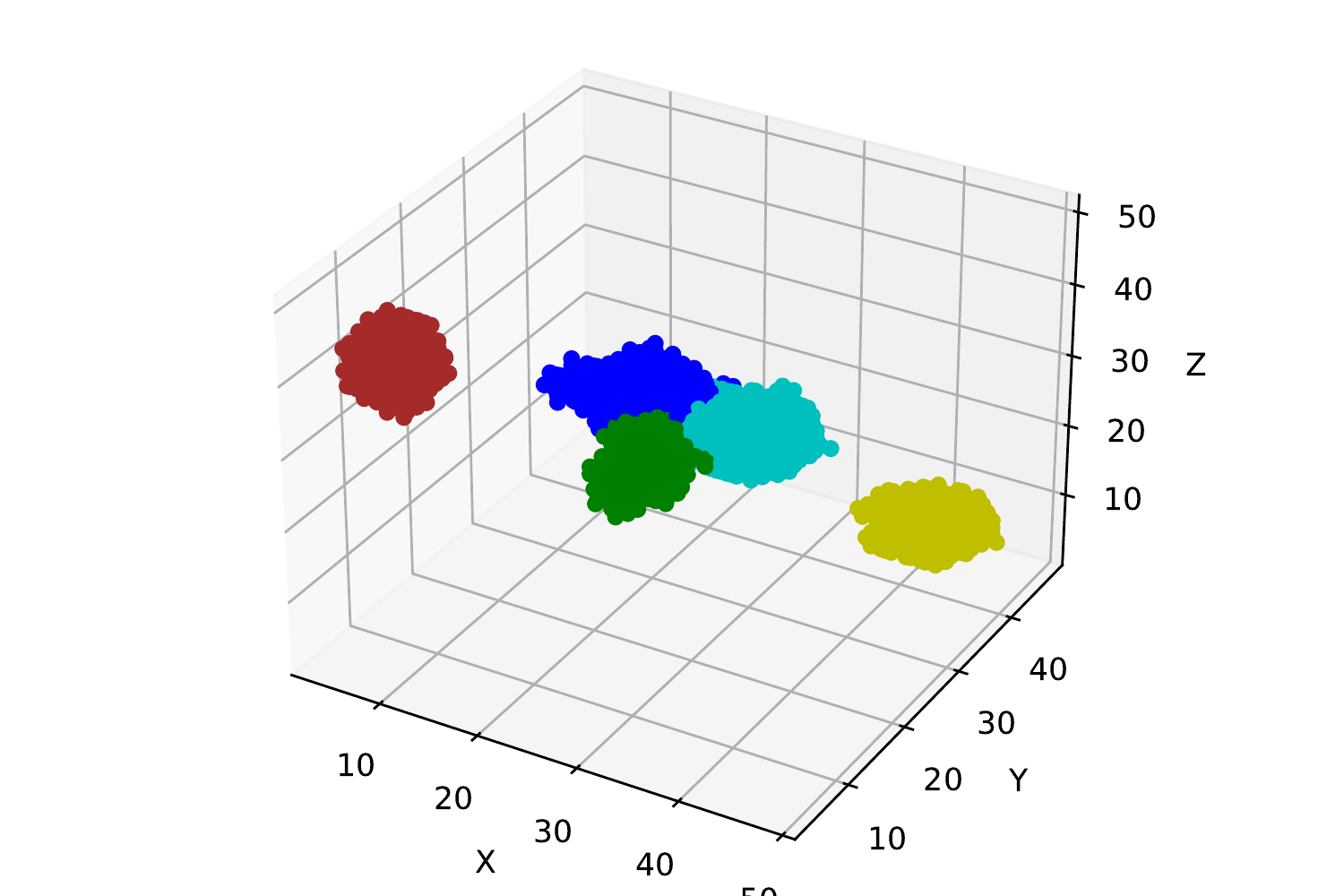}}
	\caption{The PPV view of the detected clumps. Different coloured voxels denote different clumps.}
	\label{Fig_Result_3D}
\end{figure}

Thirdly, the filtered local regions are merged to the target regions. Connectivity is the necessary condition for merging local regions. In the connected case, a filtered local region is merged into a target region whose clump center is closest to its local center. Since different signal regions are disconnected, the following operations are performed in each signal region. When there are unmerged local regions in any signal region, we introduce a three-layer main loop to merge these local regions, as shown in Figure \ref{Fig_Three_Loop}. In the first layer, a clump center of filtered clump centers, denoted as $CC1$, is selected in the order obtained by recursion to search for the $N$ nearest local centers, which are arranged in ascending order of distance. In the second layer, a local center of the sorted local centers, denoted as $LC1$, is successively selected to search for the $N$ nearest clump centers, which are also arranged in ascending order of distance. In the third layer, a clump center of the sorted clump centers, denoted as $CC2$, is successively selected to judge whether the local region of $LC1$ is connected to the target region of $CC2$, with the connectivity type being $Type \uppercase\expandafter{\romannumeral3}$. If the local region is connected, it will be merged into the target region of $CC2$ and the target region will be updated. The current local center $LC1$ and its local region will be removed from the record. If not, the local center $LC1$ and its local region are retained and await the next judgment. ($N$ is a continuously updated member of the Fibonacci sequence and its initial value is one, which corresponds to the beginning of the sequence. Following each three-layer iteration, the position index of $N$ increases by one. The updated values of $N$ can also be inferred from Figure \ref{Fig_Three_Loop}. If there are fewer items than $N$, all of them will be searched.) This approach can ensure the stability and accuracy of merging while also improving the efficiency of FacetClumps. 

The target regions that ultimately satisfy the condition for the parameter $SRecursionLBV$ will be identified. At this point, we obtain all the clump centers and corresponding regions. An integral mask of the result is shown in Figure \ref{Fig_Result_2D}, where each outline delineates the boundary of a clump. PPV graph is shown in Figure \ref{Fig_Result_3D}, in which different coloured voxels denote different clumps. It is evident that the separated clumps are properly delineated and the connected clumps are accurately segmented. 

\begin{table}
	\centering
	\caption{The input parameters of FacetClumps.}
	\begin{tabular}{|p{2.2cm}|p{5.5cm}|}\hline
		Parameters&Explanation\\\hline
		RMS&The noise RMS of the data.\\\hline
		Threshold&The minimum intensity used to truncate the signals.\\\hline
		SWindow&The scale of the window function, in pixels.\\\hline
		KBins&The coefficient used to calculate the number of eigenvalue bins, see formula (\ref{FormulaKBins}).\\\hline
		FwhmBeam&The FWHM of the instrument beam, in pixels.\\\hline
		VeloRes&The velocity resolution of the instrument, in channels.\\\hline
		SRecursionLBV&The minimum area of a region in the spatial direction (SRecursionLB) and the minimum length of a region in the velocity channels (SRecursionV) when a recursion terminates. The region of a clump also need to satisfy the conditions. See formula (\ref{SRecursionLBV}), in pixels. 
		\\\hline
	\end{tabular}
	\begin{tablenotes}
		\item \textbf{Note.} SRecursionLBV consists of SRecursionLB and SRecursionV, i.e. [SRecursionLB, SRecursionV]. The relationship between $SRecursionLBV$ and $FwhmBeam$ and $VeloRes$ is presented in Appendix \ref{Resample}.
	\end{tablenotes}
	\label{InputPar}
\end{table}

\begin{table}
	\centering
\caption{The output parameters of a clump.}
	\begin{tabular}{|p{2.2cm}|p{5.5cm}|}\hline
		Parameters&Explanation\\\hline
		PeakI&The peak intensity.\\\hline
		$\boldsymbol{PeakL}$&A vector of the peak location, in pixels.\\\hline
		$\boldsymbol{Cen}$&A vector of the clump center, see formula (\ref{Cen}), in pixels.\\\hline
		$\boldsymbol{Size}$&A vector of the sizes, see formula (\ref{size}), in pixels.\\\hline
		Sum&The sum of the voxels intensity within the clump.\\\hline
		Volume&The total number of voxels within the clump, in pixels.\\\hline
		Angle&The angle of the clump, see formula (\ref{angle}), in degree.\\\hline
		Edge&Whether the clump touches the edges, 1-Yes, 0-No.\\\hline
	\end{tabular}
	\begin{tablenotes}
		\item \textbf{Note.} The output tables include tables in both the pixel coordinate system and the WCS coordinate system. The units in the table under WCS coordinate system are consistent with those in the header file. 
	\end{tablenotes}
	\label{CoreTable}
\end{table}

\subsection{The parameters}
All the output parameters necessary for the clump tables are computed from the input data, the clump centers, and regions. The regional information is identified by a mask, and the index (starting from one) of each clump corresponds to the same number in the mask. The input parameters of FacetClumps are listed in Table \ref{InputPar}, while the output parameters of a single clump are presented in Table \ref{CoreTable}. 

The size \citep{CUPID2,ConBased} of a clump is defined as (\ref{size}):

\begin{equation}\label{size}
\it{size=\sqrt{\frac{\sum I_i\cdot \boldsymbol{u_i}^2}{\sum I_i}-\bigg(\frac{\sum I_i\cdot \boldsymbol{u_i}}{\sum I_i}\bigg)^2}}
\end{equation}

\noindent where $I_i$ is the intensity of voxel $i$ minus the minimum intensity in the clump, and $\boldsymbol{u_i}$ is the coordinate of voxel $i$. 

The angle and axis ratio of a clump are calculated by diagonalizing the moment of inertia matrix \citep{Angle}, as shown in equation (\ref{angle}):

\begin{equation}\label{angle}
\it{R_{-\theta}\begin{pmatrix}
	\sum T_{i}\alpha_{i}^2 & -\sum T_{i}\alpha_{i}\beta_{i} \\
	-\sum T_{i}\alpha_{i}\beta_{i} & \sum T_{i}\beta_{i}^2 \\
	\end{pmatrix}R_\theta=\begin{pmatrix}
	S_{xx} & 0 \\
	0 & S_{yy}  \\
	\end{pmatrix}}
\end{equation}

\noindent where $T$ is the velocity-integrated intensity map of a clump, $T_{i}$ is the intensity at coordinate $\boldsymbol{u_i}$, $\boldsymbol{u_i}=\{x, y\}$, $\alpha_{i}$ and $\beta_{i}$ are the Euclidean distances from the clump center to the coordinate $\boldsymbol{u_i}$ in the $l-$ and $b-$ direction, respectively. $R_\theta$ is a rotation matrix with rotation angle $\theta$, which is the angle between the major axis and the negative direction of $l$, ranging from $-90^\circ$ to $90^\circ$, and $\theta$ along the positive direction of $b$ is $90^\circ$. The axis ratio of a clump is given by the square root of the ratio of the lengths of its major and minor axes, i.e. $(S_{xx}/S_{yy})^{1/2}$ or $(S_{yy}/S_{xx})^{1/2}$.

\section{Experiments and discussions}
\subsection{Evaluation Metrics}
To quantitatively assess the performance of different algorithms, we introduce the evaluation metrics Recall rate ($R$, \ref{R}), Precision rate ($P$, \ref{P}), $F_1-score$ ($F_1$, \ref{F1}), location error ($\Delta X$, \ref{Delta X}), flux fluctuation ($\Delta Flux$, \ref{Delta Flux}), and regional intersection-over-union ($IOU$, \ref{IOU}) \citep{ConBased}.

\begin{equation}\label{R}
\it{R=\frac{TP}{TP+FN}}
\end{equation}

\begin{equation}\label{P}
\it{P=\frac{TP}{TP+FP}}
\end{equation}
\begin{equation}\label{F1}
\it{F_1=\frac{2TP}{2TP+FP+FN}}
\end{equation}

\begin{equation}\label{Delta X}
\it{\Delta X =\frac{1}{N}\sum_{i}^{N}||\boldsymbol{Cen_{d_i}}-\boldsymbol{Cen_{s_i}}||,i=1,2,...,N}
\end{equation}

\begin{equation}\label{Delta Flux}
\it{\Delta Flux=\frac{1}{N}\sum_{i}^{N}\frac{Sum_{d_i}-Sum_{s_i}}{Sum_{s_i}},i=1,2,...,N}
\end{equation}

\begin{equation}\label{IOU}
\it{IOU=\frac{1}{N}\sum_{i}^{N}\frac{\boldsymbol{Region_{d_i}}\bigcap \boldsymbol{Region_{s_i}}}{\boldsymbol{Region_{d_i}}\bigcup \boldsymbol{Region_{s_i}}},i=1,2,...,N}
\end{equation}

\noindent where $TP$ is the number of clumps detected correctly, $FN$ is the number of missed clumps, and $FP$ is the number of clumps detected wrongly. $\boldsymbol{Cen}$ represents the clump center, $Sum$ represents the flux of a clump, $\boldsymbol{Region}$ represents the region of a clump, and $N$ is the number of clumps detected correctly. $s_i$ is the $i$th simulated parameter, and $d_i$ refers to the corresponding parameter that is detected correctly. If $\Delta X$ between a detected clump and a simulated clump is not greater than 2 voxels, the clump is considered to be correctly detected.

The metrics $R$, $P$, and $F_1$ are utilized to assess the completeness and precision of the detections. $\Delta X$ and $\Delta Flux$ reflect the location error and flux loss of the sources, respectively. $IOU$ measures the similarity between the simulated regions and the detected regions, which is crucial for calculating physical parameters, analyzing the morphological characteristics, and other applications. The Signal-to-Noise Ratio (SNR) is defined as the ratio of peak to the noise level, and the error bar is the standard deviation of the statistic in each SNR interval. 

\subsection{Experiments with simulated clumps}\label{sec:3.2}
\subsubsection{Simulated clumps}
The model presented in \cite{ConBased} is adopted to generate 100 PPV data cubes with $100\times 100\times 100$ voxels and each data cube contains 100 Gaussian clumps. Gaussian noise with a RMS of 0.22 K is added to mimic realistic observational data. The peak intensities range from 0.44 to 4.4 K, ensuring a distribution of SNR between 2 and 20. Sizes vary between 2 and 4 voxels and rotation angles vary randomly. The distance between the truncated edge and the clump center is 3 times the sizes of the clump. 

\begin{figure}
	\centering
	\centerline{\includegraphics[width=3.5in]{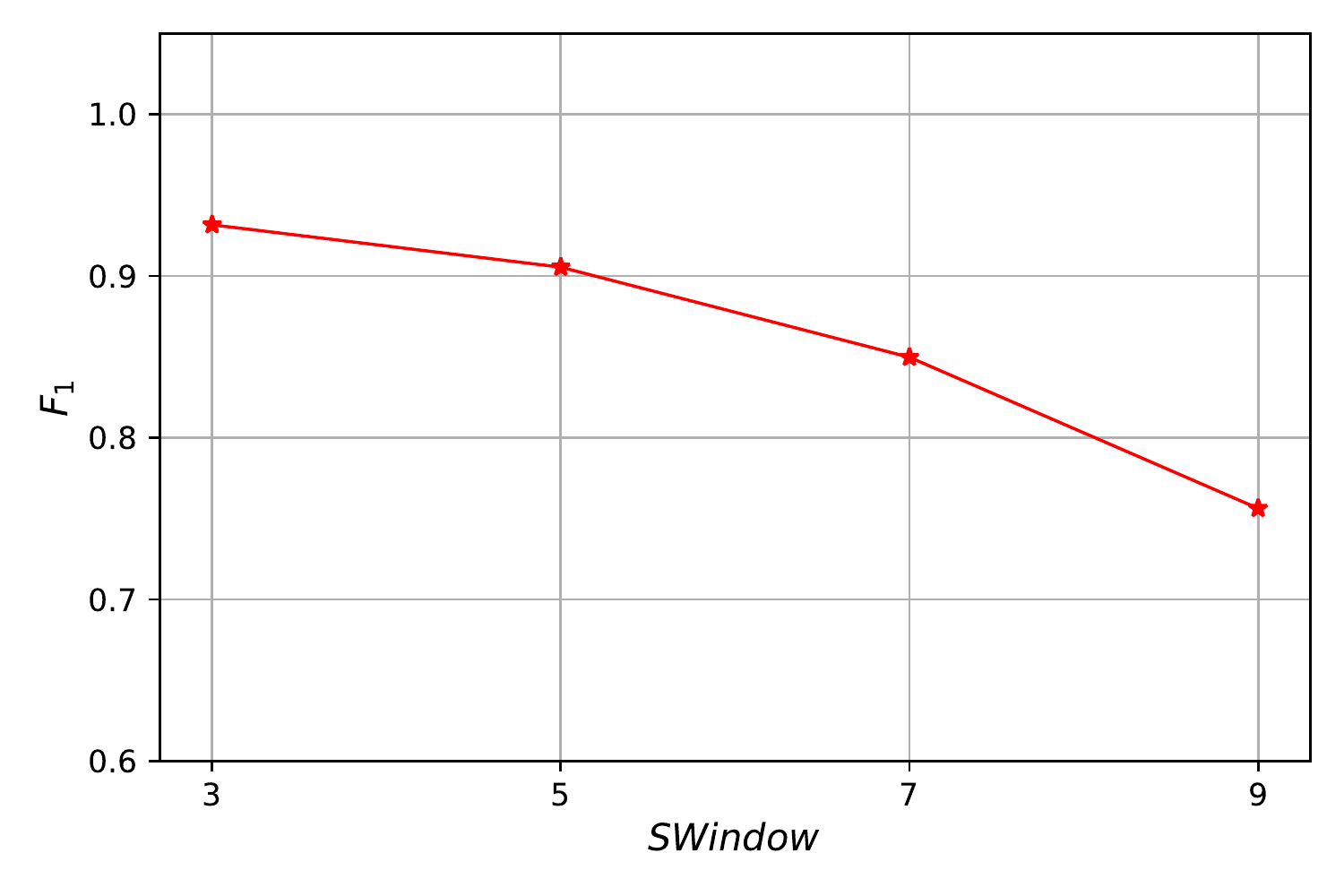}}
	\caption{The variation trend graph of the mean value of $F_1$ with respect to $SWindow$.}
	\label{Fig_Pra_SWindow}
\end{figure}

\begin{figure}
	\centering
	\centerline{\includegraphics[width=3.5in]{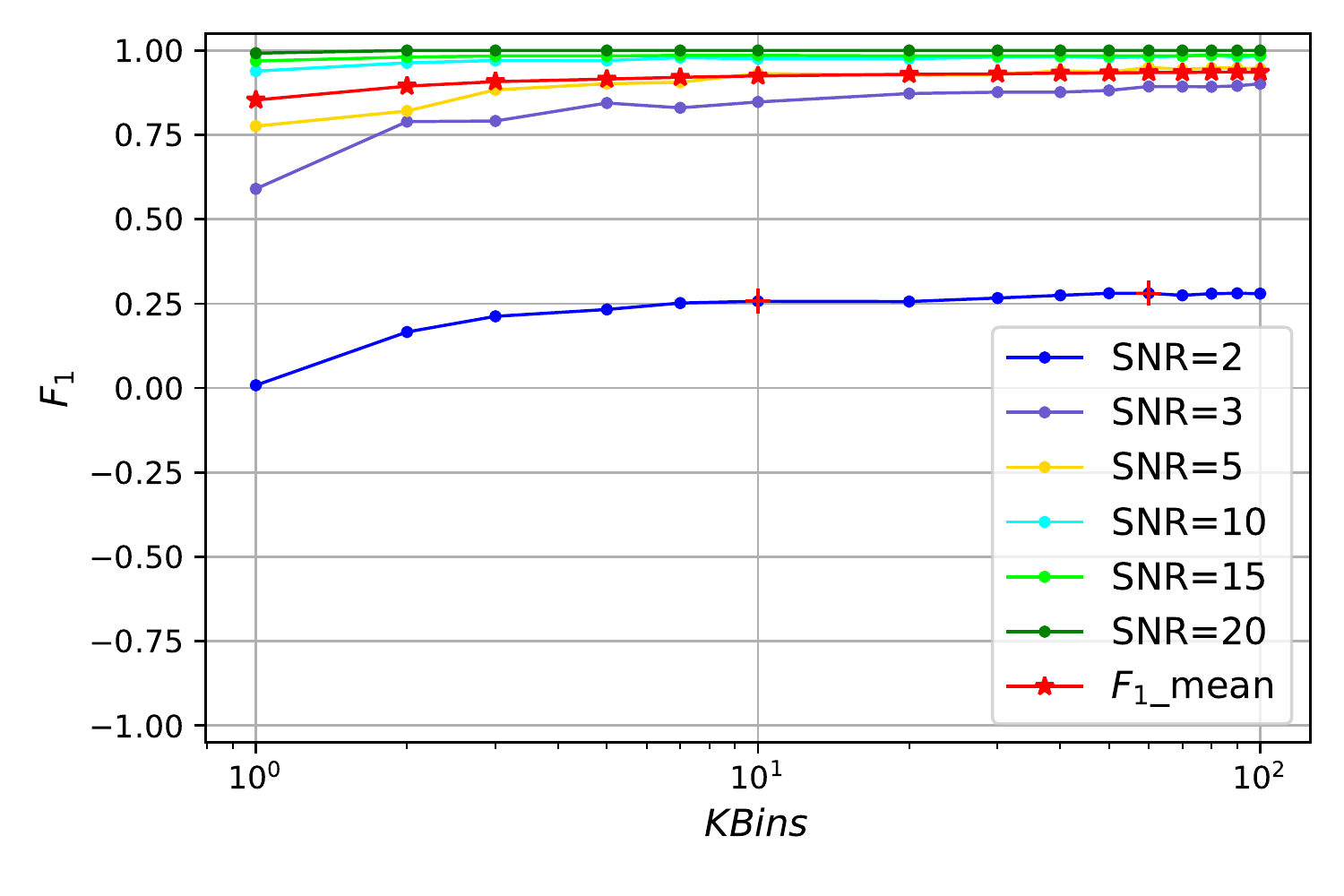}}
	\caption{The variation trend graph of $F_1$ with respect to $KBins$ under different SNRs. The red curve is the mean value of $F_1$ for all SNRs. }
	\label{Fig_Pra_KBins}
\end{figure}

\begin{figure*}
	\centering
	\includegraphics[width=7in]{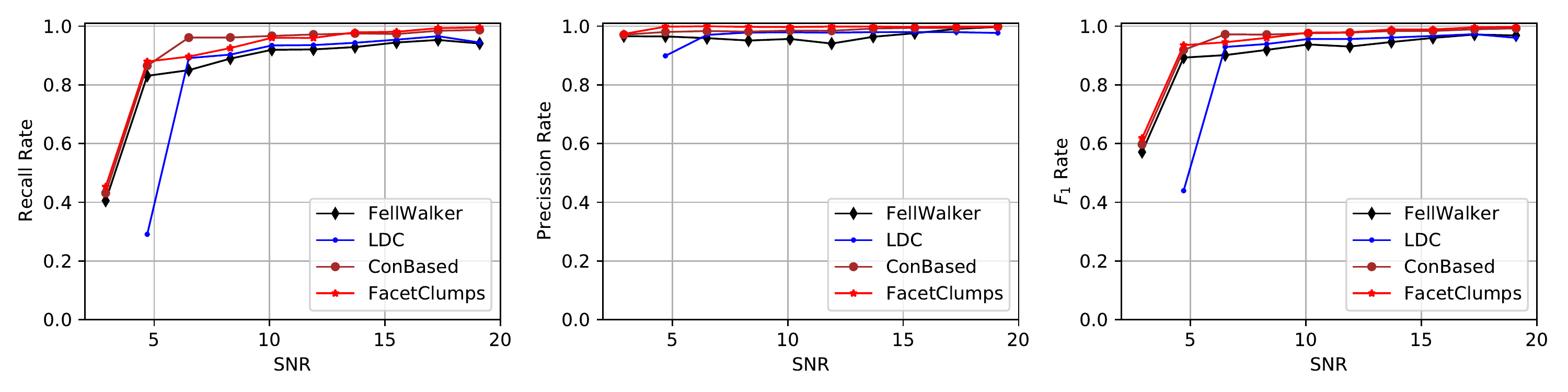}
	\caption{The statistics of $R$ (left), $P$ (middle), and $F_1$ (right) curvers of FellWalker, LDC, ConBased, and FacetClumps for the simulated clumps. Black being FellWalker, blue being LDC, brown being ConBased and red being FacetClumps.}
	\label{Fig_RPF}
\end{figure*}

\begin{figure*}
	\centering
	\includegraphics[width=7in]{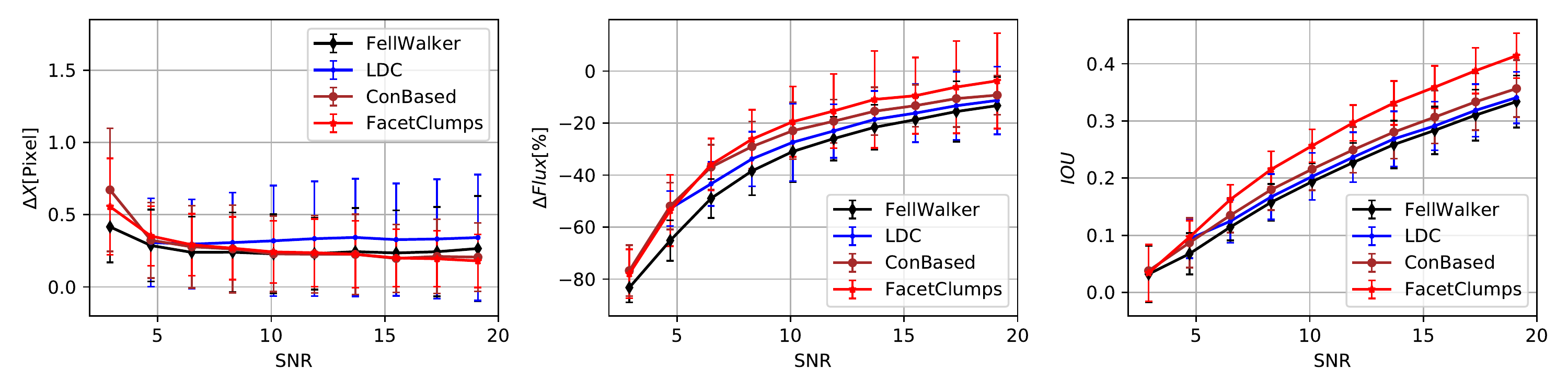}
	\caption{The statistics of $\Delta X$ (left), $\Delta Flux$ (middle), and $IOU$ (right) curvers of FellWalker, LDC, ConBased, and FacetClumps for the simulated clumps.}
	\label{Fig_DFI}
\end{figure*}

\subsubsection{Determine appropriate values for the parameters of FacetClumps}\label{sec:3.2.2}
To determine appropriate parameter values for FacetClumps, we investigate the influence of two parameters, namely $SWindow$ (the scale of the window function) and $KBins$ (the coefficient used to calculate the number of eigenvalue bins), on $F_1$ in the simulated clumps. In addition, we also consider the instrument-related parameter $SRecursionLBV$, which is discussed in Appendix \ref{Resample}. The simulated clumps generated by the simulation method proposed by \cite{ConBased} represent the theoretically detectable limit of all algorithms mentioned in this paper, i.e., these algorithms are only possible to detect clumps with valid peaks. Thus these experiments serve as an effective reference for choosing parameter values. As $F_1$ is a comprehensive metric that evaluates both recall and precision rates, it is used as the evaluation metric in this study.

Figure \ref{Fig_Pra_SWindow} shows the variation trend of the average $F_1$ with respect to the parameter $SWindow$. As $SWindow$ increases from 3 to 9, $F_1$ gradually decreases from 0.932 to 0.756, indicating that a smaller $SWindow$ value leads to better detection performance. Therefore, the default value of $SWindow$ is 3 (the minimum scale), and the recommended values are 3, 5, and 7. When dealing with poor quality observational data or large target sources, it is advisable to increase the value of $SWindow$.

Figure \ref{Fig_Pra_KBins} shows the variation trend of $F_1$ with respect to the parameter $KBins$ under different SNRs. $KBins$ varies from 1 to 100. The red line is the mean value of $F_1$, which ranges from 0.853 to 0.936. When SNR is less than 5, $F_1$ is much lower than the average value, and the curve is a bit volatile, indicating that $KBins$ has a small effect on the detection of faint clumps. When $KBins$ is greater than 10 and SNR is greater than 5, $F_1$ are almost straight lines, indicating that the change of $KBins$ has a little effect on the detection of clumps with high SNR. The mean value of $F_1$ increases gradually with the increase of $KBins$, and then tends to plateau, indicating that larger $KBins$ detect more clumps correctly. 

As $KBins$ increases from 1 to 10, there is a significant increase in $F_1$, which then gradually stabilizes as $KBins$ continues to increase. As $KBins$ increases, $F_1$ for SNR of 2 first increases and then decreases, with the turning point occurring when $KBins$ reaches 60. Therefore, the recommend values of $KBins$ range from 10 to 60, with a default value of 35. The recommended and default values for all parameters are shown in appendix Table \ref{Facet parameters}. 

\begin{figure*}
	\centering
\centerline{\includegraphics[width=10in]{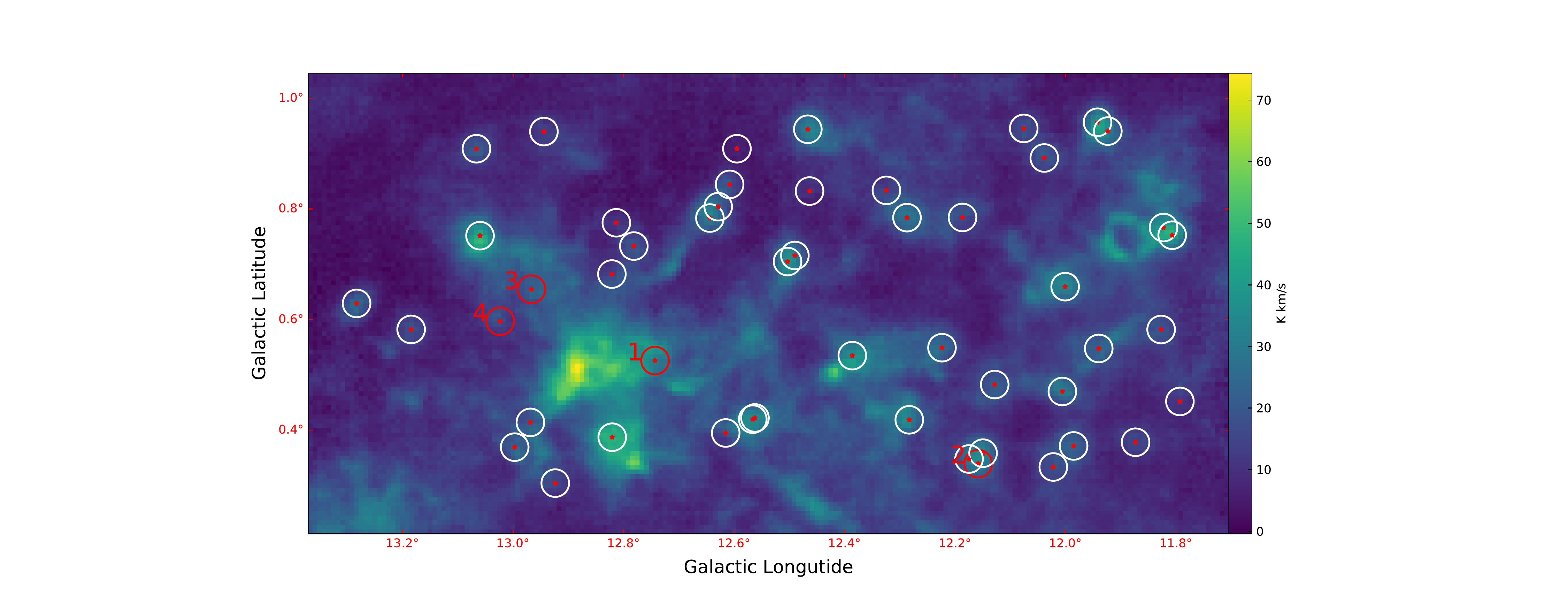}}
	\caption{An example of the synthetic data of Data1. It is superimposed by the $^{13}CO$ emission within $11.7^{\circ} \leq l \leq 13.4^{\circ}$, $0.22^{\circ} \leq b \leq 1.05^{\circ}$ and 5 km s$^{-1}$ $\leq v \leq$ 35 km s$^{-1}$ and simulated clumps. The cube size is $180\times100\times200$ voxels. The number of simulated clumps is 50, evenly distributed in peak intensities ranging from 0.44 to 16.8 K, sizes ranging from 2 to 4 voxels, and with random angles. The red asterisks denote the central locations of the simulated clumps. The white circles denote the simulated clumps detected by FacetClumps and the red circles denote the missed simulated clumps. In total, 680 clumps are detected.}
	\label{Fig_EData}
\end{figure*}

\begin{figure*}
	\centering
	\vspace{0cm}
	\begin{minipage}[t]{0.24\textwidth}
		\centering
		\centerline{\includegraphics[width=2in]{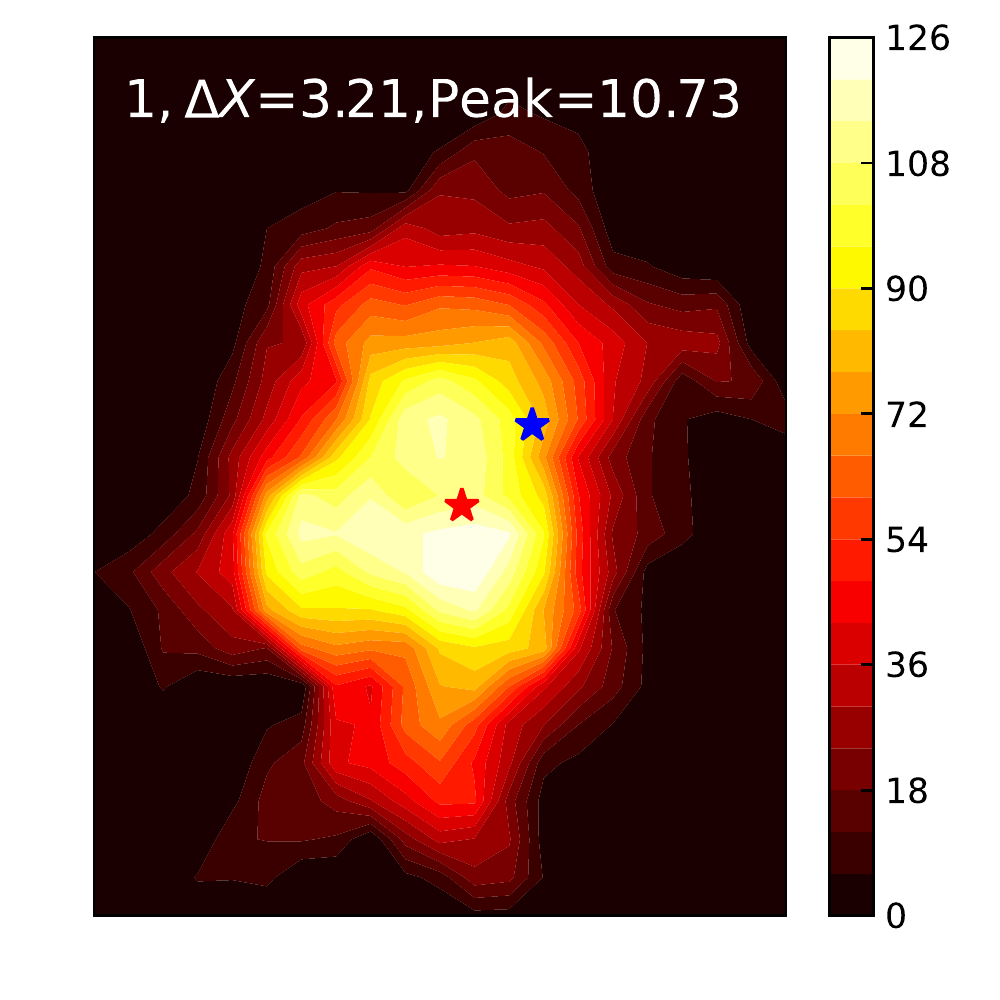}}
\end{minipage}\begin{minipage}[t]{0.24\textwidth}
		\centering
		\centerline{\includegraphics[width=2in]{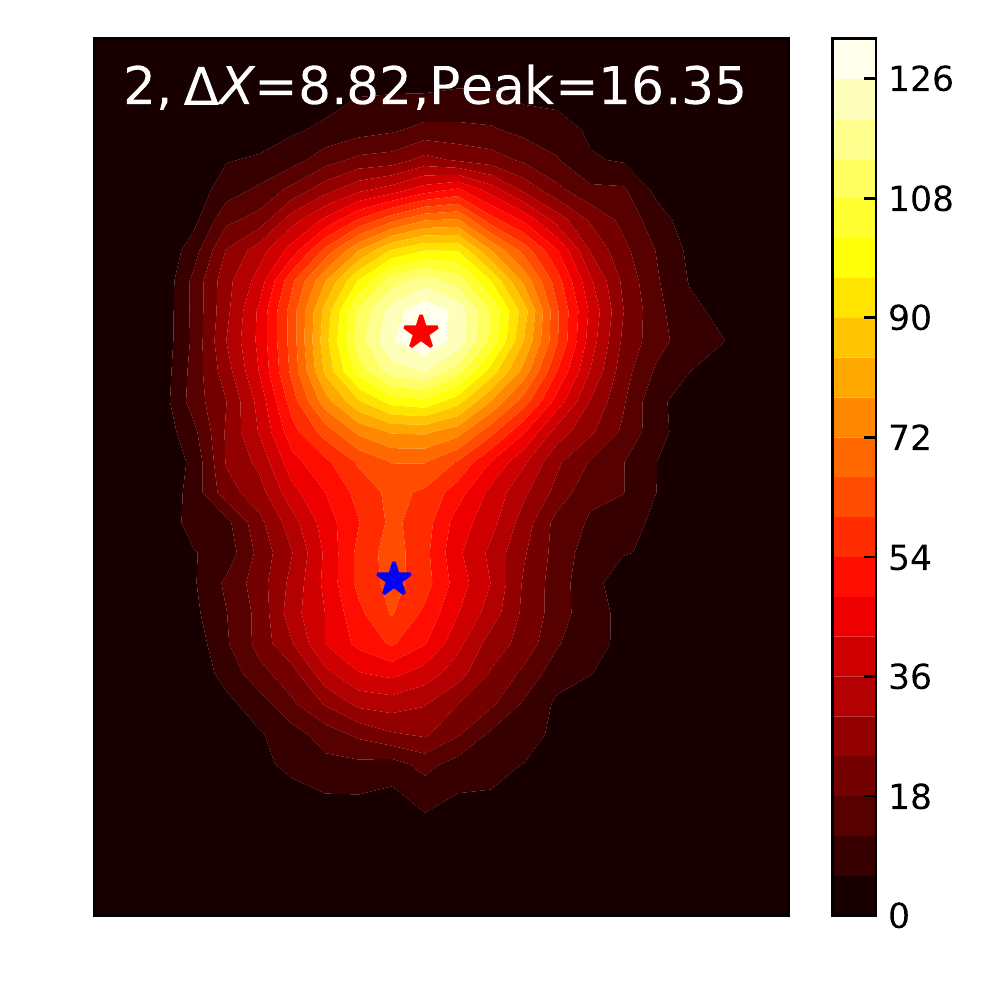}}
\end{minipage}\begin{minipage}[t]{0.24\textwidth}
		\centering
		\centerline{\includegraphics[width=2in]{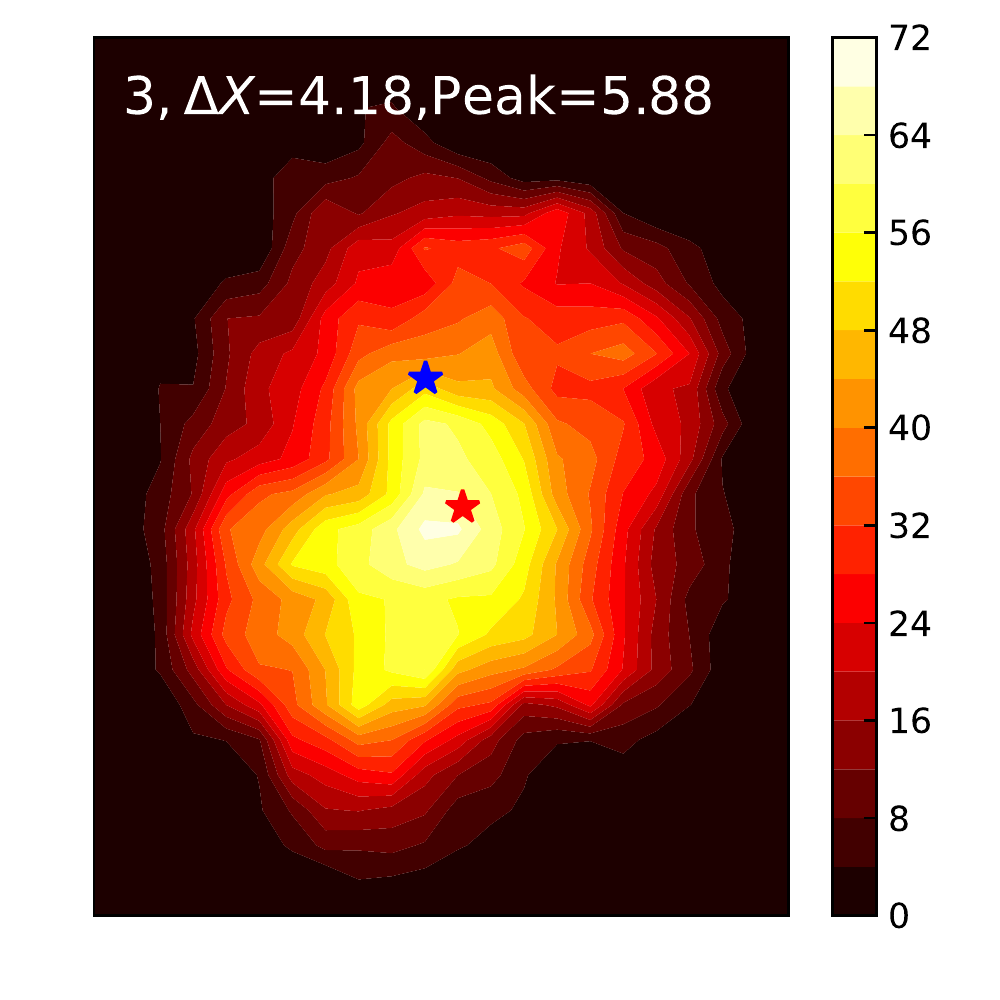}}
\end{minipage}\begin{minipage}[t]{0.24\textwidth}
		\centering
		\centerline{\includegraphics[width=2in]{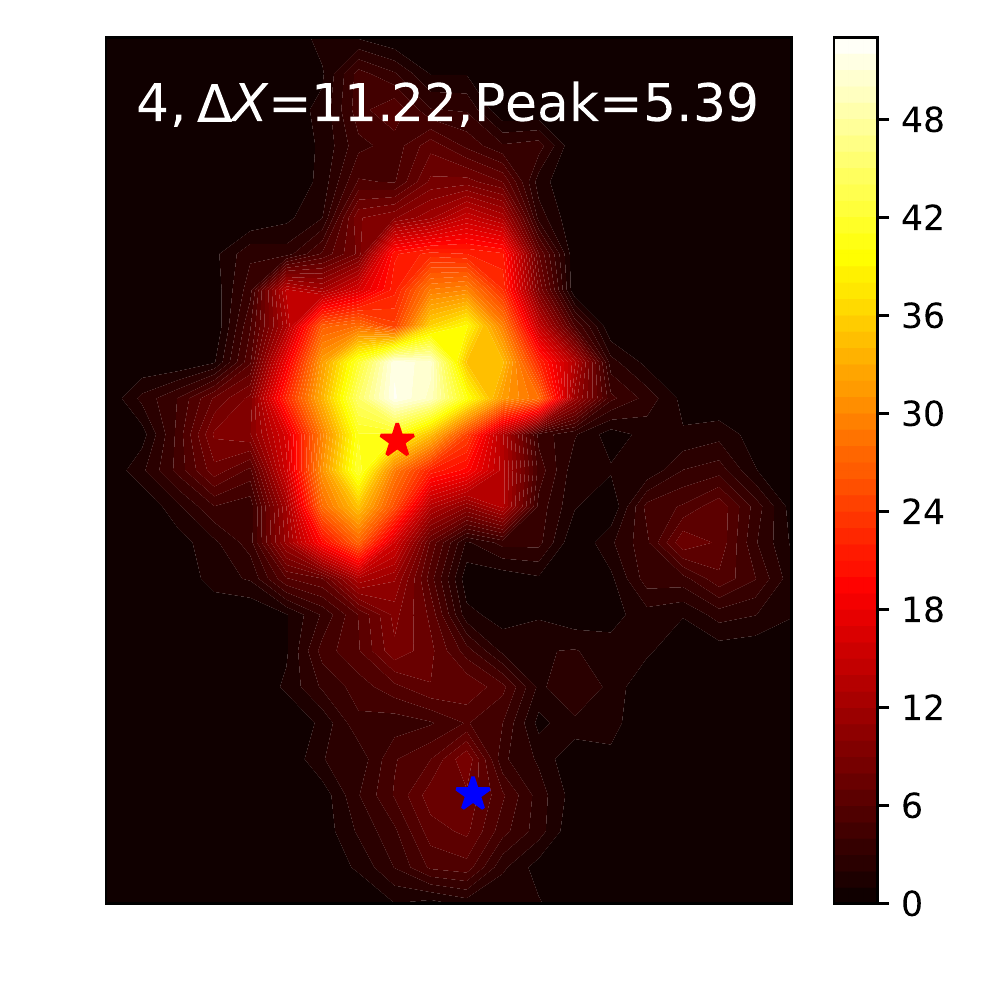}}
\end{minipage}

	\begin{minipage}[t]{0.24\textwidth}
		\centering
		\centerline{\includegraphics[width=2in]{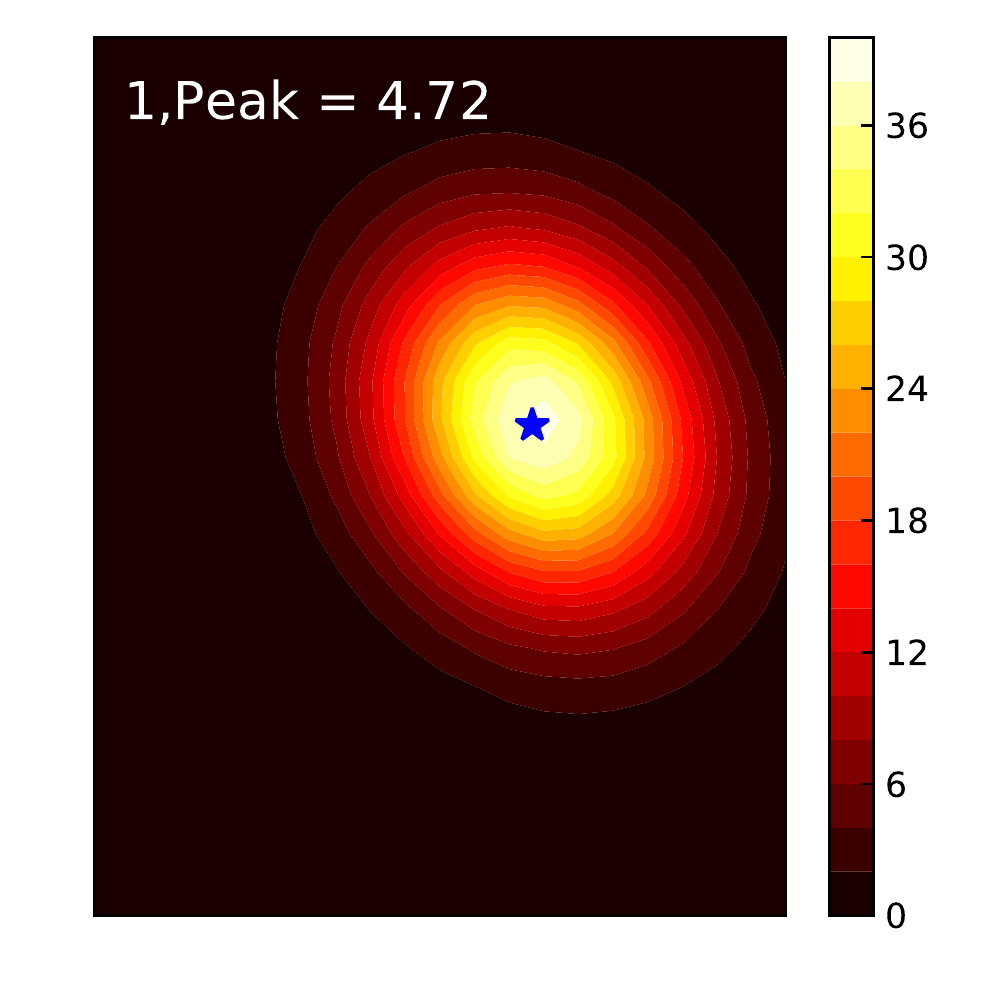}}
\end{minipage}\begin{minipage}[t]{0.24\textwidth}
		\centering
		\centerline{\includegraphics[width=2in]{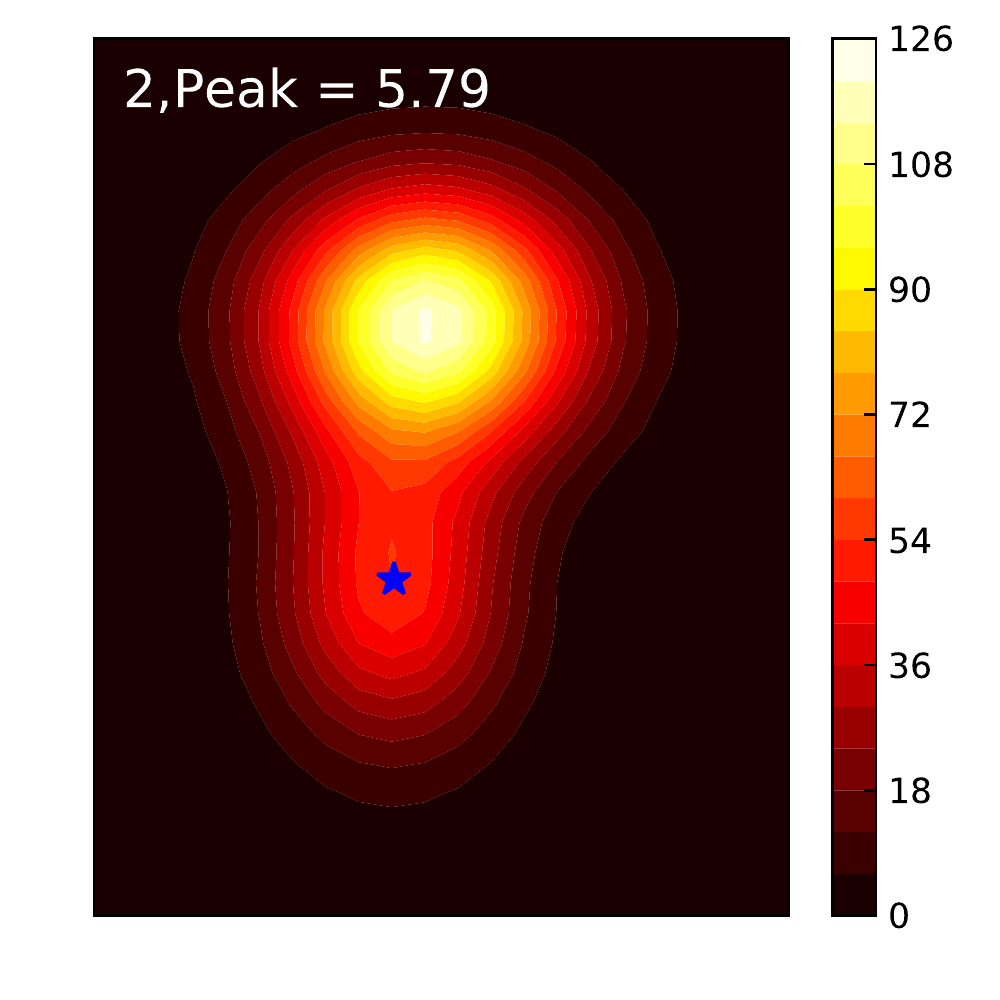}}
\end{minipage}\begin{minipage}[t]{0.24\textwidth}
		\centering
		\centerline{\includegraphics[width=2in]{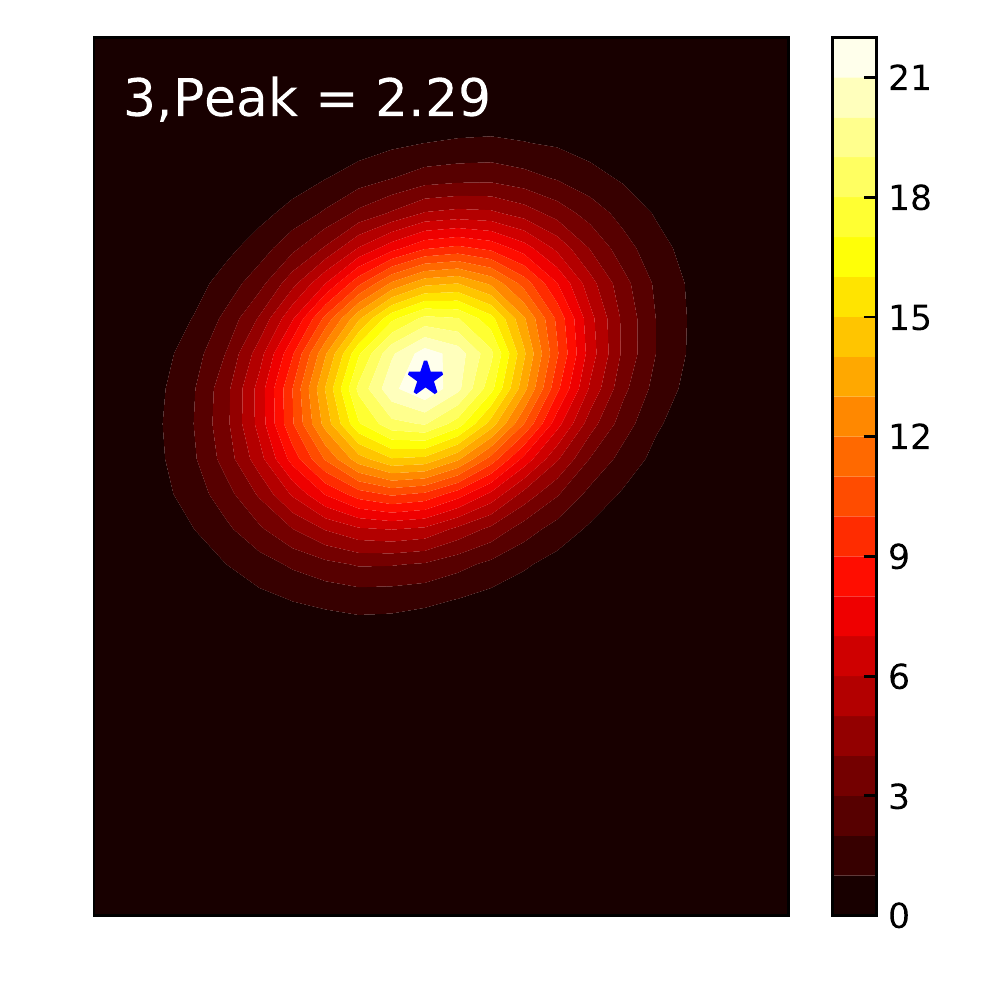}}
\end{minipage}\begin{minipage}[t]{0.24\textwidth}
		\centering
		\centerline{\includegraphics[width=2in]{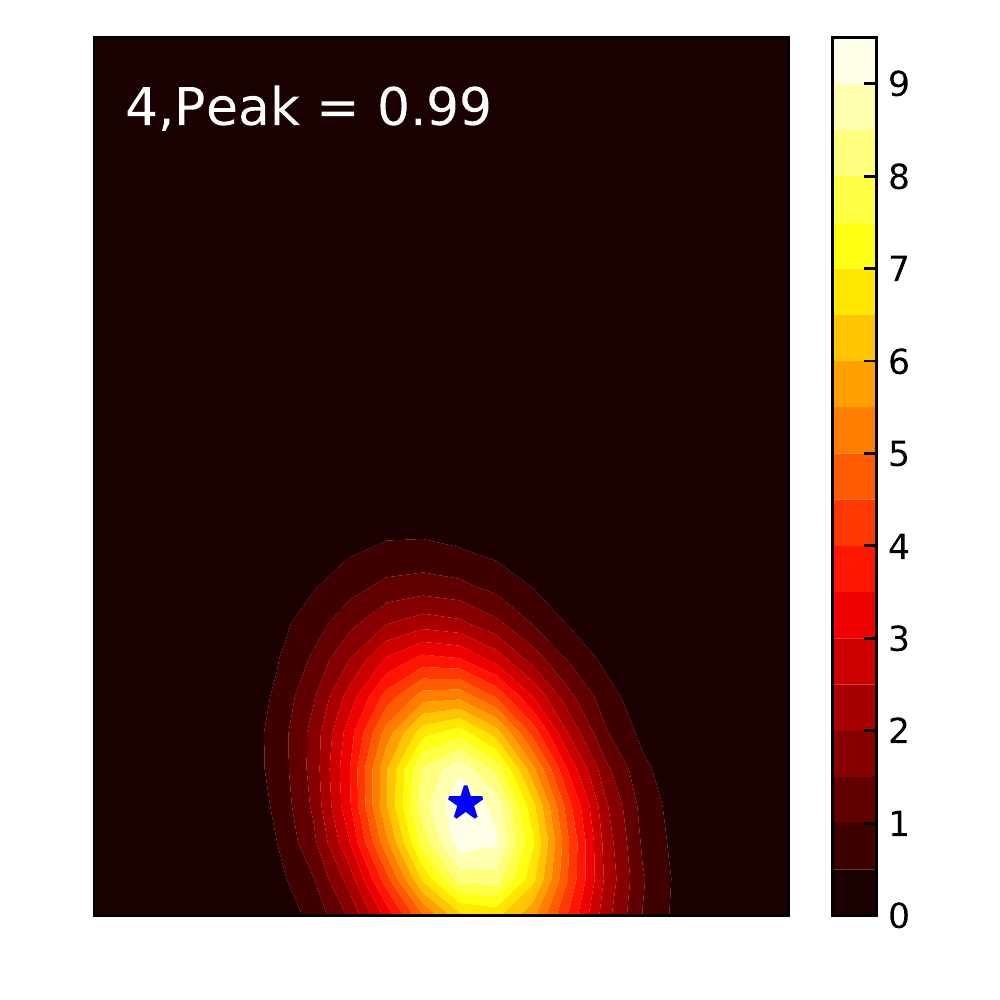}}
\end{minipage}\caption{The upper panels show the nearest detected clumps of the missed clump centers shown in Figure \ref{Fig_EData}. The lower panels show the undetected simulated clumps. The blue asterisks denote the location of the simulated clumps and the red asterisks denote the location of the closest detected clumps. $\Delta X$ is the distance between the two locations in the graph. The unit of intensity is K.}
	\label{Fig_ELoss}
\end{figure*}

\subsubsection{Compare with other algorithms in simulated clumps} \label{sec:3.2.3}
The comprehensive performance of FellWalker, LDC, and ConBased are better than GaussClumps, ClumpFind and ReinHold \citep{FellWalkerBetter,FellWalker,LiChong,LDC,ConBased}, making them suitable as comparators for our study. The changes in each evaluation metric for different algorithms with respect to SNR are statistically analyzed and presented in Figure \ref{Fig_RPF} and Figure \ref{Fig_DFI}. $\Delta X$ and $\Delta Flux$ of LDC are based on the direct detections, and the Multiple Gaussian Model \citep{LDC} is not used in this paper. 

The left, middle, and right panels of Figure \ref{Fig_RPF} show $R$, $P$, and $F_1$, respectively. $R$ and $F_1$ for each algorithm increase gradually with the increase of SNR. $R$ of FacetClumps ranges from 0.453 to 0.996, with a mean value of 0.902, which is greater than that of FellWalker and LDC. $P$ of FacetClumps varies between 0.973 to 1, with a mean value of 0.996. $F_1$ of FacetClumps ranges from 0.619 to 0.997, with a mean value of 0.939. $P$ and $F_1$ of FacetClumps are greater than those of FellWalker, LDC and ConBased, and have minimal fluctuations. 

The left, middle, and right panels of Figure \ref{Fig_DFI} show $\Delta X$, $\Delta Flux$, and $IOU$, respectively. It can be seen that clumps with lower SNR have higher error in locations, larger loss in measured fluxes, and lower $IOU$. $\Delta X$ of FacetClumps decreases from 0.56 to 0.18 voxel. The average $\Delta X$ of FacetClumps is similar to that of FellWalker, and less than that of ConBesed and LDC. $\Delta Flux$ of FacetClumps increases from -78\% to -3.7\%, with a flux loss less than other algorithms. The low bias of the fluxes can be corrected using methods such as extrapolation proposed by \citet{FluxCorrect}. $IOU$ of FacetClumps increases from 0.03 to 0.41, which is higher than that of other algorithms. 

In summary, FacetClumps exhibits greater $F_1$, indicating its better anti-noise performance; its flux loss is smaller and $IOU$ is higher, demonstrating its superior capacity for both the flux recovery of clumps and the segmentation of overlapping clumps.

\begin{table*}
	\centering
	\caption{\centering The average number of clumps, the SNR-weighted average $R$ of the varying SNR, the corresponding flux and SNR when the $R$ is equal to 0.9, the corresponding flux and SNR when $R$ is equal to 0.8, the SNR-weighted average $\Delta X$ in the spatial direction and in the velocity channels, the SNR-weighted average $\Delta Flux$ and $IOU$.}
	\begin{tabular}{c|cccccccc}\hline\hline
		Algorithm&$N$&$R_{mean}(SNR)$&$R_{0.9}(Flux/SNR)$&$R_{0.8}(Flux/SNR)$&$\Delta X_{LB}$&$\Delta X_{V}$&$\Delta Flux$&$IOU$\\\hline
		FellWalker & 546 &74.4\% & -/- & 560/30 &0.31&0.2&10.7\%&0.42\\
		LDC & 692 &80.3\%& -/- & 180/15 & 0.33&0.19 &9.4\%&0.41\\
		ConBased & 657 & 83.1\% &400/30 & 170/10 &0.27&0.18 &10.7\%&0.44\\
		FacetClumps & 671 &90.2\% & 190/14 & 100/8 &0.17&0.12&30.7\%&0.5\\\hline
	\end{tabular}
	\label{Evaluate_Infor}
\end{table*}

\begin{figure*}
	\centering
	\centerline{\includegraphics[width=6.5in]{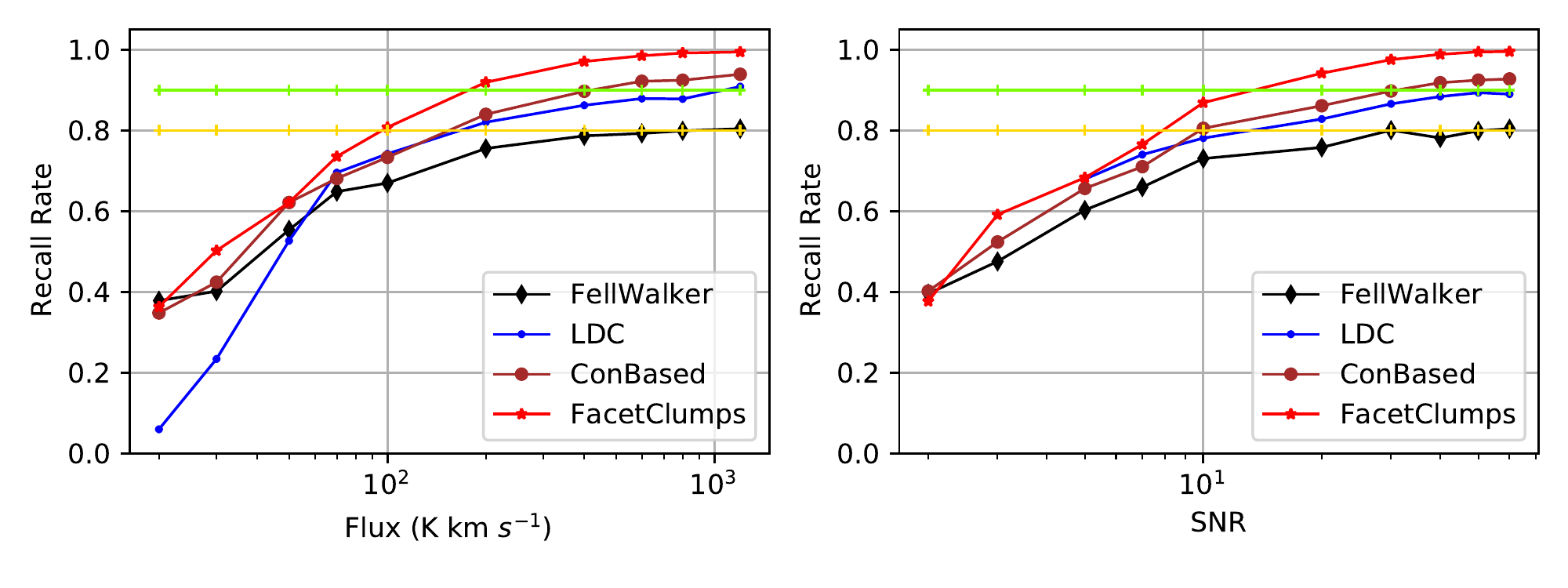}}
	\caption{The statistics of $R$ of Fellwalker, LDC, ConBased, and FacetClumps for the synthetic data. The left panel shows $R$ as a function of flux, and the right panel shows $R$ as a function of SNR. The lawngreen line is equal to 0.9, and the gold line is equal to 0.8.}
	\label{Fig_Recall}
\end{figure*}

\begin{figure*}
	\centering
	\centerline{\includegraphics[width=6.5in]{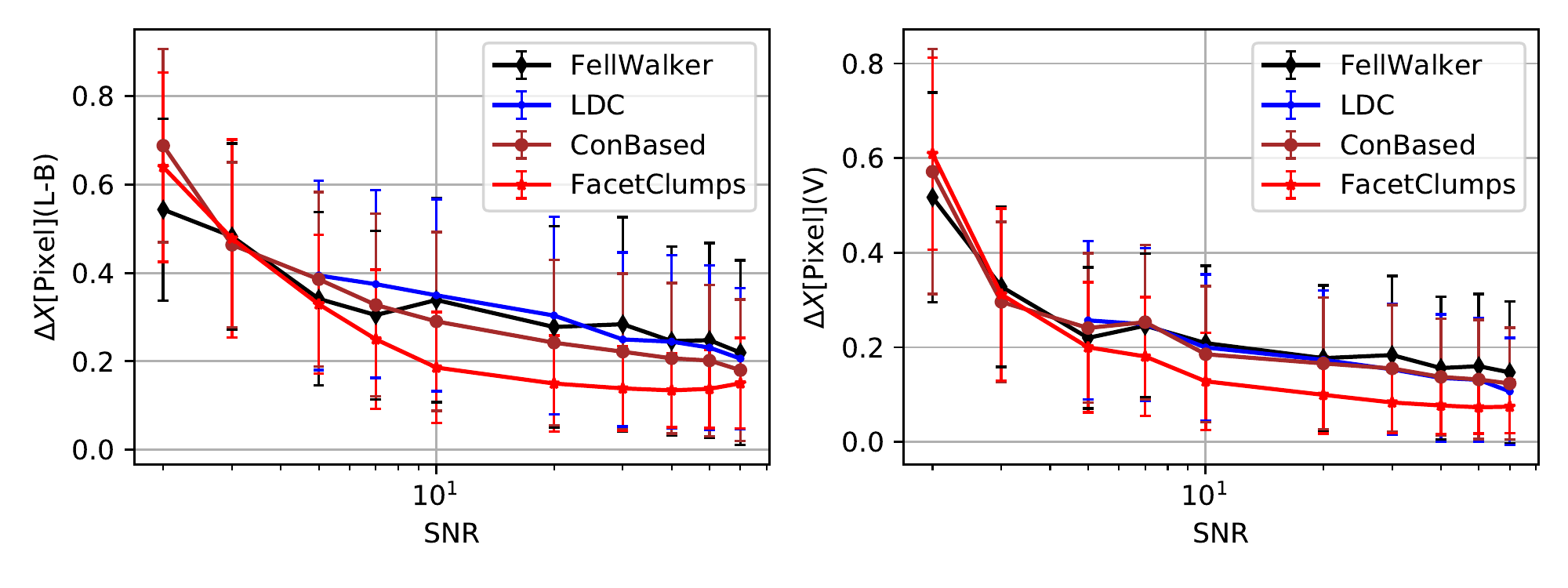}}
	\caption{The statistics of $\Delta X$ of FellWalker, LDC, ConBased, and FacetClumps for the synthetic data. $\Delta X$ is a function of SNR. The left panel shows $\Delta X$ in the spatial direction, and the right panel shows $\Delta X$ in the velocity channels.}
	\label{Fig_Dist}
\end{figure*}

\begin{figure*}
	\centering
	\centerline{\includegraphics[width=6.5in]{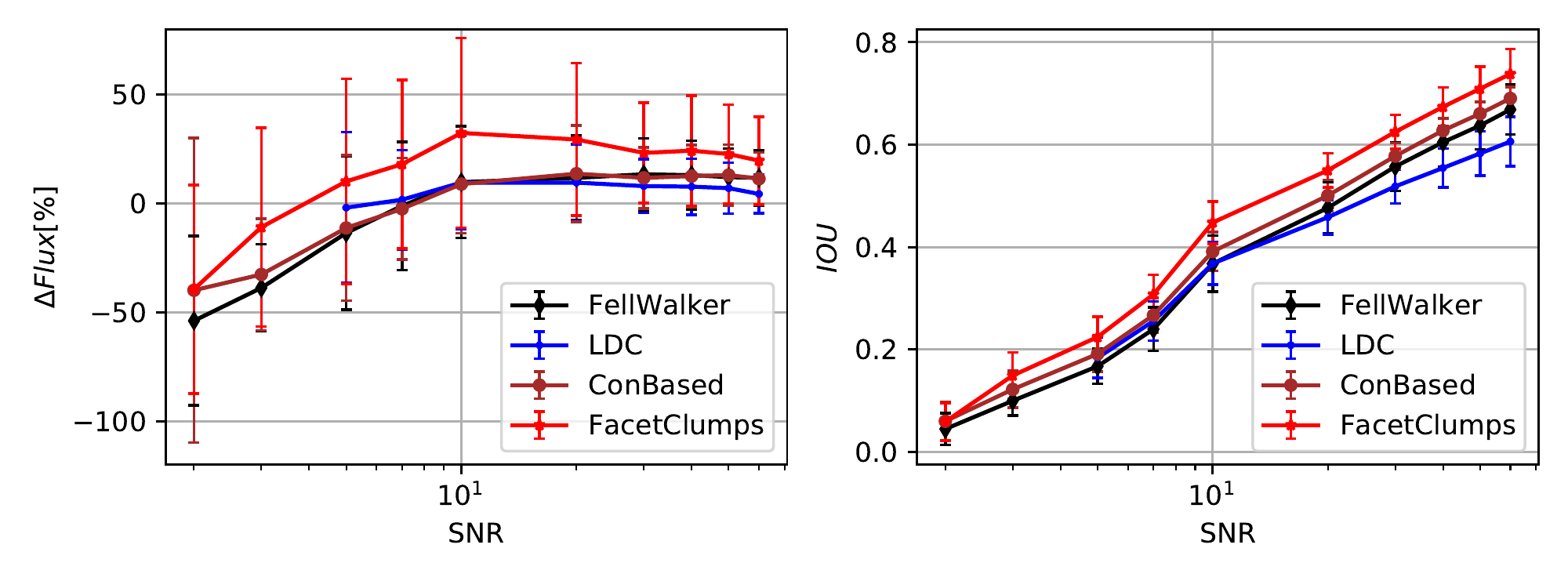}}
	\caption{The statistics of $\Delta Flux$ (left) and $IOU$ (right) of FellWalker, LDC, ConBased and FacetClumps for the synthetic data. $\Delta Flux$ and $IOU$ are functions of SNR.}
	\label{Fig_FluxIOU}
\end{figure*}

\subsection{Experiments with observational data} \label{sec:3.3}
\subsubsection{The observational and synthetic data} \label{sec:3.3.1}
The Milky Way Imaging Scroll Painting (MWISP) survey \citep{MWISP} conducted by the Purple Mountain Observatory (PMO) is a large-field survey of $^{12}CO$, $^{13}CO$, and $C^{18}O$ ($J = 1-0$) emission lines. The optical thickness of $^{13}CO$ ($J = 1-0$) is generally lower than that of $^{12}CO$ ($J = 1-0$), and the chemical properties of the isotope molecule itself are stable. $C^{18}O$ haves a lower abundance than $^{13}CO$, and the optical thickness of $C^{18}O$ ($J = 1-0$) in the same line of sight is lower than that of the $^{13}CO$ ($J = 1-0$) transition lines, making it an ideal tool for detecting regions of higher density. 

The half power beam width (HPBW) of $^{13}CO$ and $C^{18}O$ emission in MWISP is 52", with a grid spacing of 30". However, due to undersampling, the spatial sampling rate is set to $FwhmBeam=2$. Similarly, the velocity sampling rate is $VeloRes=2$. The spectral resolution of the observational data is approximately 0.166 km s$^{-1}$, and the total fluxes are multiplied by this factor to obtain physically meaningful values. The $^{13}CO$ and $C^{18}O$ ($J = 1-0$) emission of MWISP within $11.7^{\circ} \leq l \leq 13.4^{\circ}$, $0.22^{\circ} \leq b \leq 1.05^{\circ}$ and 5 km s$^{-1}$ $\leq v \leq$ 35 km s$^{-1}$  where many notable star-forming activities have been discovered \citep[e.g.][]{M17Star4,M17Star5,M17Star6} are used to examine the performance of FacetClumps. The noise level of the $^{13}CO$ and $C^{18}O$ emissions are approximately 0.22 K and 0.2 K, respectively. The maximum intensity of the $^{13}CO$ emission is approximately 16.8 K. 

We generate 200 synthetic data cubes named Data1, which are superimposed with the $^{13}CO$ emission and simulated clumps. Each synthetic data cube contains 50 simulated clumps, with uniform peak intensities ranging from 0.44 ($2\times RMS$) to 16.8 K (maximum intensity) to ensure that the statistical magnitude of clumps is consistent under different SNRs, uniform sizes ranging from 2 to 4 voxels, and random angles. The flux is primarily distributed between 20 and 1200 K km s$^{-1}$, which closely resembles the distribution observed in actual molecular clumps.  

To evaluate the performance of the FacetClumps in a wider range of signal densities and environments, we have selected a high-density and a low-density area of $^{13}CO$ emission to create more diverse synthetic datasets (Data2 and Data3, see Appendix \ref{Larger} for the details). In addition, to investigate the robustness of FacetClumps parameters under various sampling conditions, we have resampled the synthetic data of Data1 using different sampling factors in different directions (see Appendix \ref{Resample} for the details).

An example of the synthetic data of Data1 is shown in Figure \ref{Fig_EData}. The red asterisks in the white circle denote simulated clumps detected by FacetClumps, while the red circles denote missed simulated clumps. Figure \ref{Fig_ELoss} shows the four missed clumps, where the blue asterisks denote the locations of the simulated clumps and the red asterisks denote the locations of the closest detected clumps. It can be seen that the undetected simulated clumps usually have large overlaps with real clumps of higher intensity (e.g., Nos. 1 and 3), are affected by other simulated clumps (e.g., Nos. 2), or have a low SNR and have become part of real ones (e.g., Nos. 4). 

\begin{figure*}
	\centering
\centerline{\includegraphics[width=10in]{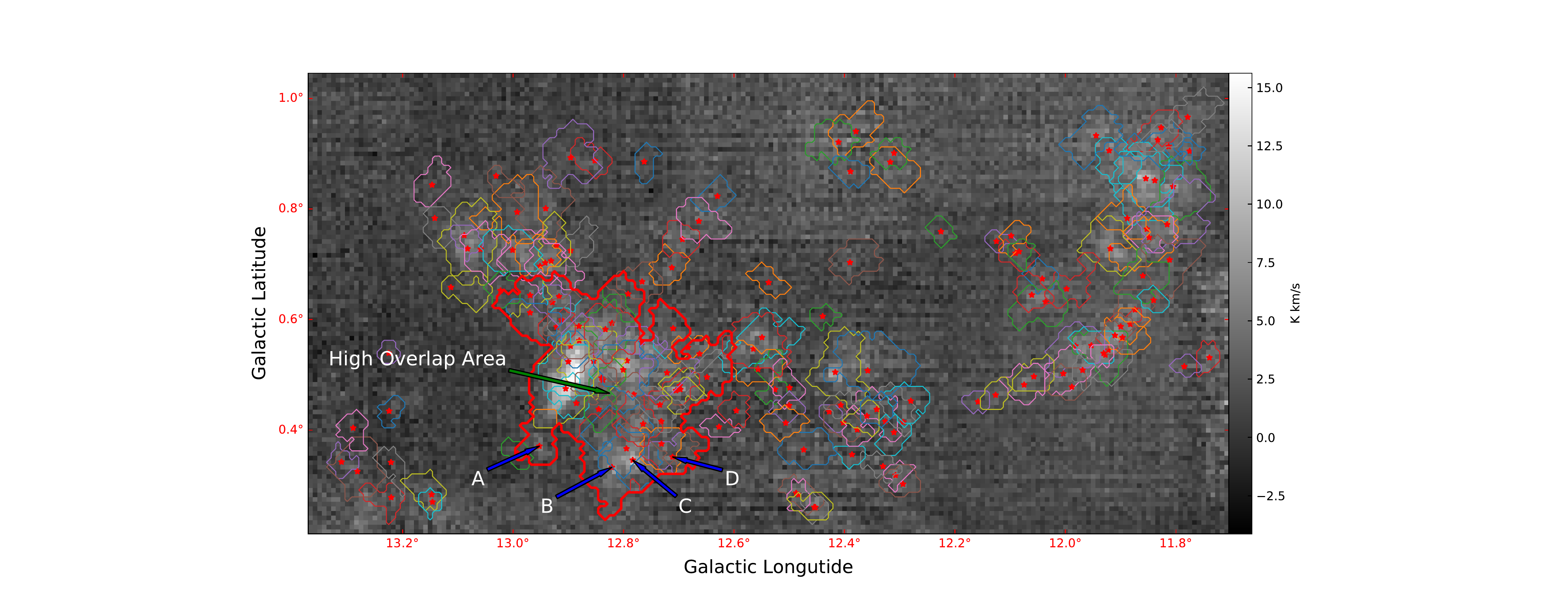}}
	\caption{The result of the application of FacetClumps in observational data of $C^{18}O$ emission within $11.7^{\circ} \leq l \leq 13.4^{\circ}$, $0.22^{\circ} \leq b \leq 1.05^{\circ}$ and 5 km s$^{-1}$ $\leq v \leq$ 35 km s$^{-1}$. The total number of clumps is 185, of which 163 did not touch the edge. The red asterisks denote the central locations of the clumps which do not touch the edge, and different thin outlines delineate the boundaries of different clumps. The thick red outline circles an area of high overlap, with four clumps (A, B, C, and D) at the bottom of the area, as shown in Figure \ref{Fig_C18O_1}, and the other clumps are shown in Figures \ref{Fig_C18O_2} and \ref{Fig_C18O_3}.}
	\label{Fig_C18O_0}
\end{figure*}

\begin{figure*}
	\centering
	\vspace{0cm}
	\begin{minipage}[t]{0.24\textwidth}
		\centering
		\centerline{\includegraphics[width=1.9in]{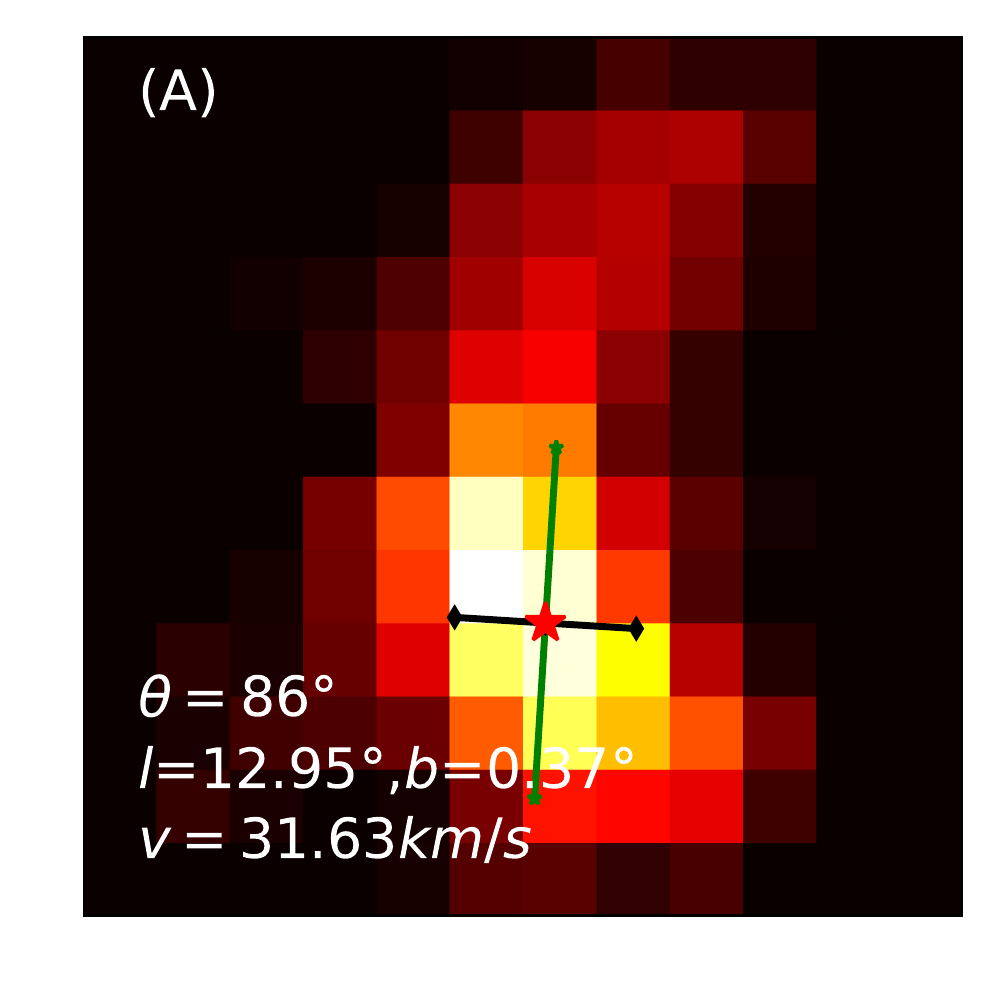}}
\end{minipage}\begin{minipage}[t]{0.24\textwidth}
		\centering
		\centerline{\includegraphics[width=1.9in]{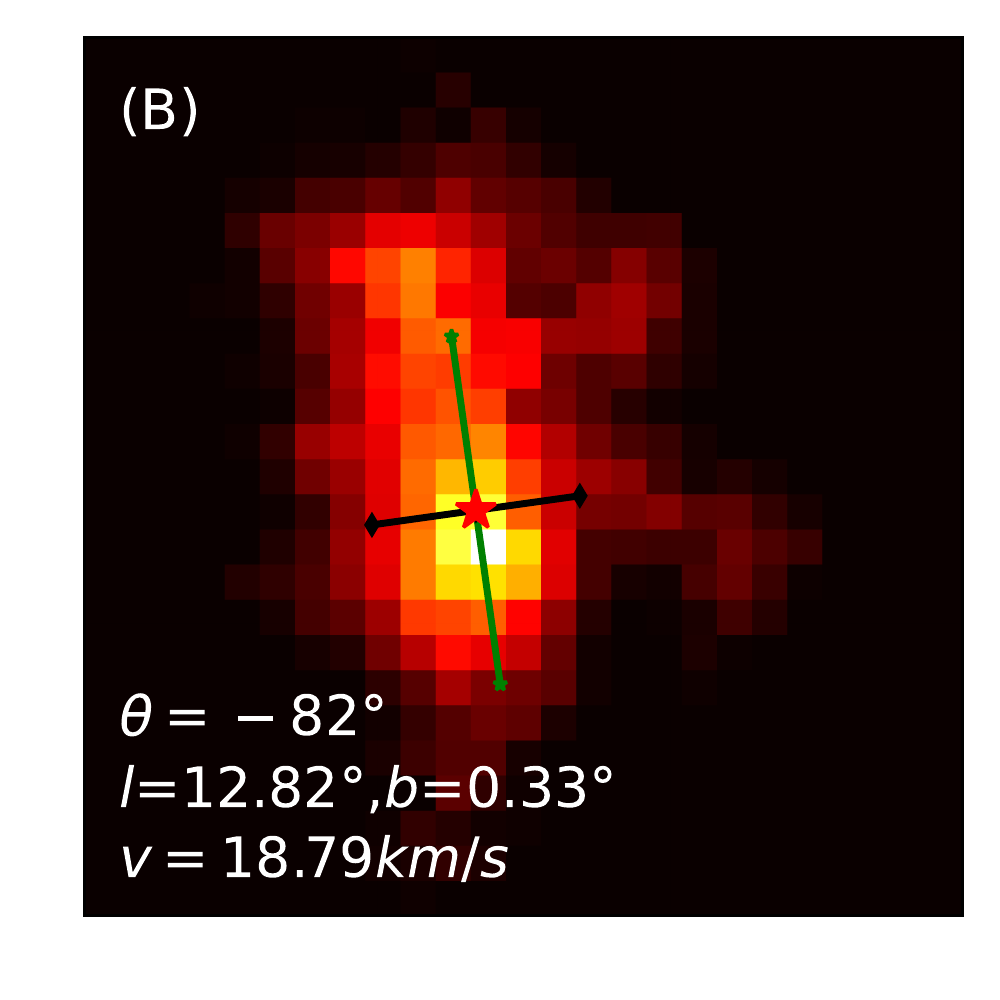}}
\end{minipage}\begin{minipage}[t]{0.24\textwidth}
		\centering
		\centerline{\includegraphics[width=1.94in]{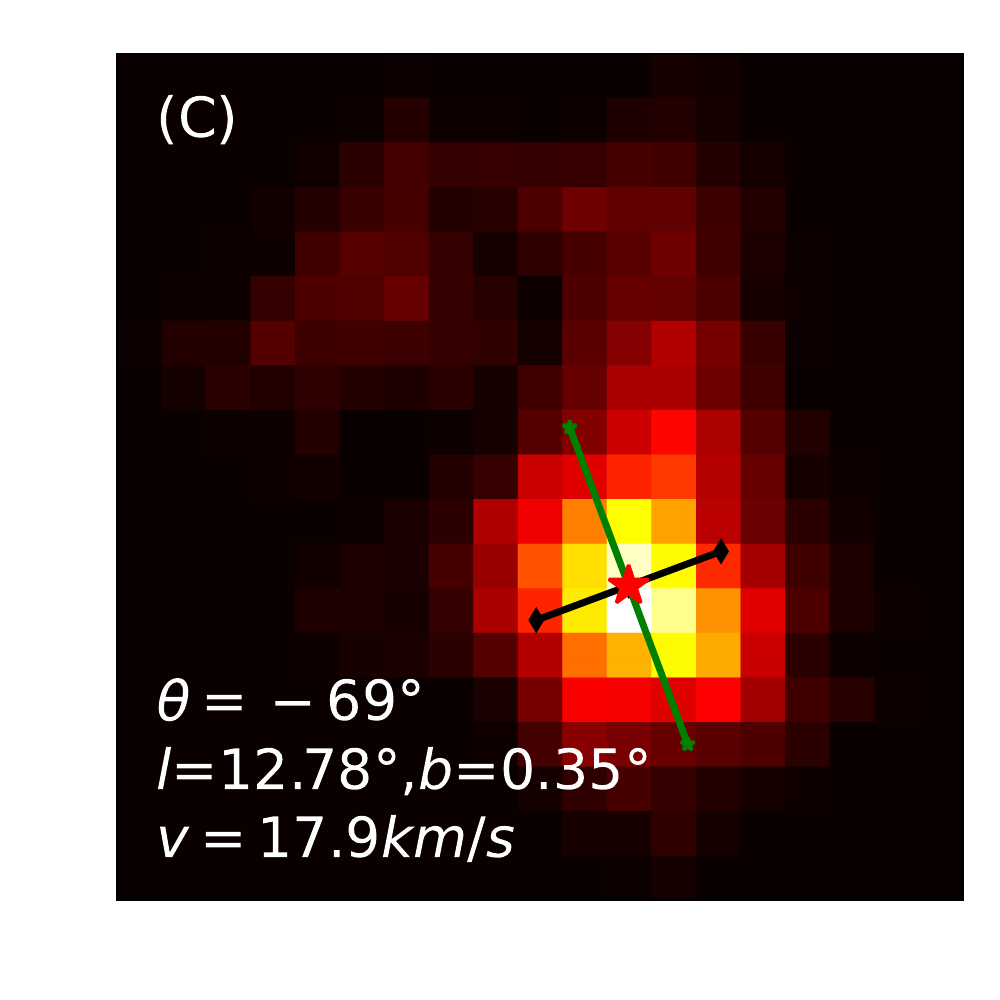}}
\end{minipage}\begin{minipage}[t]{0.24\textwidth}
		\centering
		\centerline{\includegraphics[width=1.9in]{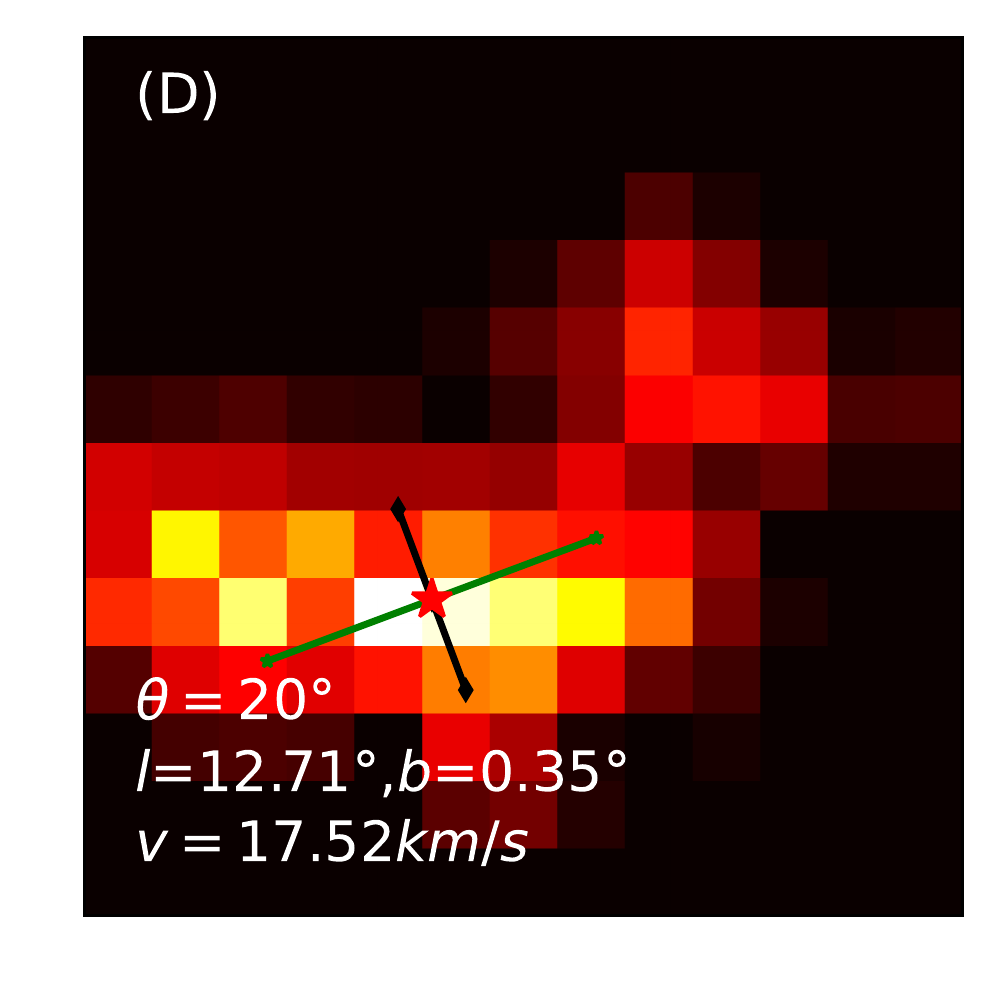}}
\end{minipage}\caption{Velocity-integrated intensity images of four clumps in the high overlap area as shown in Figure \ref{Fig_C18O_0}. A, B, C, and D correspond to the markers in Figure \ref{Fig_C18O_0}. The red asterisks denote the central locations of the clumps. The green lines denote the principal axis and the black lines denote the secondary axis. The ratio of the lengths of the principal and secondary axes is equivalent to the ratio of their respective axes. $\theta$ denotes the angle between the principal axis and the negative direction of galactic longitude. $(l, b, v)$ is the central coordinate.}
	\label{Fig_C18O_1}
\end{figure*}

\subsubsection{Compare with other algorithms in synthetic data}\label{sec:3.3.2}
To test the performance of the algorithms in observational environments, we apply them to detect the synthetic data Data1. The average number of detected clumps by FellWalker, LDC, ConBased, and FacetClumps is 546, 692, 657, and 671, respectively. We analyze several evaluation metrics, including $R$ as a function of flux and SNR, $\Delta X$ in the spatial direction and in the velocity channels as a function of SNR, and $\Delta Flux$ and $IOU$ as functions of SNR. The results of these analyses are shown in Figure \ref{Fig_Recall}, Figure \ref{Fig_Dist}, and Figure \ref{Fig_FluxIOU}, respectively. 

When computing the average metrics, we take into account the weighted mean value of the intervals with SNR greater than 5. The weights are obtained by using FellWalker to detect the observational data used to construct the synthetic data, and then calculating the corresponding weights of the SNR distribution for each interval. The average number of clumps, the SNR-weighted average $R$ of the varying SNR, the corresponding flux and SNR when $R$ is equal to 0.9 and 0.8, the SNR-weighted average $\Delta X$ in the spatial direction and in the velocity channels, and the SNR-weighted average $\Delta Flux$ and $IOU$ are presented in Table \ref{Evaluate_Infor}. 

As shown in Figure \ref{Fig_Recall}, $R$ of FacetClumps ranges from 0.38 to 0.99, which is higher than that of other algorithms, especially when flux and SNR are low. This indicates that FacetClumps detects more correct clumps and can better detect faint clumps in observational background. When $R$ reaches 0.9, the corresponding flux/SNR of ConBased and FacetClumps are about 400 K km s$^{-1}$/30 and 190 K km s$^{-1}$/14, respectively. When $R$ reaches 0.8, the corresponding flux/SNR of FellWalker, LDC, ConBased and FacetClumps are about 560 K km s$^{-1}$/30, 180 K km s$^{-1}$/15, 170 K km s$^{-1}$/10 and 100 K km s$^{-1}$/8, respectively. $R$ of ConBased is higher than that of FellWalker and LDC, and $R$ of LDC is higher than that of FellWalker. The SNR-weighted average $R$ of FellWalker, LDC, ConBased, and FacetClumps is 74.4\%, 80.3\%, 83.1\%, and 90.2\%, respectively. 

\begin{table}
	\centering
	\caption{\centering The number of clumps, the cross-matching rate of $C^{18}O$/$^{13}CO$, and the relative time unit of the $^{13}CO$ emission.}
	\begin{tabular}{c|cccc}\hline\hline
		Algorithm&$^{13}CO$&$C^{18}O$&$C^{18}O$/$^{13}CO$&$T_{^{13}CO}$\\\hline
		FellWalker & 538 & 134 & 48.5\% & /\\
		LDC & 680 & 138 & 62.3\% & 3.2 \\
		ConBased & 629 & 105 & 68.6\% & 1.9 \\
		FacetClumps & 692 & 185 & 56.8\% & 1 \\\hline
	\end{tabular}
	\label{CrossMatch}
\end{table}

The left panel of Figure \ref{Fig_Dist} shows $\Delta X$ in the spatial direction, while the right panel shows $\Delta X$ in the velocity channels. $\Delta X$ of FacetClumps in the spatial direction decreases from 0.64 to 0.14 voxel, with a SNR-weighted mean value of 0.17 voxel, and that in the velocity channels decreases from 0.61 to 0.07 voxel, with a SNR-weighted mean value of 0.12 voxel. $\Delta X$ of FacetClumps in the spatial direction and in the velocity channels are smaller, indicating that the clumps detected by FacetClumps in the real environment have more precise locations.

$\Delta Flux$ and $IOU$ are shown in the left and right panels of Figure \ref{Fig_FluxIOU}, respectively. $\Delta Flux$ of FacetClumps is between -39.4\% and 32.4\%. The SNR-weighted average $\Delta Flux$ of FellWalker, LDC, and ConBased is 10.7\%, 9.4\%, 10.7\%, and 30.7\%, indicating that FacetClumps has less flux loss. $IOU$ of FacetClumps increases from 0.06 to 0.74, and its SNR-weighted mean value is 0.5. The SNR-weighted average $IOU$ of FellWalker, LDC, and ConBased are 0.42, 0.41, and 0.44, indicating that FacetClumps can better segment the regions of simulated clumps from real signals. 

In summary, FacetClumps exhibits a greater $R$ and a smaller $\Delta X$, indicating that it is better suited for the observational environments and can locate clumps more accurately in complex backgrounds. The $\Delta Flux$ and $IOU$ statistics show that FacetClumps performs slightly better in detecting useful signals and segmenting different clumps. Furthermore, the $R$ curve of FacetClumps is smoother and the error bars for $\Delta X$ are shorter, indicating its superior stability.

\subsubsection{Experiments to observational data} \label{sec:3.3.3}
To evaluate the usability of FacatClumps on observational data, we apply FellWalker, LDC, ConBased, and FacetClumps to detect the data cubes of $^{13}CO$ and $C^{18}O$ emission. The number of clumps, the cross-matching rate, and the relative time unit $T$ in $^{13}CO$ emission are presented in Table \ref{CrossMatch}. The cross-matching rate of $Line1$/$Line2$ is defined as the percentage of $Line2$ clumps that coincide with $Line1$ clumps \citep{LiChong, ConBased}. The results show that the cross-matching rate of FacetClumps is higher than that of FellWalker, and ConBased has the highest cross-matching rate. Since the time spent in different programming languages is not comparable, the minimum time spent by the algorithm in the same language is recorded as 1 unit, and the time spent by the other algorithms in the same language is recorded as a multiple of this unit. LDC, ConBased, and FacetClumps use the same programming language, while FellWalker uses a different one. As shown in Table \ref{CrossMatch}, the time taken by LDC and ConBased is 3.2 and 1.9 times that of FacetClumps, respectively. 

Figure \ref{Fig_C18O_0} shows the results of the application of FacetClumps in the $C^{18}O$ emission. A total of 185 clumps are identified, of which 163 do not touch the edges. The red asterisks denote the central locations of the clumps that do not touch the edges. Different thin outlines delineate the boundaries of different clumps, while the thick red outline circles an area of high overlap. A, B, C, and D represent the four clumps at the bottom of the high overlap area, as shown in Figure \ref{Fig_C18O_1}, and the other clumps are shown in Figures \ref{Fig_C18O_2} and \ref{Fig_C18O_3}. Most clumps contain a well-defined denser central source surrounded by weaker gas, while some may be dual-clump systems (e.g., Figure \ref{Fig_C18O_1}(B)), or some may have trailing substructures (e.g., Figure \ref{Fig_C18O_1}(C)). These images illustrate that FacetClumps effectively and accurately detect clumps, even in areas with a relatively higher degree of overlap. Further scientific analysis of the clumps will be carried out in the future. 

\section{Summary}
We propose FacetClumps, a new algorithm for detecting molecular clumps in astronomical data. Initially, the signal regions are extracted based on morphology. Then, the Gaussian Facet model operators are utilized to fit the signal regions to derive the first and second derivatives, which are used in conjunction with the maximum determination theorem to locate the clump centers. Subsequently, signal regions are segmented into local regions based on local gradients. Finally, the local regions are clustered into the clump centers based on connectivity and minimum distance, thus identifying the region of each clump.

We have conducted parametric and comparative experiments on simulated clumps with different SNRs to determine appropriate values for the instrument-independent parameters of FacetClumps and to evaluate the performance of Fellwalker, LDC, ConBased, and FacetClumps. The parametric experiments provide an effective reference for the parameters, while the comparative experiments demonstrate that FacetClumps improves the noise resistance and segmentation accuracy of overlapping clumps even further.

We have conducted a series of experiments with synthetic data. Experiments performed in synthetic data consisting of an active star-forming zone and simulated clumps demonstrate that FacetClumps has greater $R$ with 90.2\%, smaller $\Delta X$, better $\Delta Flux$, and higher $IOU$. Experiments performed in two types of larger synthetic data with different signal densities in Appendix \ref{Larger} demonstrate that FacetClumps has pronounced advantages in high-density signal environments. Experiments performed in different resampled synthetic data in Appendix \ref{Resample} indicate that the parameters of FacetClumps have self-adaptability. We have conducted some tests in the observational data, which illustrate that FacetClumps is more efficient and can be applied in observational data well.

\section*{Acknowledgements}
We are grateful to the anonymous referees for their invaluable insights and comments, which enabled us to refine and enhance this work. This work is supported by the National Natural Science Foundation of China (grants Nos. U2031202, 11903083, 11873093). This research make use of the data from the Milky Way Imaging Scroll Painting (MWISP) project, which is a multi-line survey in $^{12}CO$/$^{13}CO$/$C^{18}O$ along the northern galactic plane with PMO-13.7m telescope. We are grateful to all the members of the MWISP working group, particularly the staff members at PMO-13.7m telescope, for their long-term support. MWISP is sponsored by National Key R\&D Program of China with grant 2017YFA0402701 and CAS Key Research Program of Frontier Sciences with grant QYZDJ-SSW-SLH047. 

\bibliography{FacetClumpsBIB.bib}{}
\bibliographystyle{aasjournal}

\appendix
\section{Comparation of different algorithms on a larger range of synthetic data}\label{Larger}
\begin{figure*}
	\centering
	\vspace{0cm}
	\begin{minipage}[t]{0.24\textwidth}
		\centering
		\centerline{\includegraphics[width=8.5in]{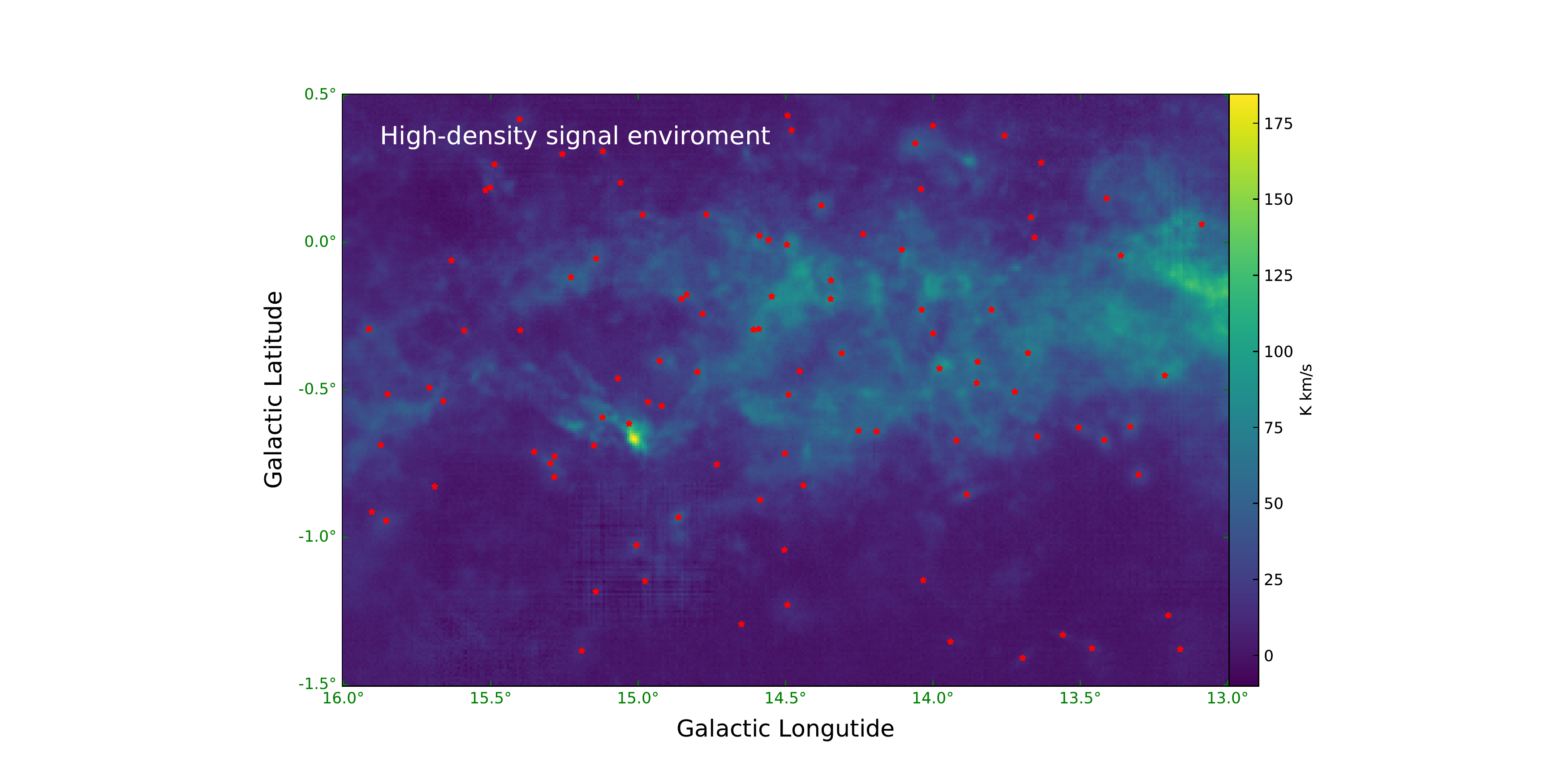}}
\end{minipage}

	\begin{minipage}[t]{0.24\textwidth}
		\centering
		\centerline{\includegraphics[width=8.5in]{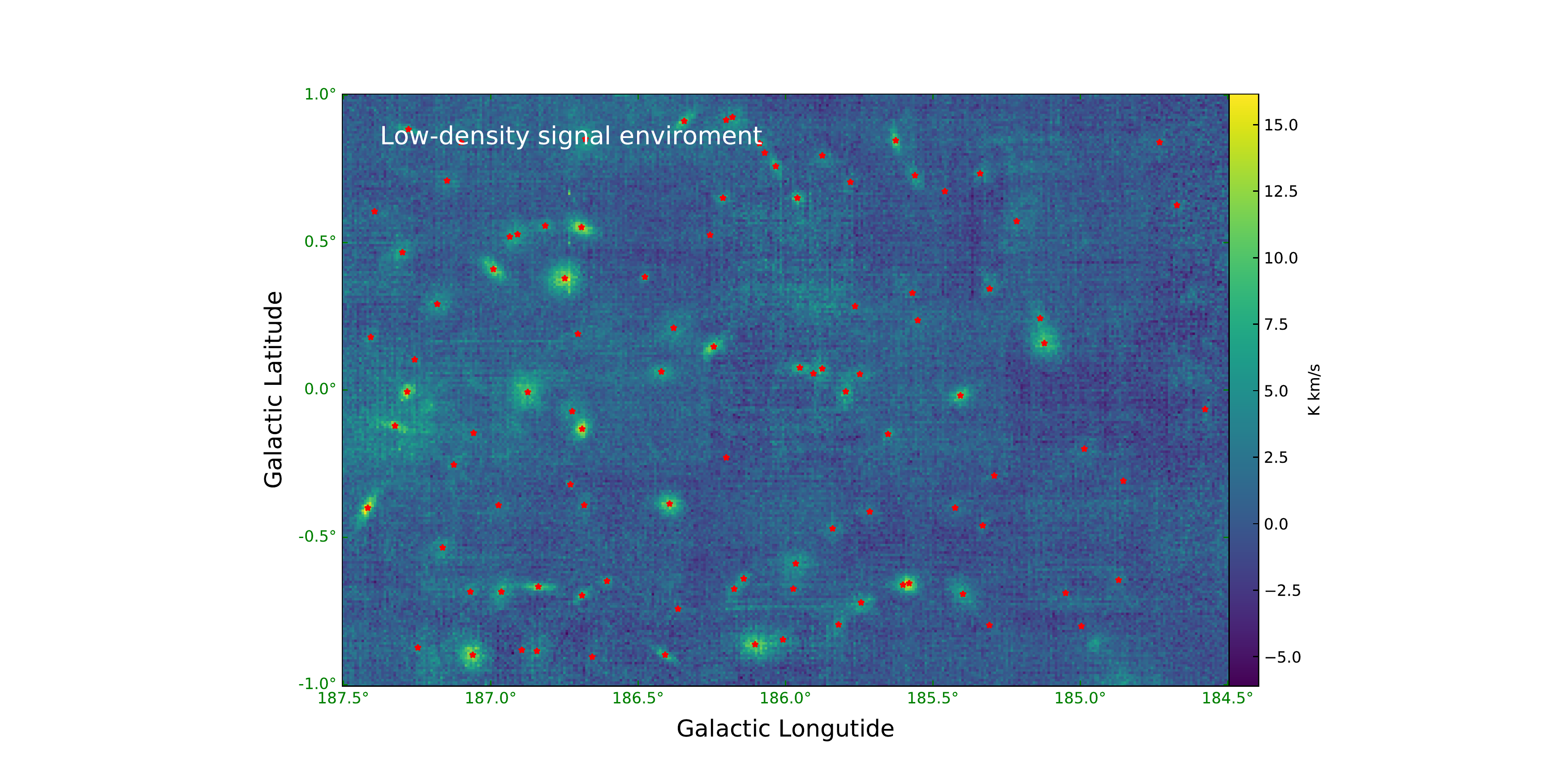}}
\end{minipage}

	\caption{Examples of the synthetic data of Data2 and Data3. The top panel is superimposed by the $^{13}CO$ emission within $13^{\circ} \leq l \leq 16^{\circ}$, $-1.5^{\circ} \leq b \leq 0.5^{\circ}$ and 0 km s$^{-1}$ $\leq v \leq$ 70 km s$^{-1}$ and simulated clumps, representing a typical high-density signal environment. The bottom panel is superimposed by the $^{13}CO$ emission within $184.5^{\circ} \leq l \leq 187.5^{\circ}$, $-1^{\circ} \leq b \leq 1^{\circ}$ and -10 km s$^{-1}$ $\leq v \leq$ 60 km s$^{-1}$ and simulated clumps, representing a typical low-density signal environment. The cube size is $423\times241\times361$ voxels, and there are 100 simulated clumps. The red asterisks denote the central locations of the simulated clumps.}
	\label{Fig_EData_R2_R16}
\end{figure*}

We select the $^{13}CO$ ($J = 1-0$) emission of MWISP within $13^{\circ} \leq l \leq 16^{\circ}$, $-1.5^{\circ} \leq b \leq 0.5^{\circ}$ and 0 km s$^{-1}$ $\leq v \leq$ 70 km s$^{-1}$ and within $184.5^{\circ} \leq l \leq 187.5^{\circ}$, $-1^{\circ} \leq b \leq 1^{\circ}$ and -10 km s$^{-1}$ $\leq v \leq$ 60 km s$^{-1}$ as high-density and low-density signal environments, respectively. They are used to construct synthetic data, namely Data2 and Data3, as shown in Figure \ref{Fig_EData_R2_R16}. Each dataset contains 100 synthetic data cubes, with each cube containing 100 simulated clumps. The range of peak intensities is from $2\times RMS$ to maximum intensity, and the range of sizes is from 1 to 5 voxels.

\begin{figure*}
	\centering
	\vspace{0cm}
	\begin{minipage}[t]{0.24\textwidth}
		\centering
		\centerline{\includegraphics[width=6.5in]{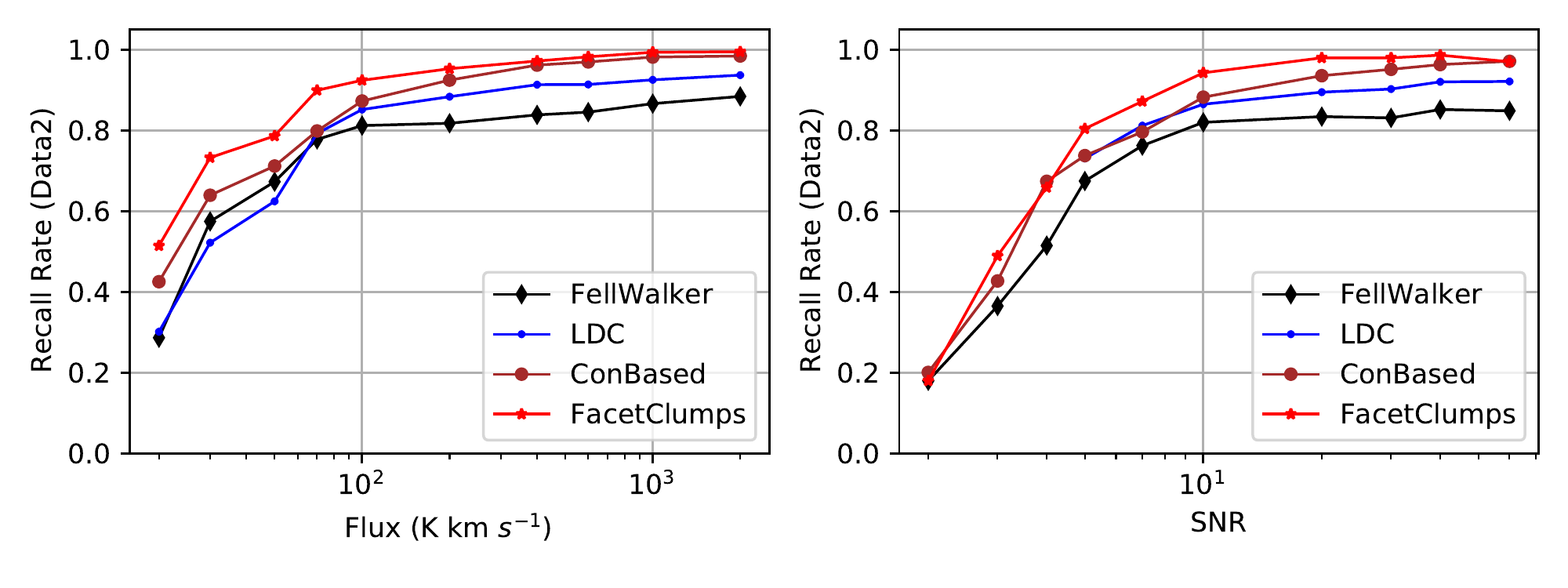}}
\end{minipage}

	\begin{minipage}[t]{0.24\textwidth}
		\centering
		\centerline{\includegraphics[width=6.5in]{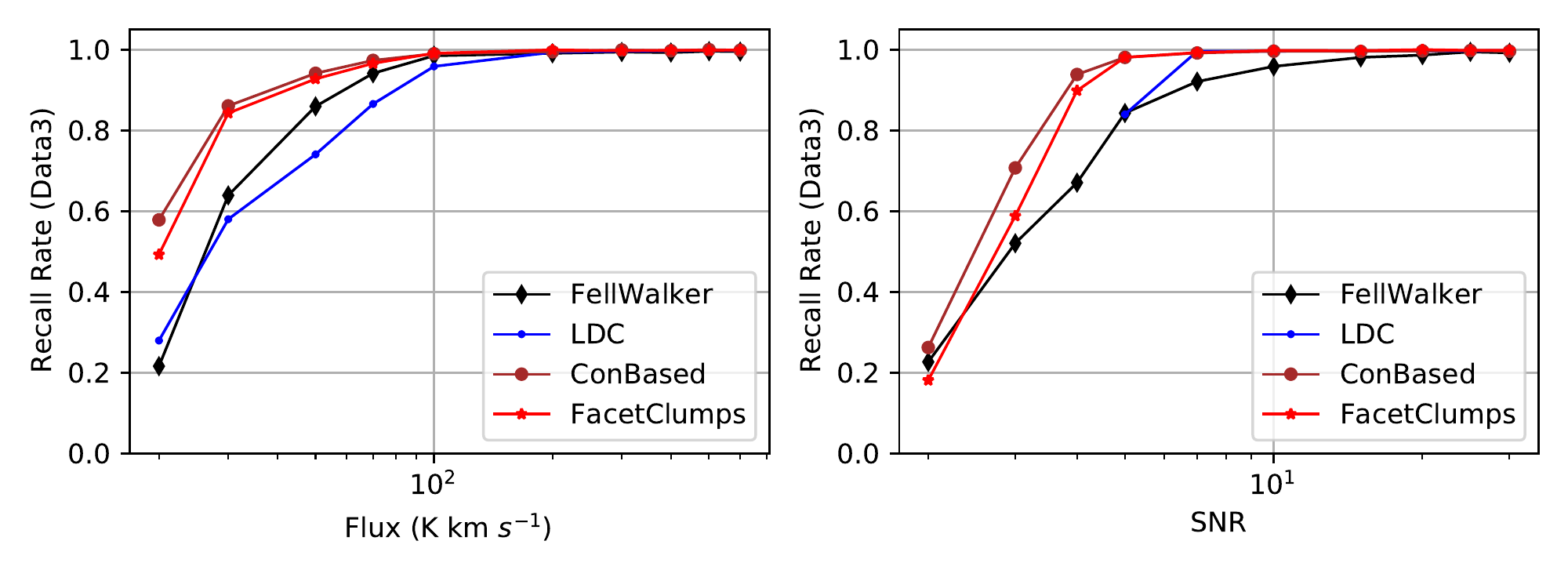}}
\end{minipage}

	\caption{The statistics of $R$ of FellWalker, LDC, ConBased, and FacetClumps for synthetic data of different density signal environments. The upper panels are the statistics of synthetic data in a high-density signal environment as shown in the top panel of Figure \ref{Fig_EData_R2_R16}. The lower panels are the statistics of synthetic data in a low-density signal environment as shown in the bottom panel of Figure \ref{Fig_EData_R2_R16}. The left panels depict $R$ as a function of flux, and the right panels depict $R$ as a function of SNR.}
	\label{Fig_Recall_R2_R16}
\end{figure*}

\begin{figure*}
	\centering
\centerline{\includegraphics[width=6.5in]{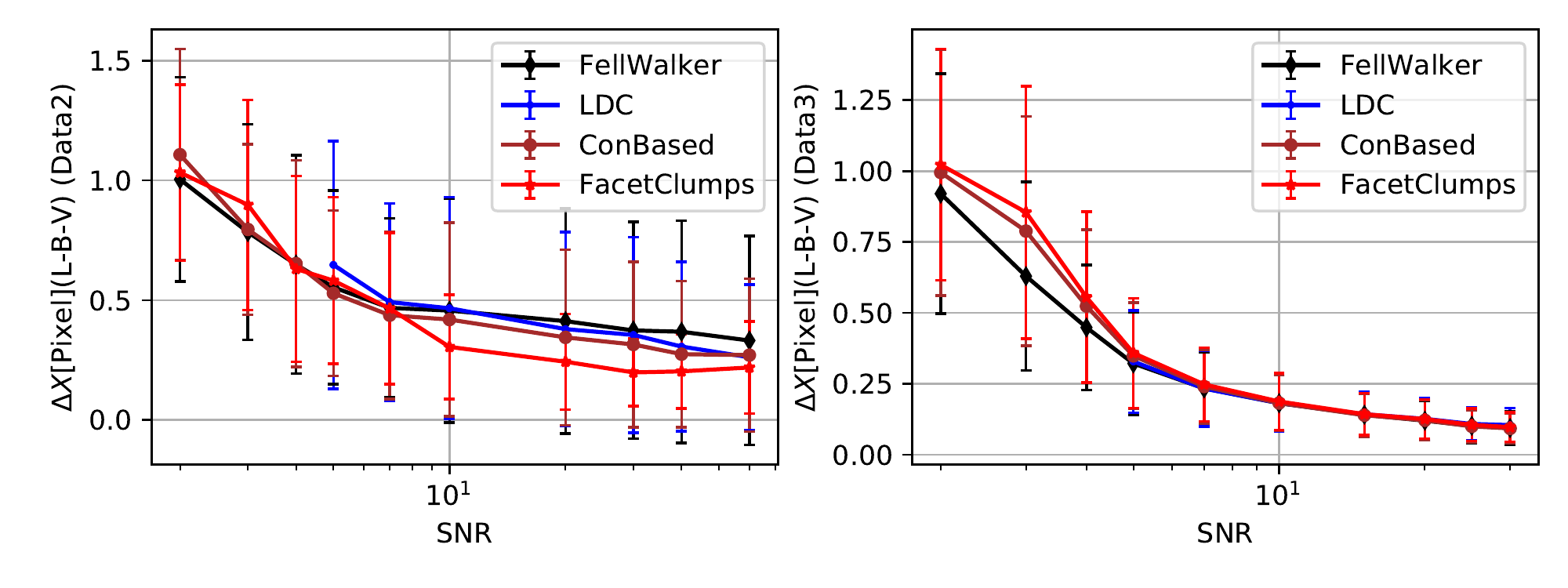}}
	\caption{The statistics of $\Delta X$ of FellWalker, LDC, ConBased, and FacetClumps for synthetic data of different density signal environments. $\Delta X$ is a function of SNR. The left panel shows the statistic of synthetic data in a high-density signal environment as shown in the top panel of Figure \ref{Fig_EData_R2_R16}. The right panel shows the statistic of synthetic data in a low-density signal environment as shown in the bottom panel of Figure \ref{Fig_EData_R2_R16}.}
	\label{Fig_Dist_R2_R16}
\end{figure*}

To evaluate the performance of completeness and the accuracy of locations for the algorithms under different densities, we analyse the variations of $R$ with flux and SNR and $\Delta X$ in PPV with SNR, which are shown in Figure \ref{Fig_Recall_R2_R16} and Figure \ref{Fig_Dist_R2_R16}, respectively. The average numbers of clumps, the SNR-weighted average $R$, and the SNR-weighted average $\Delta X$ of the varying SNR are summarized in Table \ref{Infor_R2_R16}. 

In the high-density signal environment: The average number of FacetClumps is 3769. FacetClumps has advantages in terms of $R$ (92\%) and $\Delta X$ (0.36 voxel) compared to ConBased, LDC, and FellWalker; $R$ (85.7\%) and $\Delta X$ (0.41 voxel) of ConBased is the second best only to FacetClumps; $R$ (84.9\%) of LDC is better than $R$ (79.9\%) of FellWalker, and $\Delta X$ (0.46 voxel) of LDC is greater than $\Delta X$ (0.45 voxel) of FellWalker. In the low-density signal environment: The average number of FacetClumps is 290. FacetClumps has an advantage over FellWalker and LDC in terms of $R$ (98.8\%); $R$ (98.8\%) of ConBased is higher than that of FellWalker and LDC; $R$ (93.6\%) of LDC is better than $R$ (89.6\%) of FellWalker. $\Delta X$ obtained by FacetClumps (0.28 voxel) and ConBased (0.27 voxel) is greater than that of FellWalker (0.26 voxel) and LDC (0.26 voxel).

In brief, FacetClumps has obvious advantages over other algorithms in high-density signal environments.

\begin{table*}[h]
	\centering
	\caption{\centering The average number of clumps, the SNR-weighted average $R$, and the SNR-weighted average $\Delta X$ of the varying SNR of Data2 and Data3.}
	\begin{tabular}{c|cccccccc}\hline\hline
		Algorithm&$N_{Data2}$&$N_{Data3}$&$R_{Data2}$&$R_{Data3}$&$\Delta X_{Data2}$&$\Delta X_{Data3}$\\\hline
		FellWalker & 3557 & 215 & 79.9\% & 89.6\%& 0.45 & 0.26\\
		LDC & 4492 & 202 & 84.9\% & 93.6\%& 0.46 & 0.26\\
		ConBased & 4776 & 237 & 85.7\% & 98.8\% & 0.41 & 0.27\\
		FacetClumps & 3769 & 290 & 92\% & 98.8\%& 0.36 & 0.28\\\hline
	\end{tabular}
	\label{Infor_R2_R16}
\end{table*}

\section{{Configuration parameters of FacetClumps under different resample factors}}\label{Resample}
Downsampling increase the overlap among clumps, making the peaks of some clumps disappear and become undetectable. To roughly assess the effect of downsampling on the valid peaks, we apply FacetClumps to detect the $^{13}CO$ emission within $11.7^{\circ} \leq l \leq 13.4^{\circ}$, $0.22^{\circ} \leq b \leq 1.05^{\circ}$ and 5 km s$^{-1}$ $\leq v \leq$ 35 km s$^{-1}$ to obtain the attributes of the clumps, which are used to generate a simulated data cube with the same volume and similar properties as the observed one. The simulated data is downsampled in spectral channels (hereafter $V$), spatial direction (hereafter $LB$), and simultaneously in both spatial direction and spectral channels (hereafter $LBV$) with a sampling factor of 2, respectively. We check the change in the number of peaks in each resampled data. The results show that the number of detectable clumps decreased by about 3.5\%, 6.7\%, and 9.7\%, respectively. It should be noted that these estimates only give a lower limit for reductions of detectable clumps in observed data, as some peaks in the original simulated data already have disappeared and downsampling makes it easier to eliminate peaks in irregular clumps.

To enable FacetClumps to be applied to multiple datasets, associating the parameters of FacetClumps with the parameters of the instrument will minimize systematic differences. In Section \ref{sec:3.3.2}, we have discussed the influence of instrument-independent parameters $SWindow$ and $KBins$ on the performance of FacetClumps. Here, we discuss the associations between the instrument-related parameter $SRecursionLBV$ and the FWHM of the instrument beam $FwhmBeam$ and the velocity resolution of the instrument $VeloRes$, which are expressed by equation (\ref{SRecursionLBV}). Besides, the associations between resampling and $FwhmBeam$ and $VeloRes$ are expressed by equation (\ref{SFactorLBV}).

\begin{equation}\label{SRecursionLBV}
\it{SRecursionLB=(F_0+FwhmBeam)^2, SRecursionV=V_0+VeloRes}
\end{equation}

\begin{equation}\label{SFactorLBV}
\it{FwhmBeam'=FwhmBeam\times SFactorLB, VeloRes'=VeloRes\times SFactorV}
\end{equation}

\noindent where, $SFactorLB$ and $SFactorV$ are the resample factors in spatial direction and spectral channels, respectively. $F_0$ and $V_0$ can be customized according to the minimum size of the clumps required for scientific objectives, with $F_0=2$ and $V_0=3$ as the default values. The values of $FwhmBeam$ and $VeloRes$ can be modified to suit the specific instrument parameters of the radio telescope used to acquire the molecular line data.

\begin{figure*}
	\centering
\centerline{\includegraphics[width=7in]{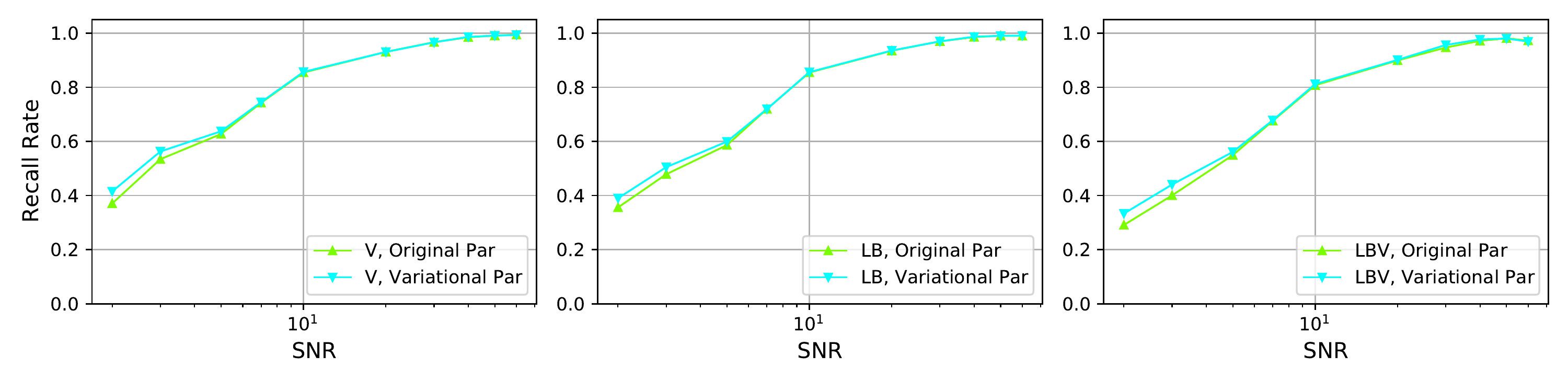}}
	\caption{The statistics of $R$ of resampling along different directions. V, LB, and LBV represent the statistics in data downsampled in the spectral channel, data downsampled in the spatial direction and data downsampled in both the spatial direction and the spectral channel, respectively. 'Original Par' means that $SRecursionLBV$ uses the same value as when detecting unresampled synthetic data, and 'Variational Par' means that $SRecursionLBV$ varies according to formula (\ref{SRecursionLBV}).}
	\label{Fig_Recall_Resample_Reduction}
\end{figure*}

To explore the robustness of $SRecursionLBV$ in different sampled data, we downsample the synthetic data of Data1 ($FwhmBeam=2,VeloRes=2$) described in Section \ref{sec:3.3.3} in $V$, $LB$, and $LBV$ with a sampling factor of 2 ($SFactorLB=0.5,SFactorV=0.5$), respectively. The statistics of $R$ of resampling along different directions are shown in Figure \ref{Fig_Recall_Resample_Reduction}. $SRecursionLBV$ adopts two different types of values, which are respectively the same value as when detecting unresampled synthetic data and the values varying with (\ref{SRecursionLBV}). Figure \ref{Fig_Recall_Resample_Reduction} shows $R$ of low SNR can be slightly improved by using variational parameters. Moreover, resampling can change the minimum size of clumps that are required, making the variational parameters even more desirable. Therefore, the following analysis is based on the statistics derived from the utilization of the variational parameters. 

\begin{figure*}
	\centering
\centerline{\includegraphics[width=7in]{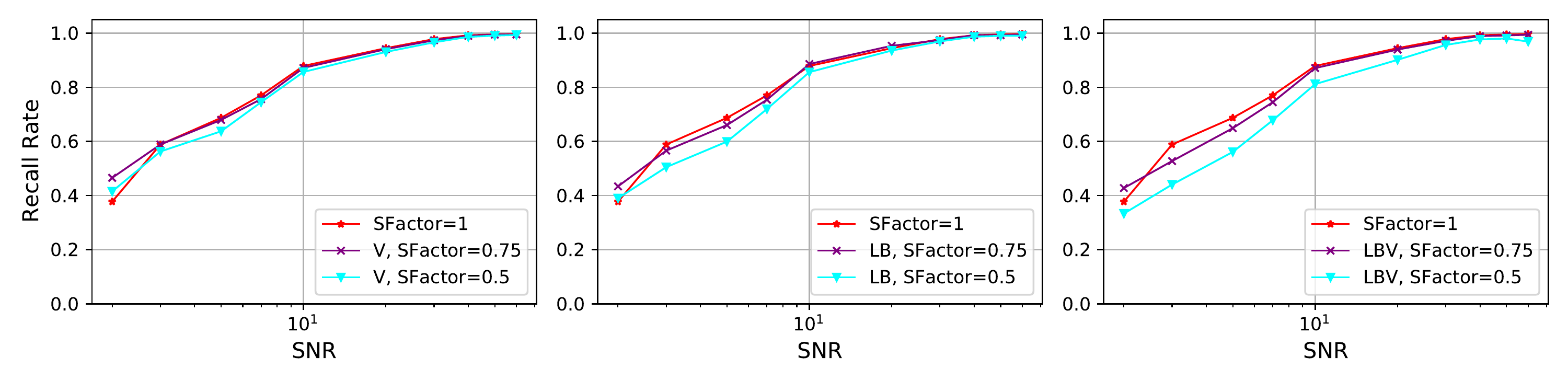}}
	\caption{The statistics of $R$ of resampling with different resample factors and along different directions. $SFactor=1$ means that no resampling is applied to the data. $V, SFactor=0.5$ means that the data is downsampled with a resample factor of 2 in the spectral channel, and other labels have similar meanings. The plots are the statistics of utilizing variational parameters.}
	\label{Fig_Recall_Resample_DSFactor}
\end{figure*}

To further investigate the robustness of parameters, we have added a new similar downsampled dataset ($SFactorLB=0.75,SFactorV=0.75$), and the statistics of $R$ of resampling with different resample factors and along different directions are shown in Figure \ref{Fig_Recall_Resample_DSFactor}. For $SFactorLB=0.75,SFactorV=0.75$: The reductions of the mean values of $R$ are all less than 0.5\%. For $SFactorLB=0.5,SFactorV=0.5$: When the SNR is greater than 10, the reductions of $R$ in $V$ sampling and $LB$ sampling are less than 1\%, and that in $LBV$ sampling is less than 4\%; The mean values of $R$ in different resampled data decrease by about 0.8\%, 1.2\%, and 4\% respectively, all of which are smaller than the corresponding reductions in the simulated data. Experimental results indicate that the parameters of FacetClumps can be effectively adapted to different sampled data, particularly for clumps with SNR of more than 10, where the parameters are reliable.

\begin{table}
	\centering
	\caption{FellWalker parameters.}
	\label{FellWalker parameters}
	\begin{tabular}{l}
FellWalker.AllowEdge=1\\
		FellWalker.CleanIter=1\\
		FellWalker.FlatSlope=2*RMS\\
		FellWalker.FwhmBeam=2\\
		FellWalker.MaxBad=0.05\\
		FellWalker.MaxJump=4\\
		FellWalker.MinDip=1*RMS\\
		FellWalker.MinHeight=3*RMS\\
		FellWalker.MinPix=27\\
		FellWalker.Noise=2*RMS\\
		FellWalker.RMS=RMS\\
		FellWalker.VeloRes=2\\
	\end{tabular}
\end{table}

\begin{table}
	\centering
	\caption{LDC parameters.}
	\label{LDC parameters}
	\begin{tabular}{l}
		LDC.RMS=RMS\\
		LDC.Threshold=2*RMS\\
		LDC.GradientMin=0.01\\
		LDC.DistanceMin=4\\
		LDC.PeakMin=5*RMS\\
		LDC.PixelMin=27\\
\end{tabular}
\end{table}

\begin{table}
	\centering
	\caption{ConBased parameters.}
	\label{ConBased parameters}
	\begin{tabular}{l}
ConBased.RMS=RMS\\
		ConBased.Threshold=2*RMS\\
		ConBased.RegionMin=27\\
		ConBased.ClumpMin=216\\
		ConBased.DIntensity=2*RMS\\
		ConBased.DDistance=8\\
\end{tabular}
\end{table}

\begin{table}
	\centering
	\caption{FacetClumps parameters.}
	\label{Facet parameters}
	\begin{tabular}{l}
FacetClumps.RMS=RMS\\
		FacetClumps.Threshold=2*RMS,[$n*RMS$]\\
		FacetClumps.SWindow=3,[3,5,7]\\
		FacetClumps.KBins=35,[10,\ldots,60]\\
		FacetClumps.FwhmBeam=2\\
		FacetClumps.VeloRes=2\\
		FacetClumps.SRecursionLBV=[16,5]\\
	\end{tabular}
\end{table}

\section{Configuration parameters of different algorithms}\label{Parameters}
The configuration parameters are presented in Table \ref{FellWalker parameters}, Table \ref{LDC parameters}, Table \ref{ConBased parameters}, and Table \ref{Facet parameters}. For FacetClumps, we provide both default and recommended values.

\section{Partial Clumps in the high overlap area in Section 3.3.3}\label{PClumps}
\begin{figure*}
	\centering
	\vspace{0cm}
	\begin{minipage}[t]{0.24\textwidth}
		\centering
		\centerline{\includegraphics[width=1.9in]{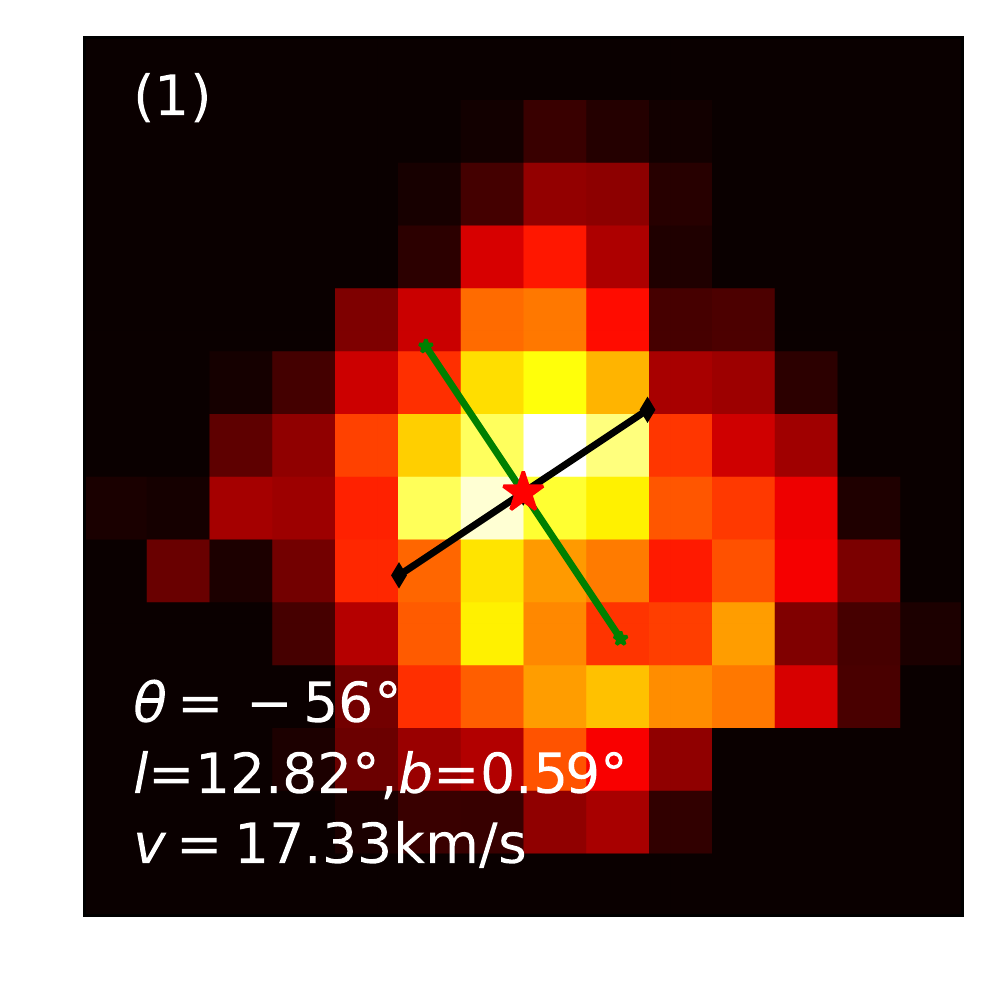}}
\end{minipage}\begin{minipage}[t]{0.24\textwidth}
		\centering
		\centerline{\includegraphics[width=1.9in]{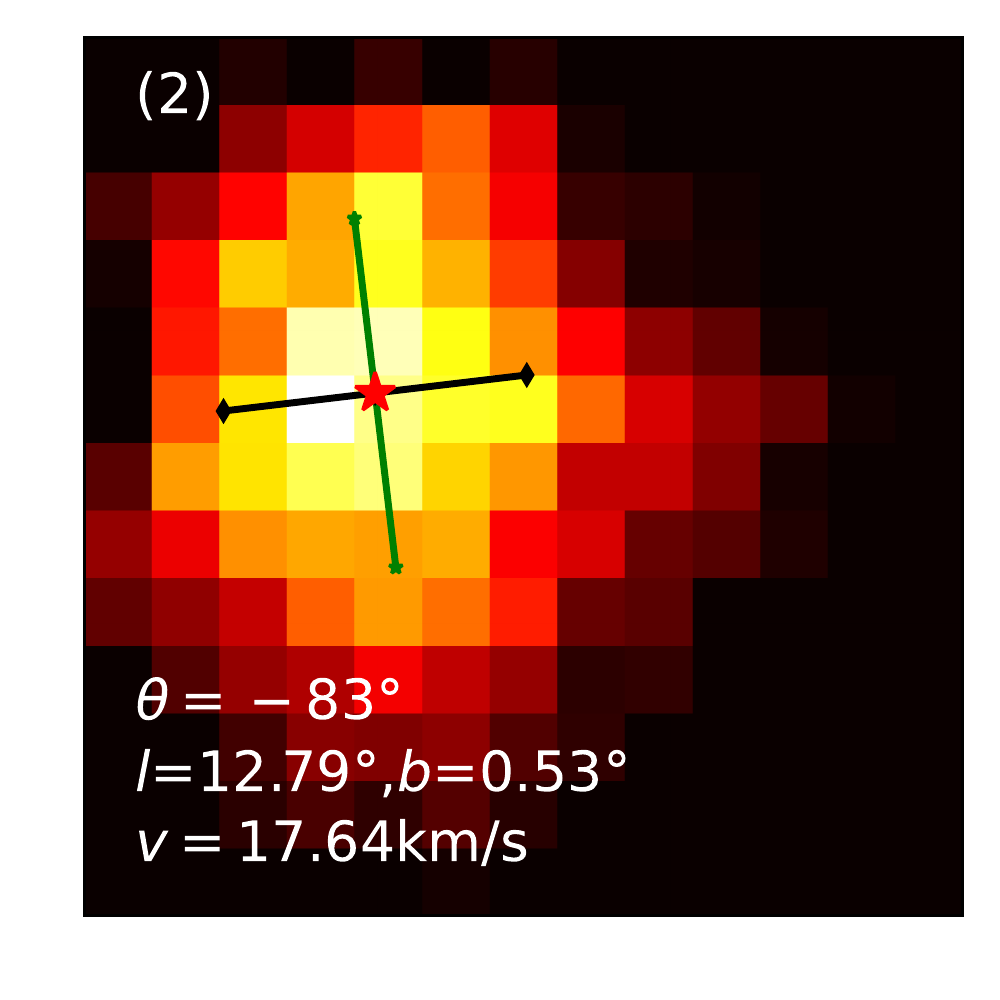}}
\end{minipage}\begin{minipage}[t]{0.24\textwidth}
		\centering
		\centerline{\includegraphics[width=1.9in]{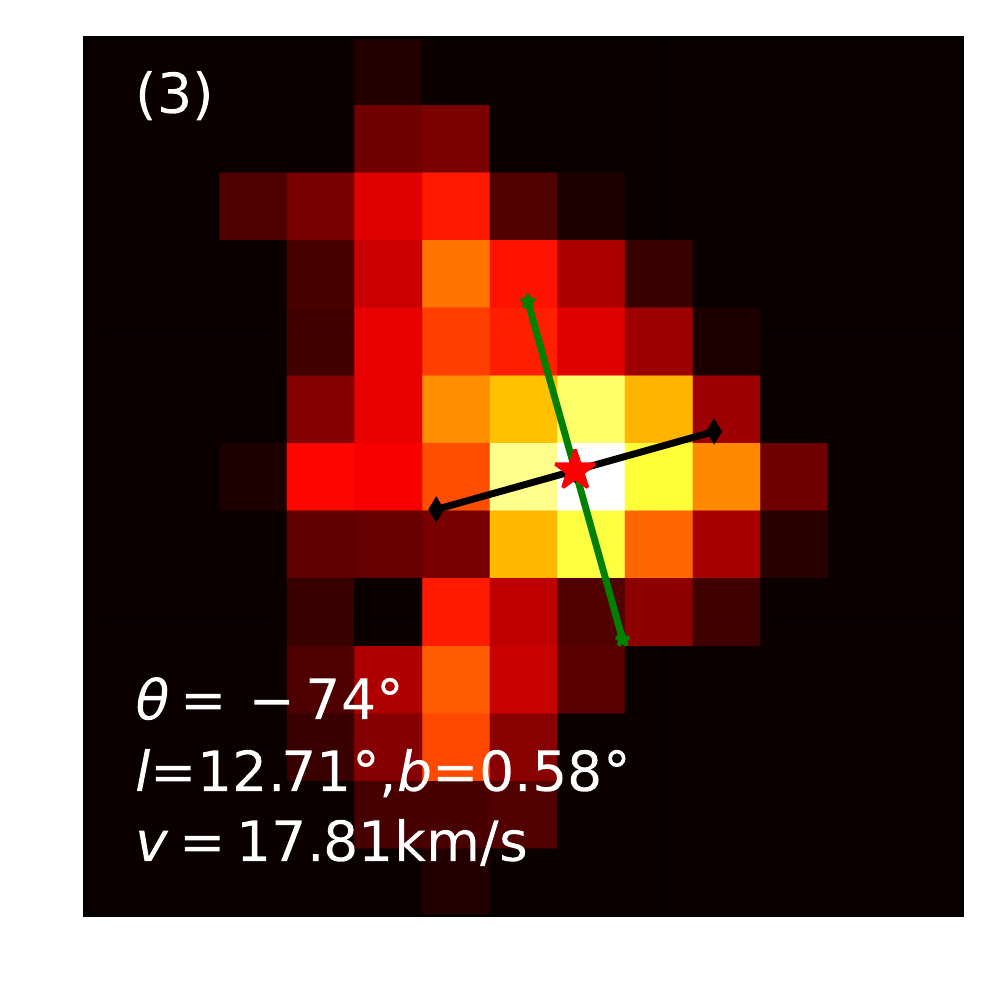}}
\end{minipage}\begin{minipage}[t]{0.24\textwidth}
		\centering
		\centerline{\includegraphics[width=1.9in]{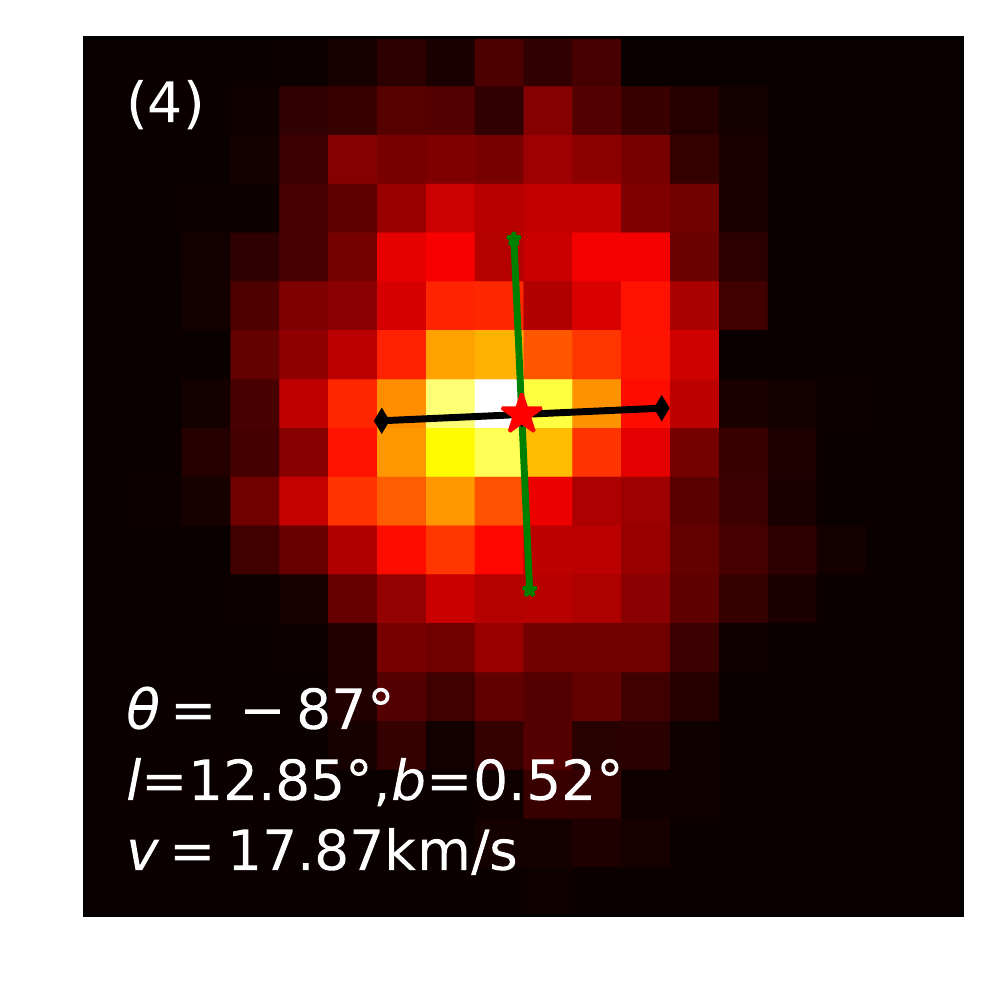}}
\end{minipage}

	\begin{minipage}[t]{0.24\textwidth}
		\centering
		\centerline{\includegraphics[width=1.9in]{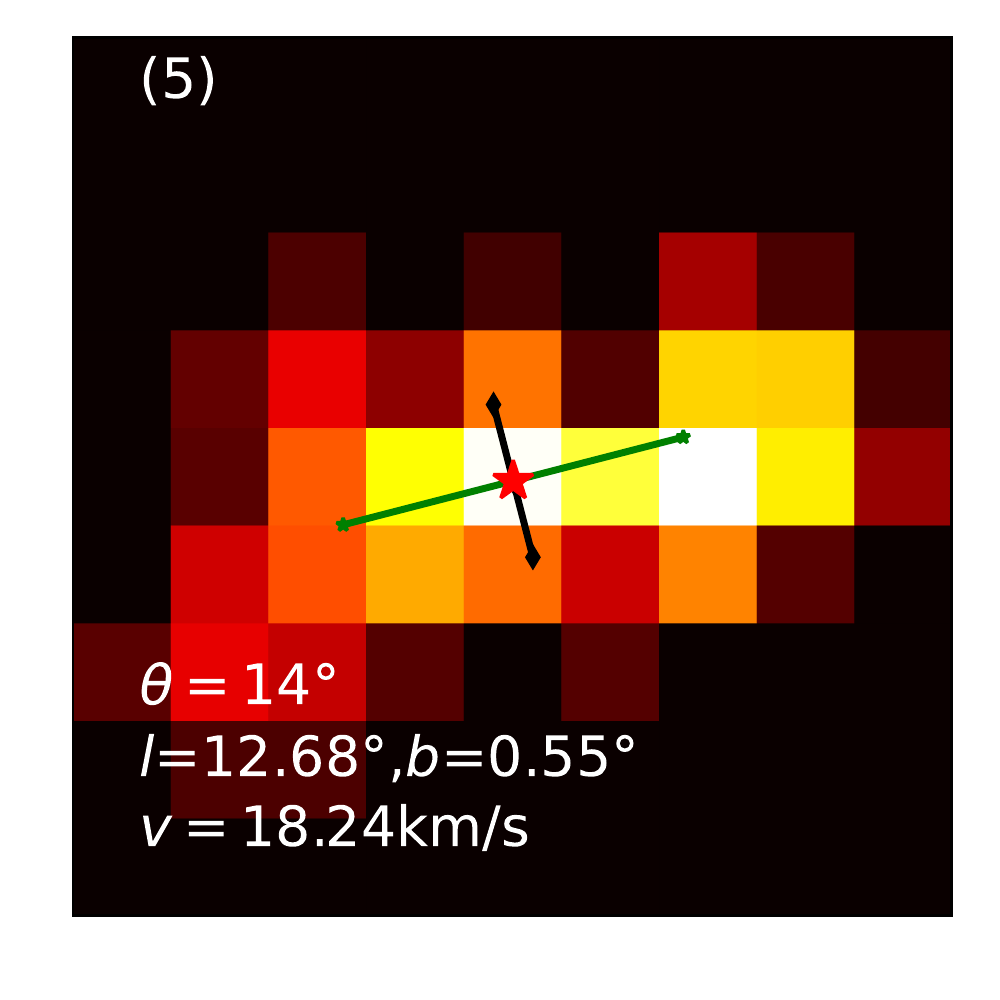}}
\end{minipage}\begin{minipage}[t]{0.24\textwidth}
		\centering
		\centerline{\includegraphics[width=1.9in]{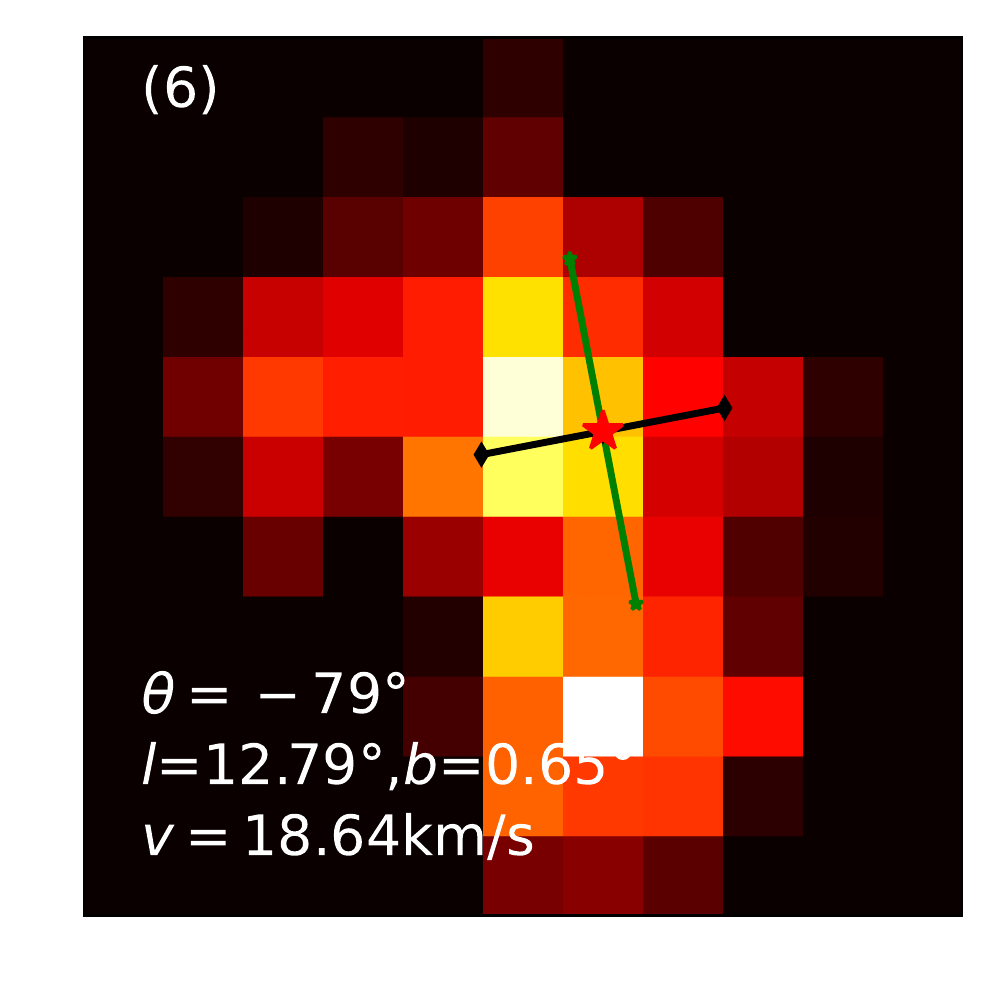}}
\end{minipage}\begin{minipage}[t]{0.24\textwidth}
		\centering
		\centerline{\includegraphics[width=1.9in]{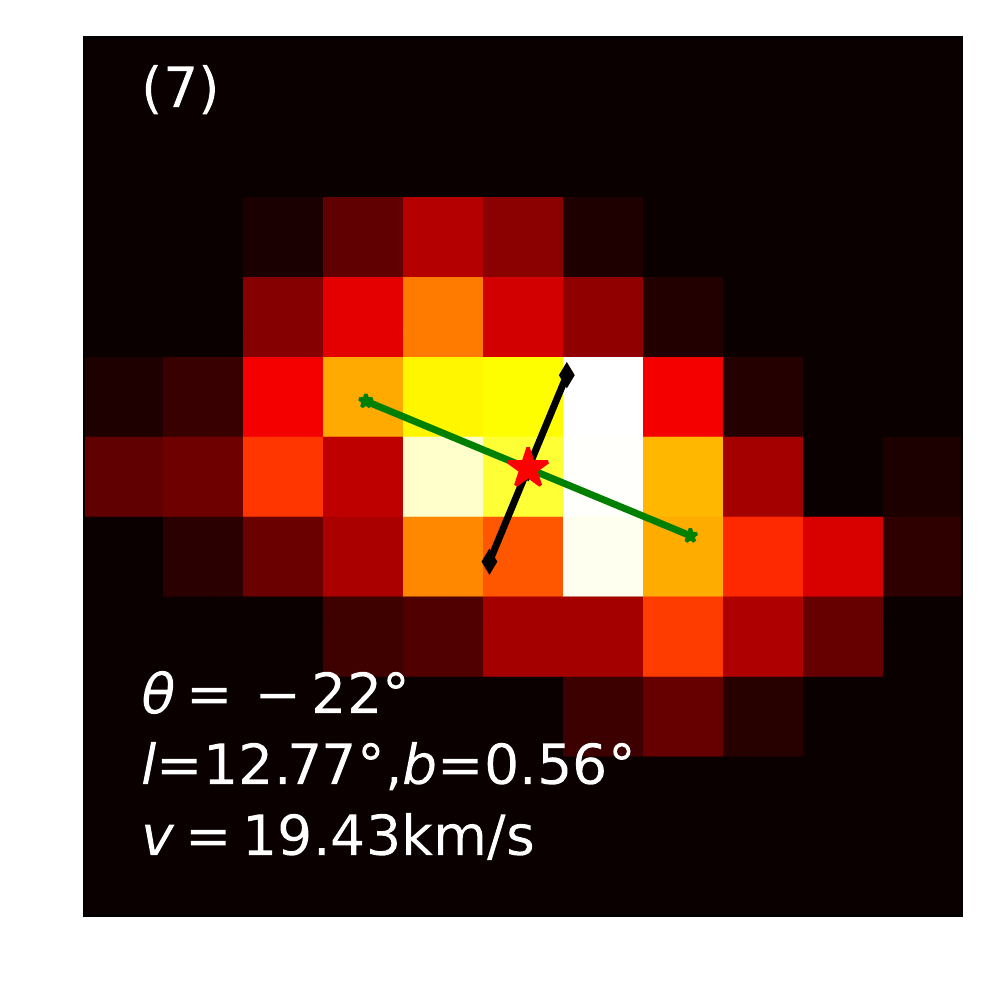}}
\end{minipage}\begin{minipage}[t]{0.24\textwidth}
		\centering
		\centerline{\includegraphics[width=1.9in]{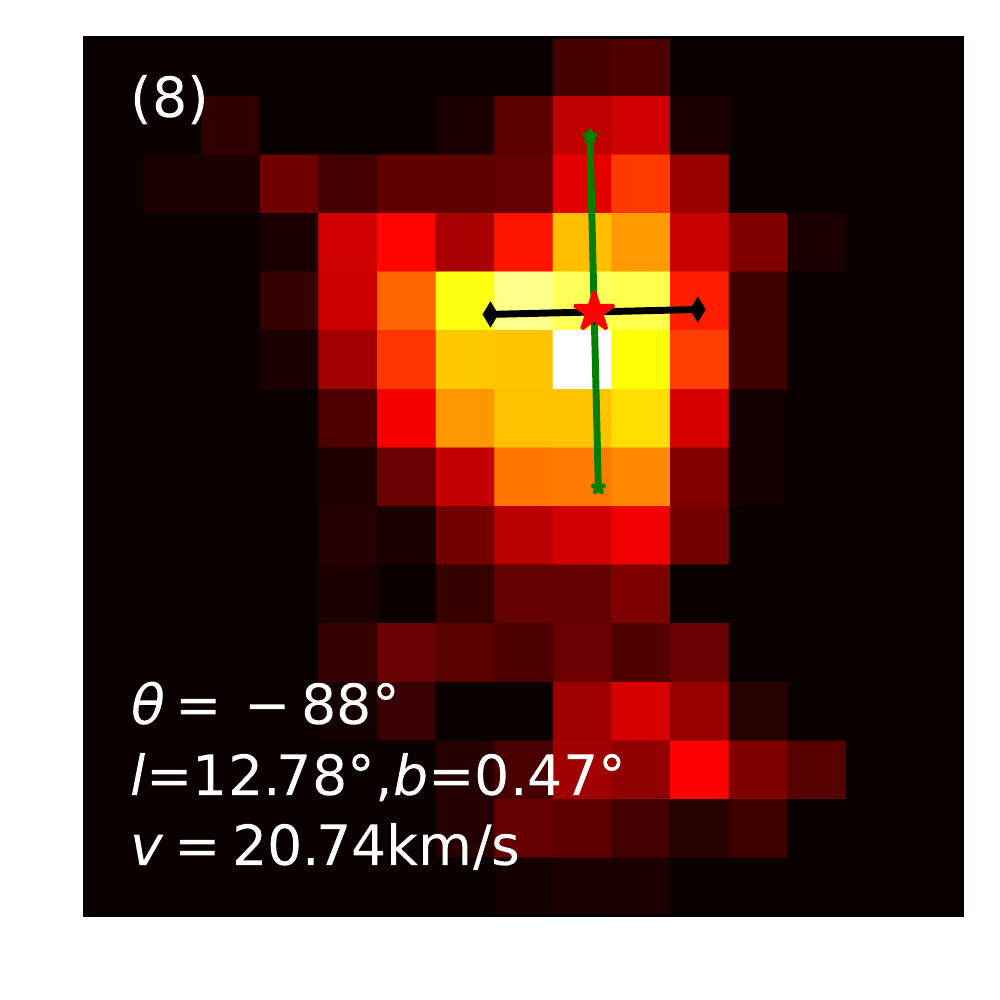}}
\end{minipage}

	\begin{minipage}[t]{0.24\textwidth}
		\centering
		\centerline{\includegraphics[width=1.9in]{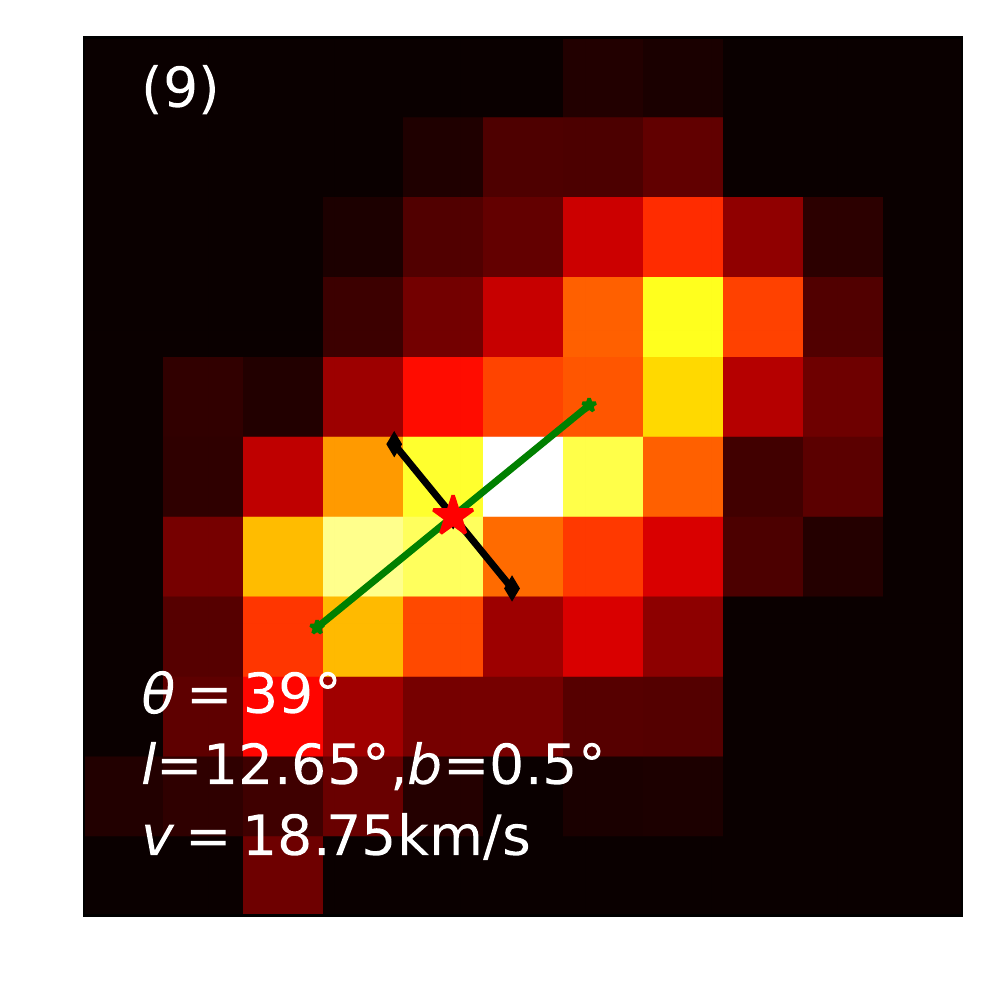}}
\end{minipage}\begin{minipage}[t]{0.24\textwidth}
		\centering
		\centerline{\includegraphics[width=1.9in]{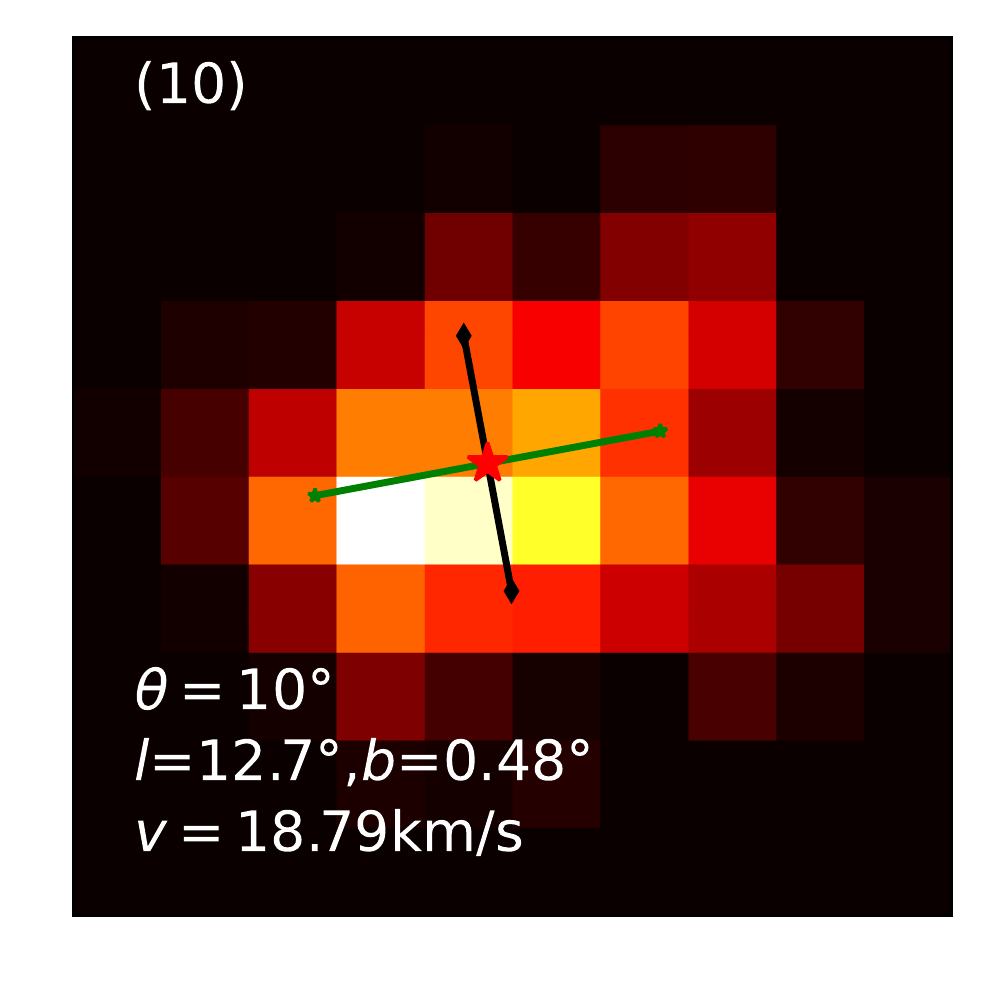}}
\end{minipage}\begin{minipage}[t]{0.24\textwidth}
		\centering
		\centerline{\includegraphics[width=1.9in]{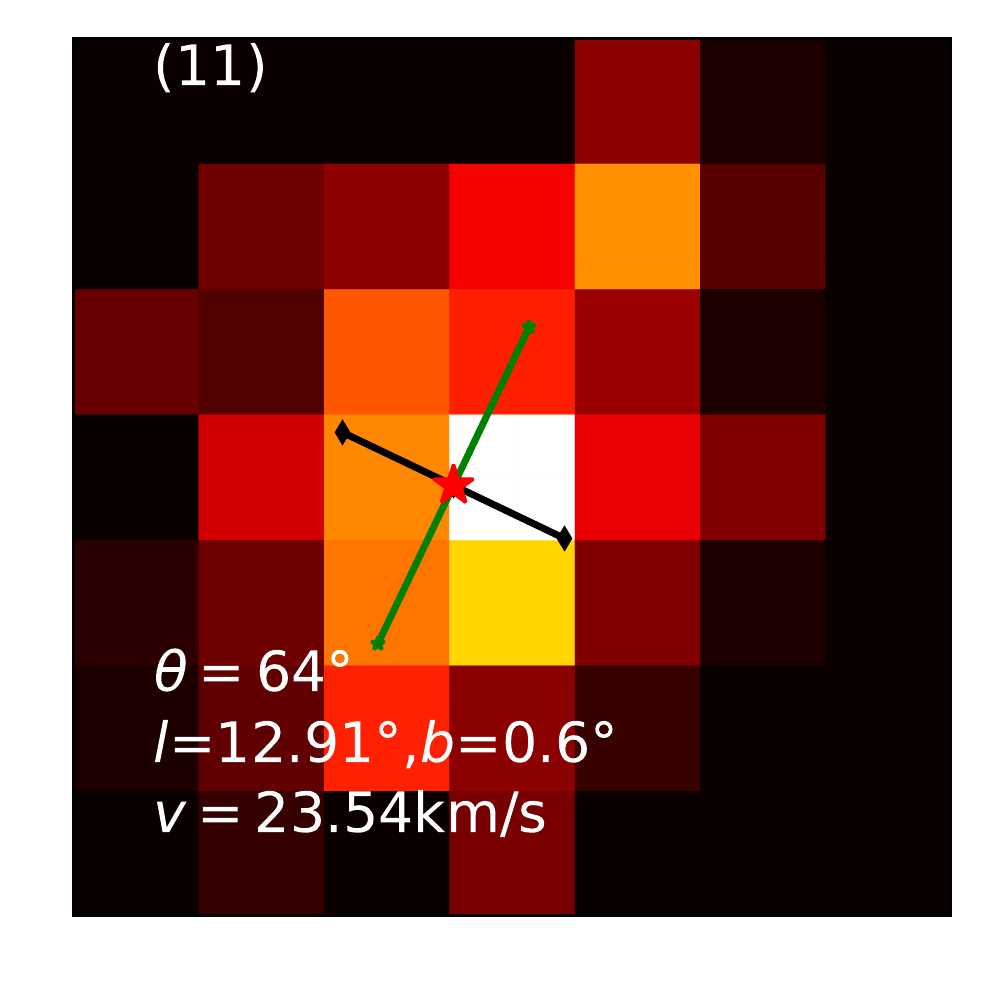}}
\end{minipage}\begin{minipage}[t]{0.24\textwidth}
		\centering
		\centerline{\includegraphics[width=1.9in]{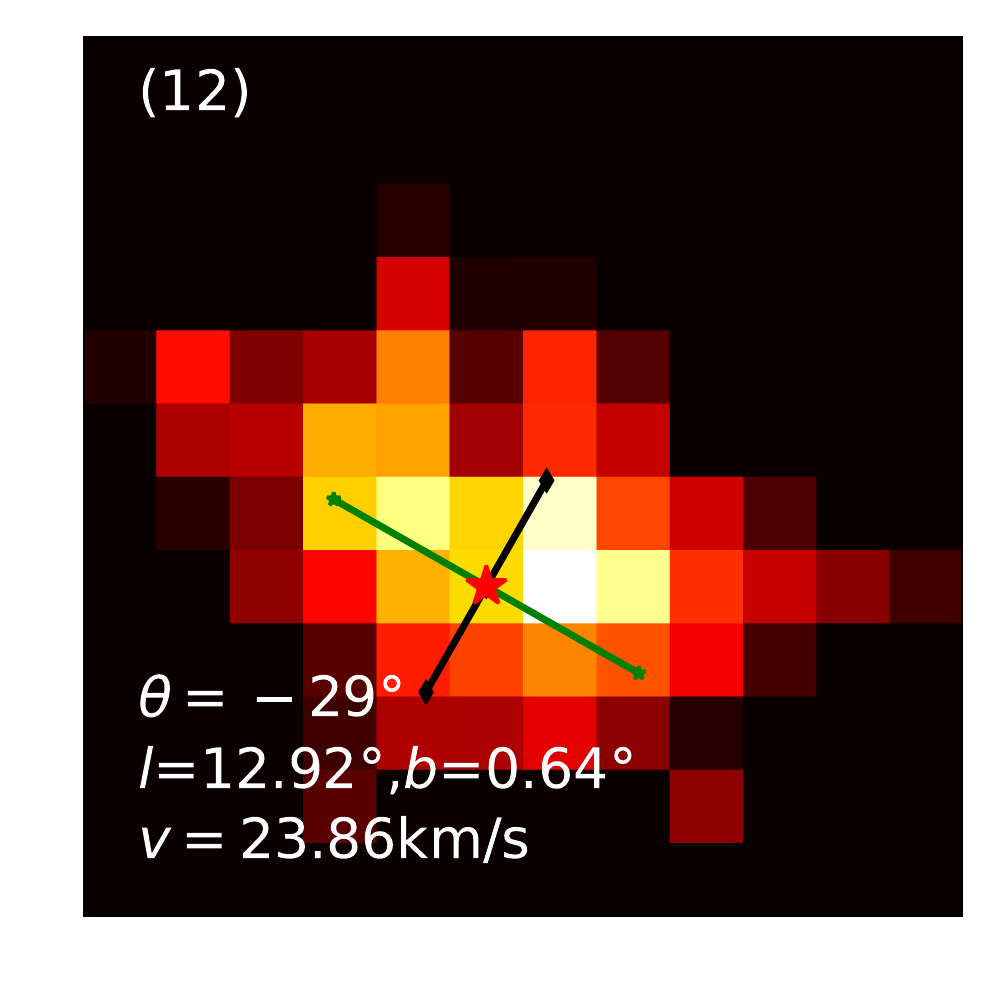}}
\end{minipage}

	\begin{minipage}[t]{0.24\textwidth}
		\centering
		\centerline{\includegraphics[width=1.9in]{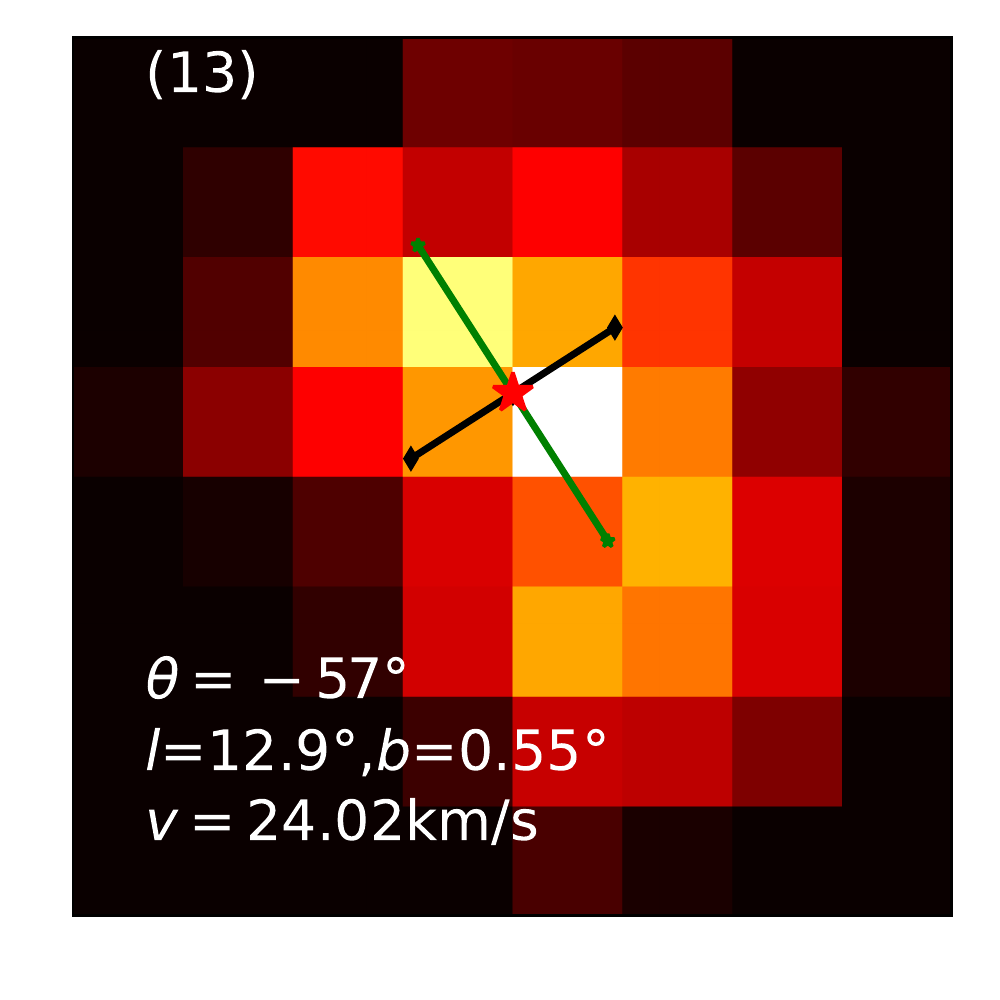}}
\end{minipage}\begin{minipage}[t]{0.24\textwidth}
		\centering
		\centerline{\includegraphics[width=1.9in]{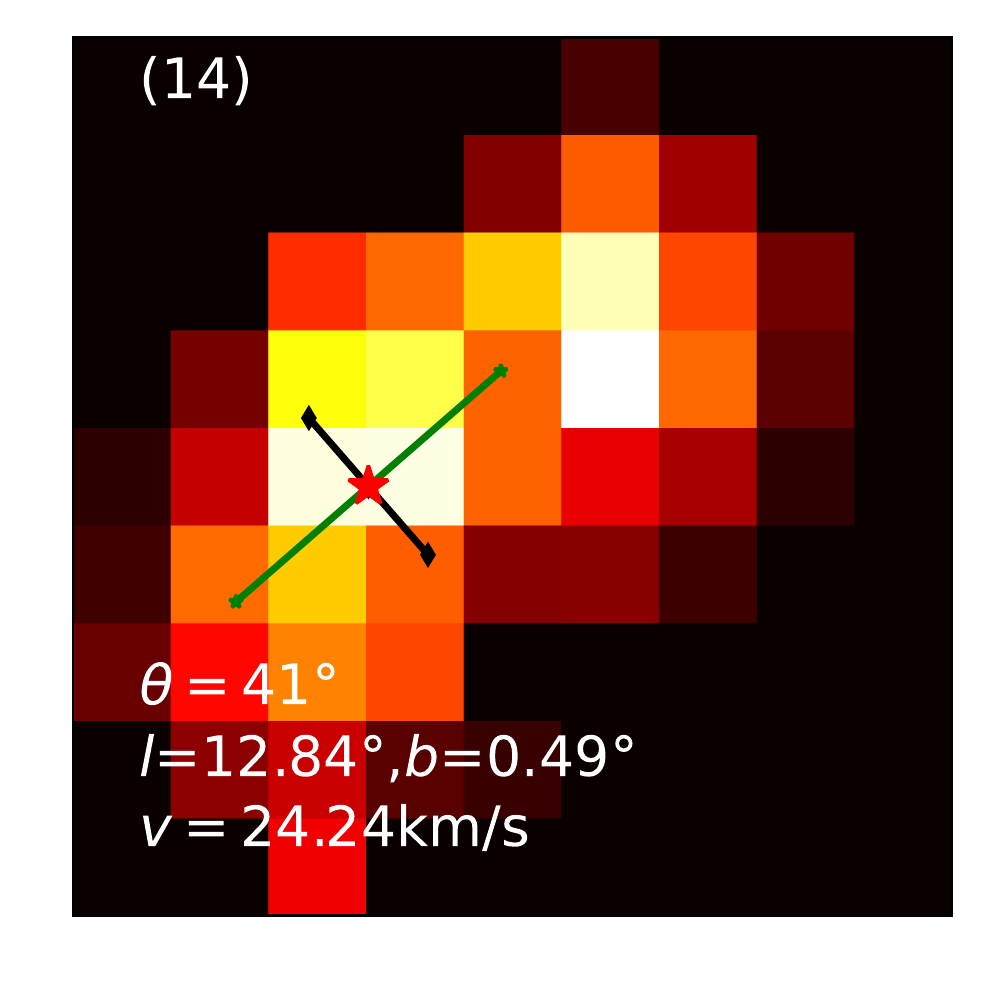}}
\end{minipage}\begin{minipage}[t]{0.24\textwidth}
		\centering
		\centerline{\includegraphics[width=1.9in]{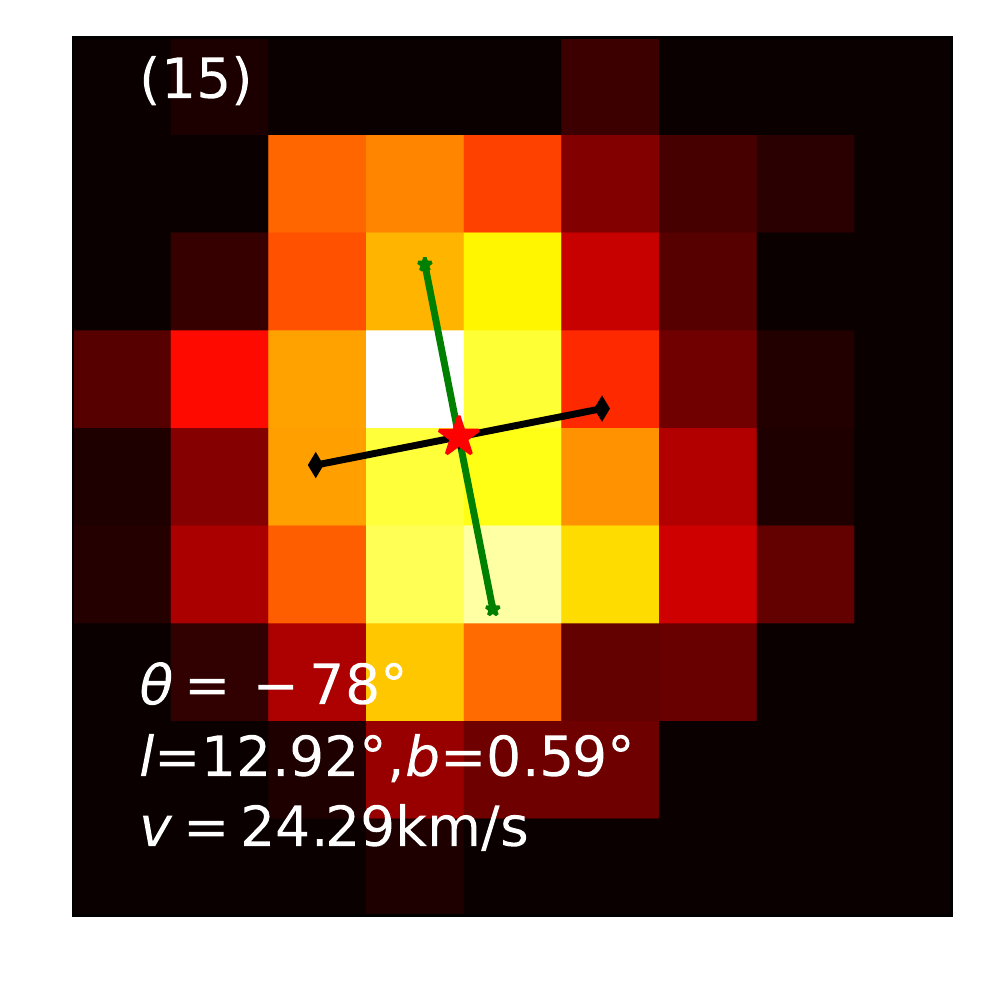}}
\end{minipage}\begin{minipage}[t]{0.24\textwidth}
		\centering
		\centerline{\includegraphics[width=1.9in]{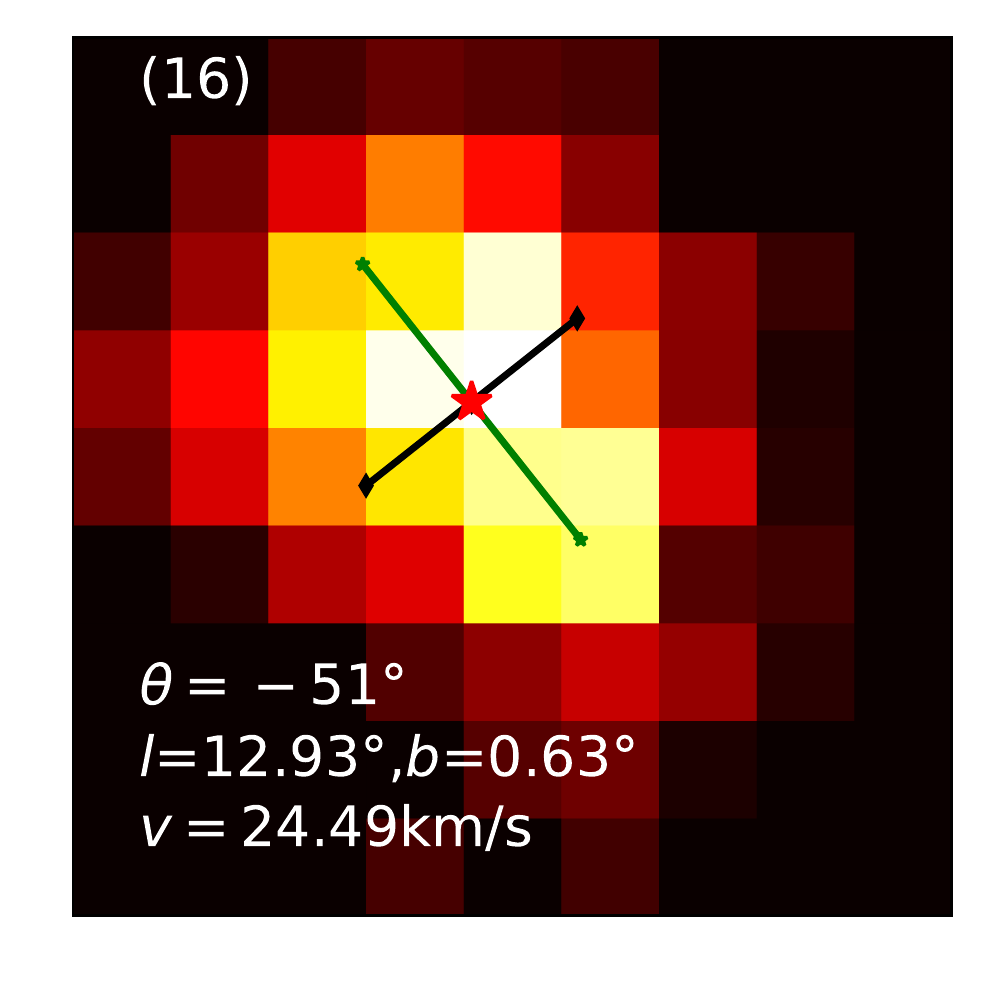}}
\end{minipage}

	\begin{minipage}[t]{0.24\textwidth}
		\centering
		\centerline{\includegraphics[width=1.9in]{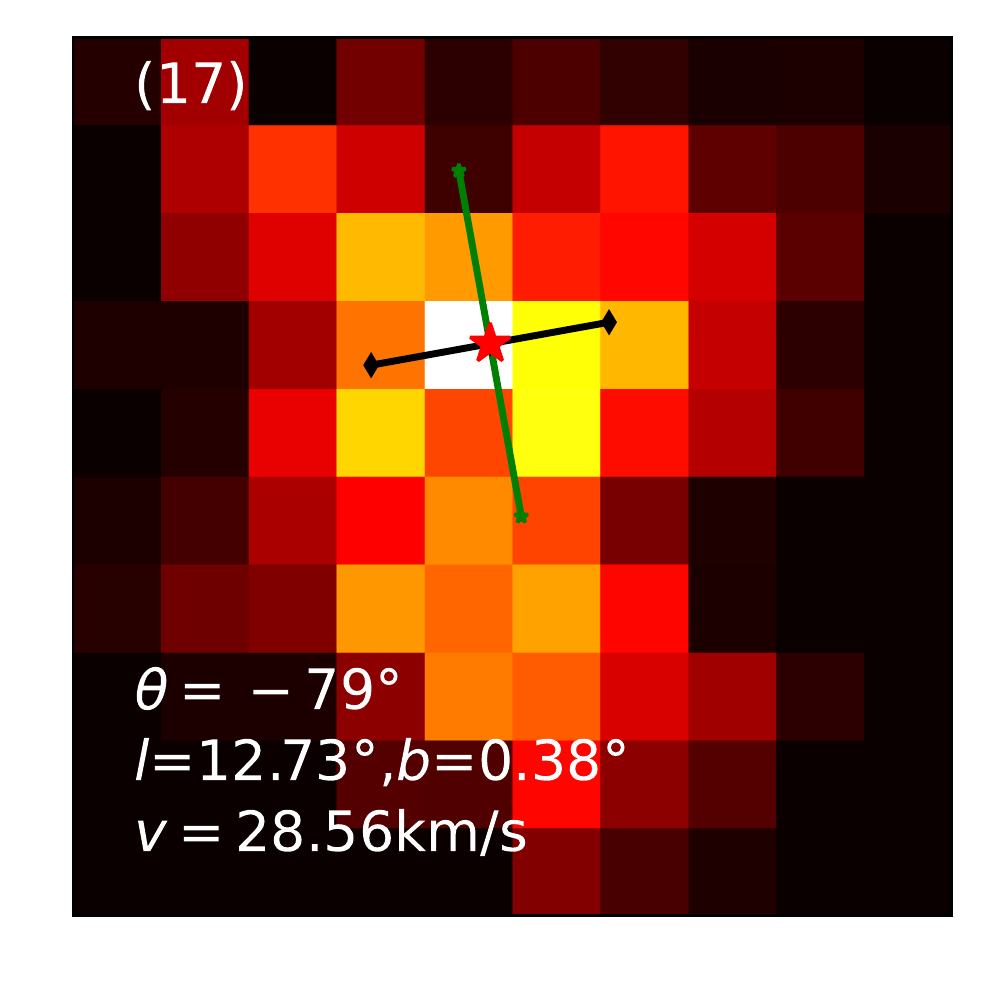}}
\end{minipage}\begin{minipage}[t]{0.24\textwidth}
		\centering
		\centerline{\includegraphics[width=1.9in]{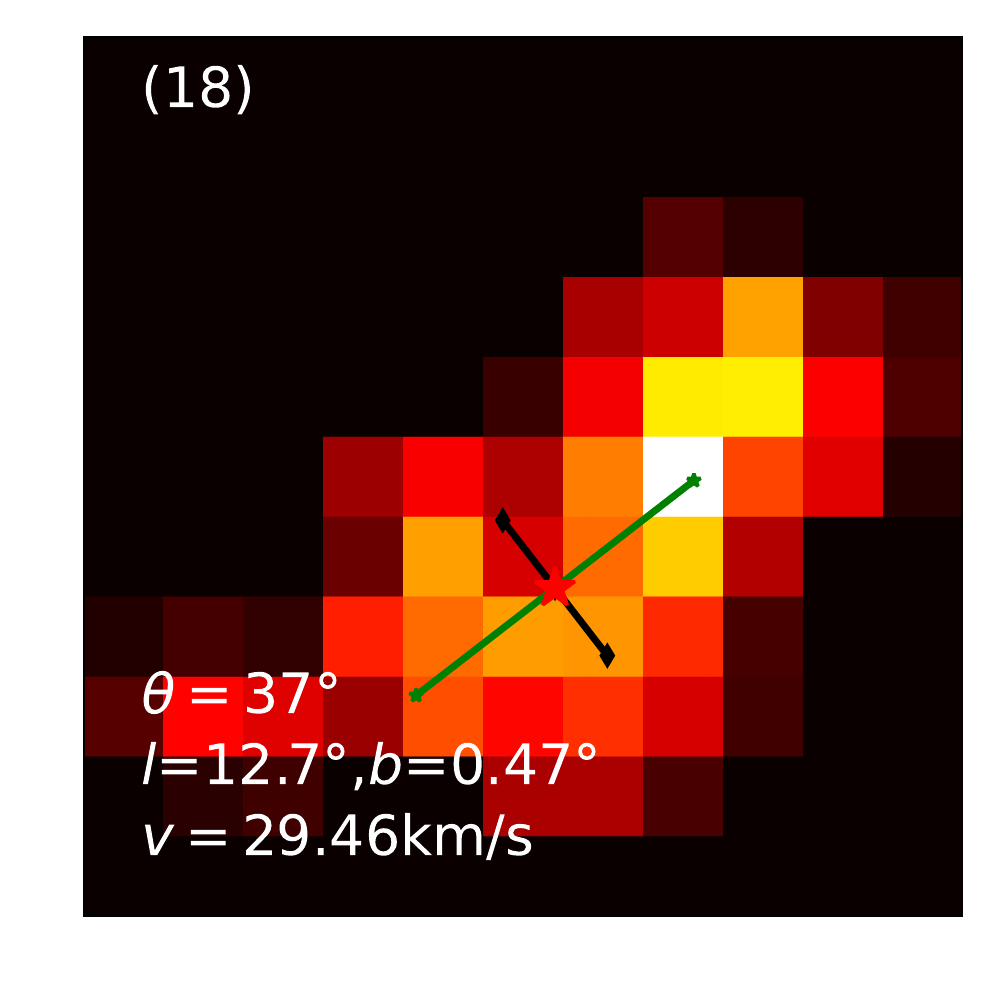}}
\end{minipage}\begin{minipage}[t]{0.24\textwidth}
		\centering
		\centerline{\includegraphics[width=1.9in]{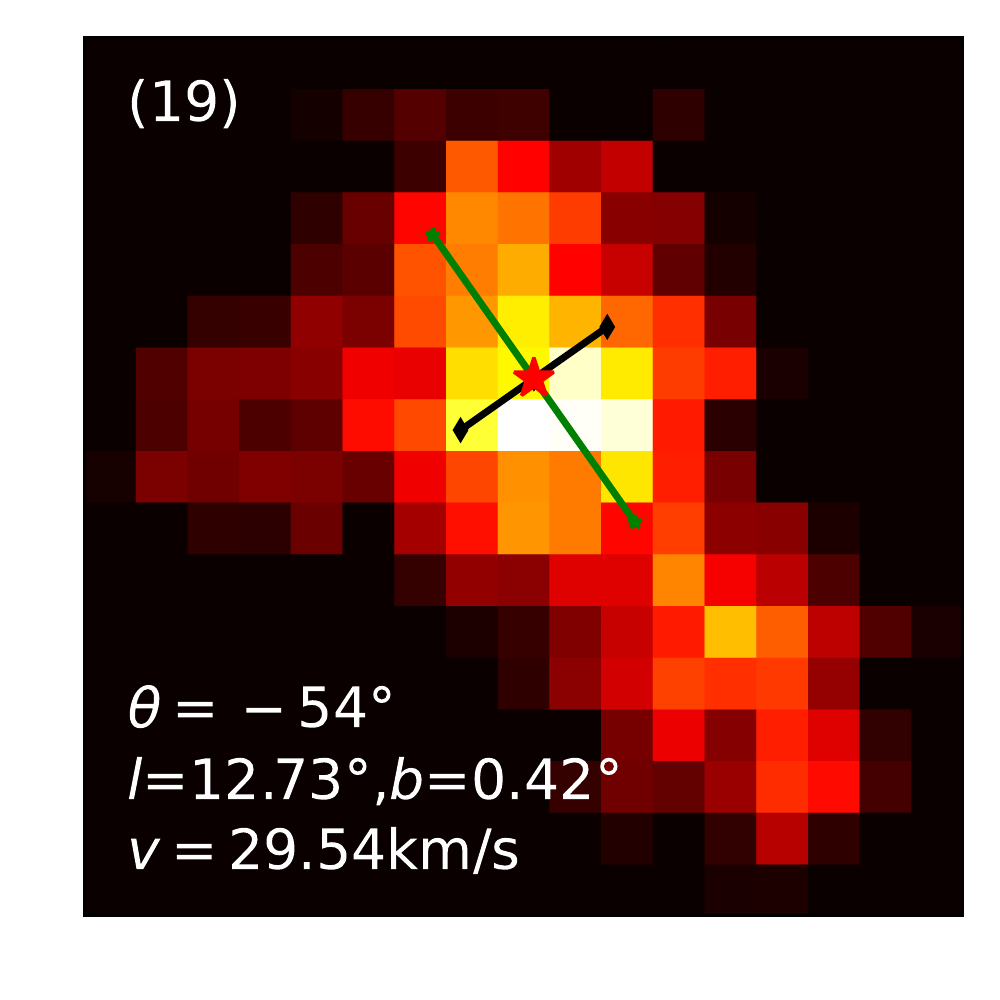}}
\end{minipage}\begin{minipage}[t]{0.24\textwidth}
		\centering
		\centerline{\includegraphics[width=1.9in]{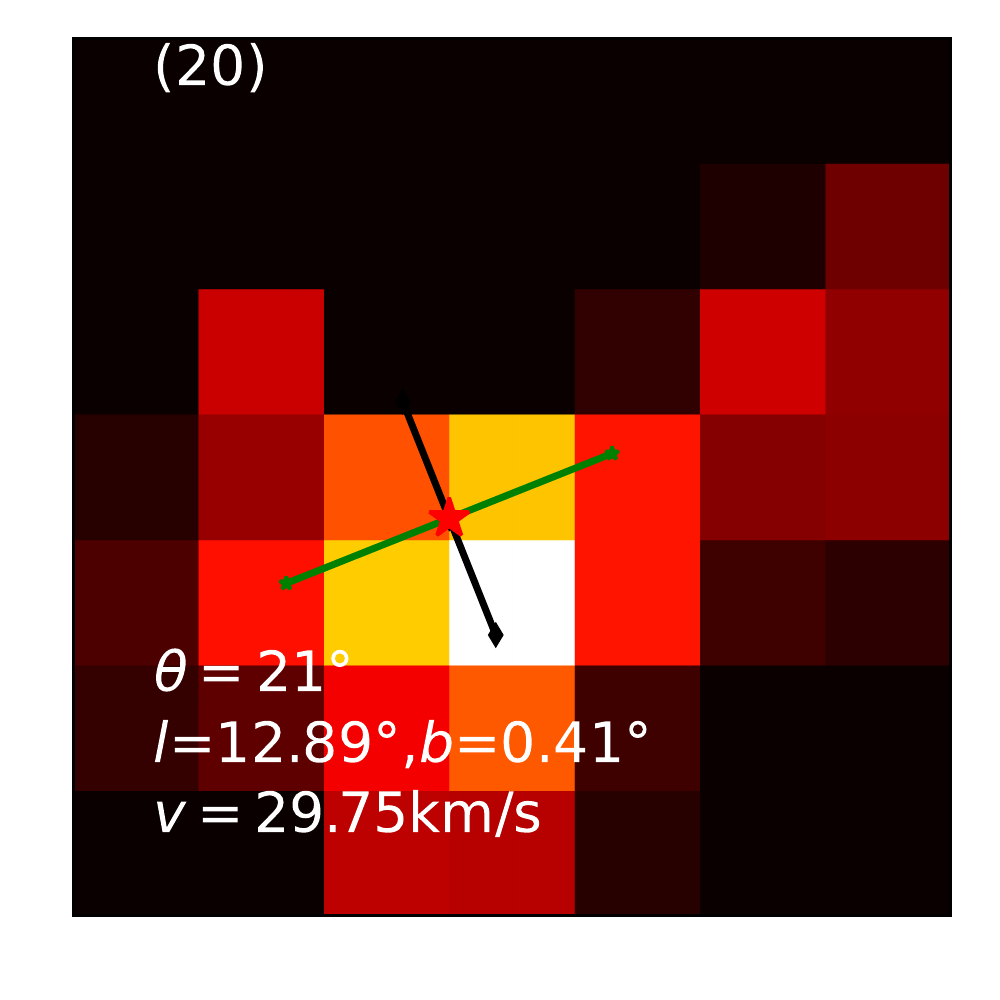}}
\end{minipage}\caption{Velocity-integrated intensity images of clumps in the high overlap area as shown in Figure \ref{Fig_C18O_0}. Each panel is presented in the same manner as in Figure \ref{Fig_C18O_1}.}
	\label{Fig_C18O_2}
\end{figure*}

\begin{figure*}
	\centering
	\vspace{0cm}
	\begin{minipage}[t]{0.24\textwidth}
		\centering
		\centerline{\includegraphics[width=1.9in]{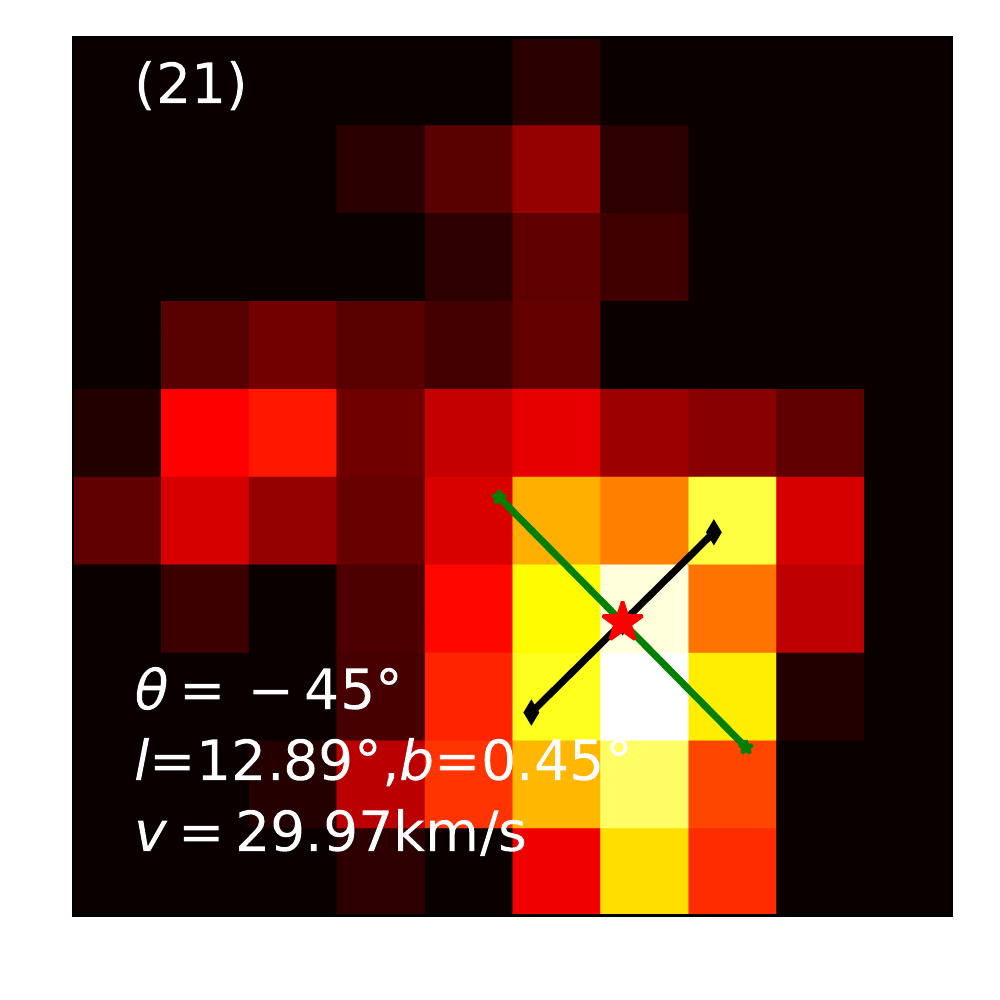}}
\end{minipage}\begin{minipage}[t]{0.24\textwidth}
		\centering
		\centerline{\includegraphics[width=1.9in]{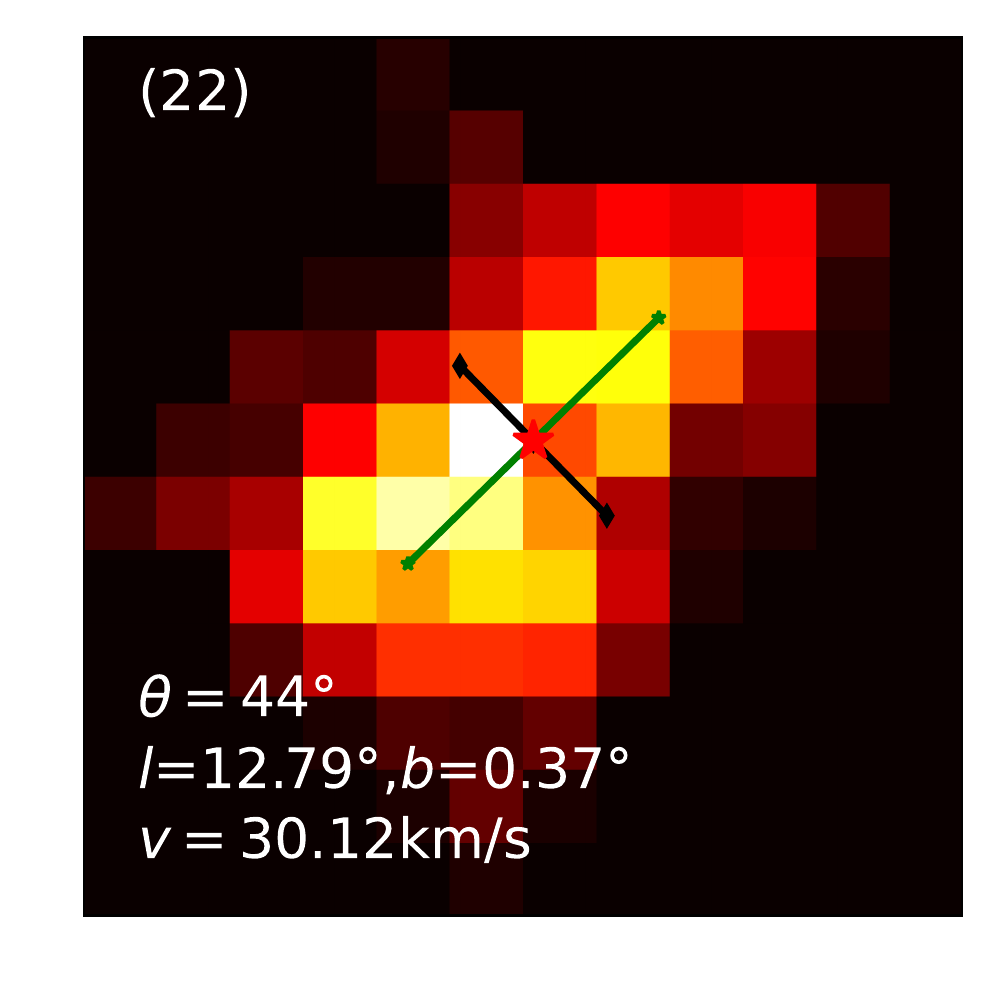}}
\end{minipage}\begin{minipage}[t]{0.24\textwidth}
		\centering
		\centerline{\includegraphics[width=1.9in]{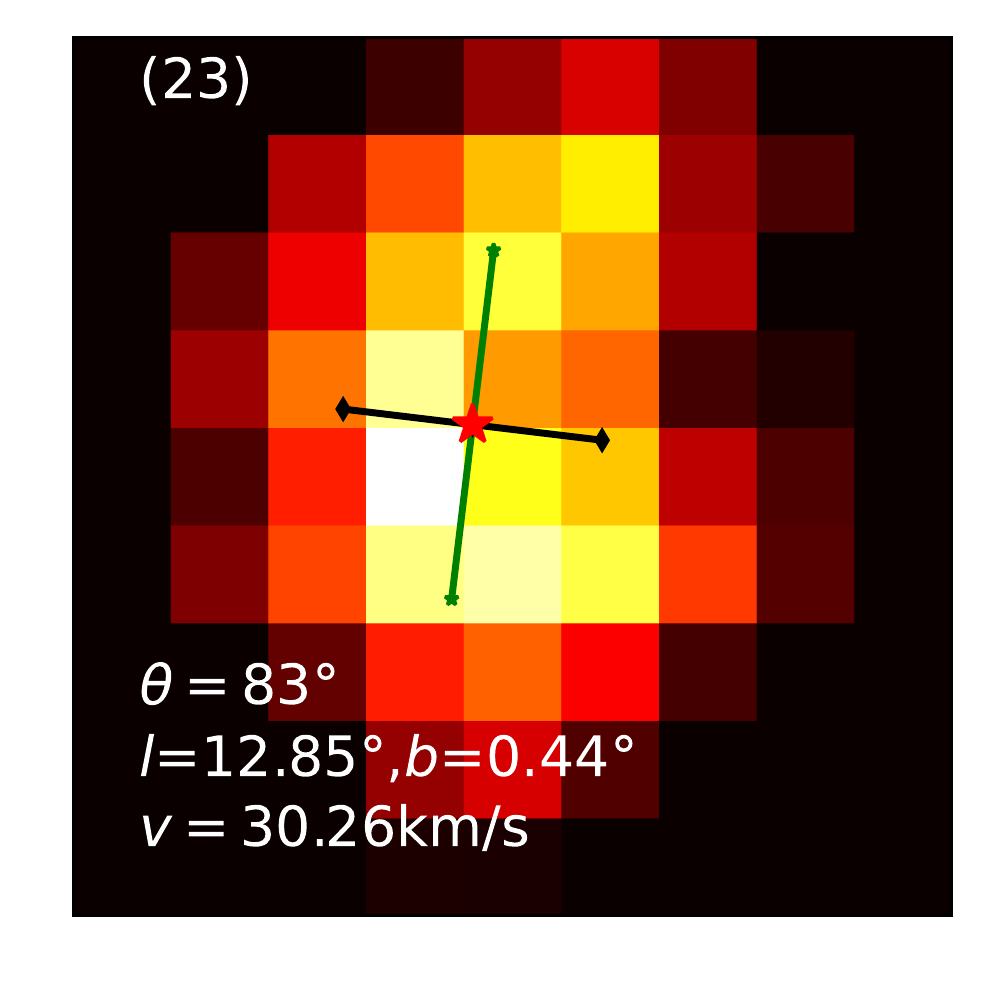}}
\end{minipage}\begin{minipage}[t]{0.24\textwidth}
		\centering
		\centerline{\includegraphics[width=1.9in]{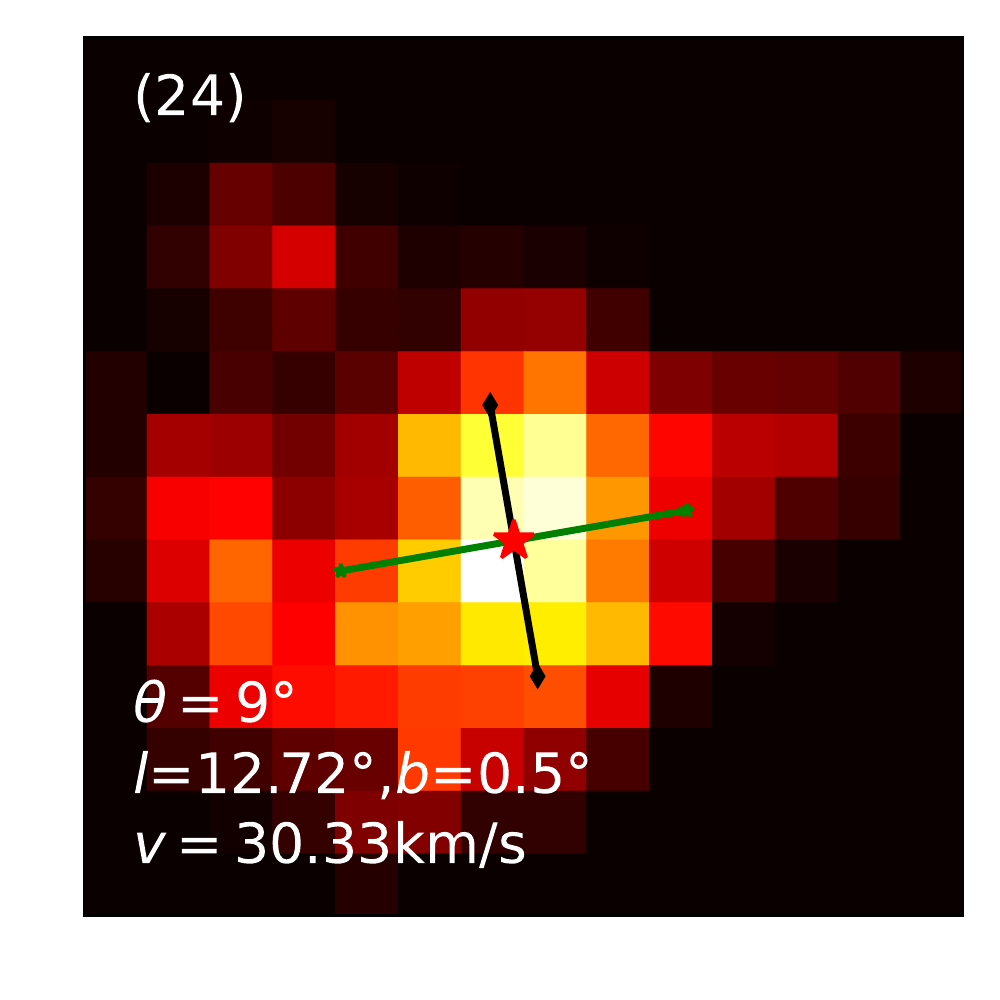}}
\end{minipage}

	\begin{minipage}[t]{0.24\textwidth}
		\centering
		\centerline{\includegraphics[width=1.9in]{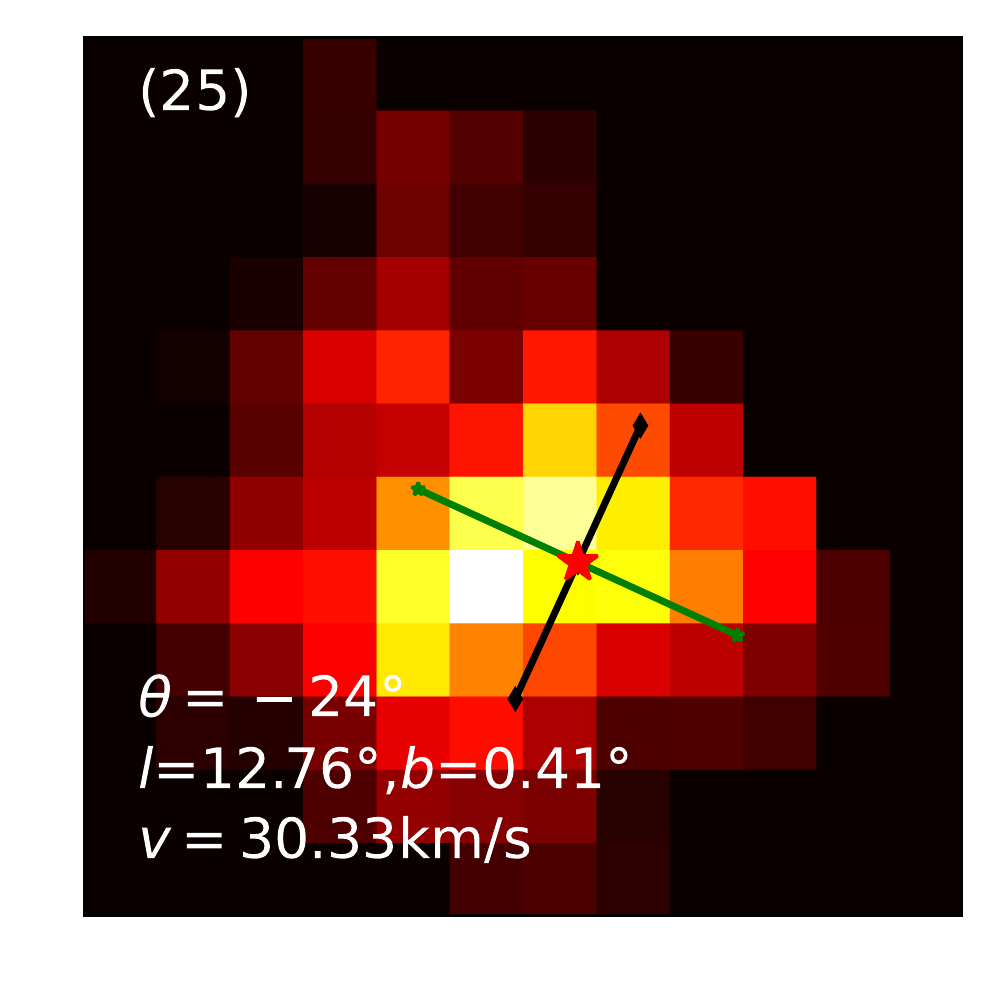}}
\end{minipage}\begin{minipage}[t]{0.24\textwidth}
		\centering
		\centerline{\includegraphics[width=1.9in]{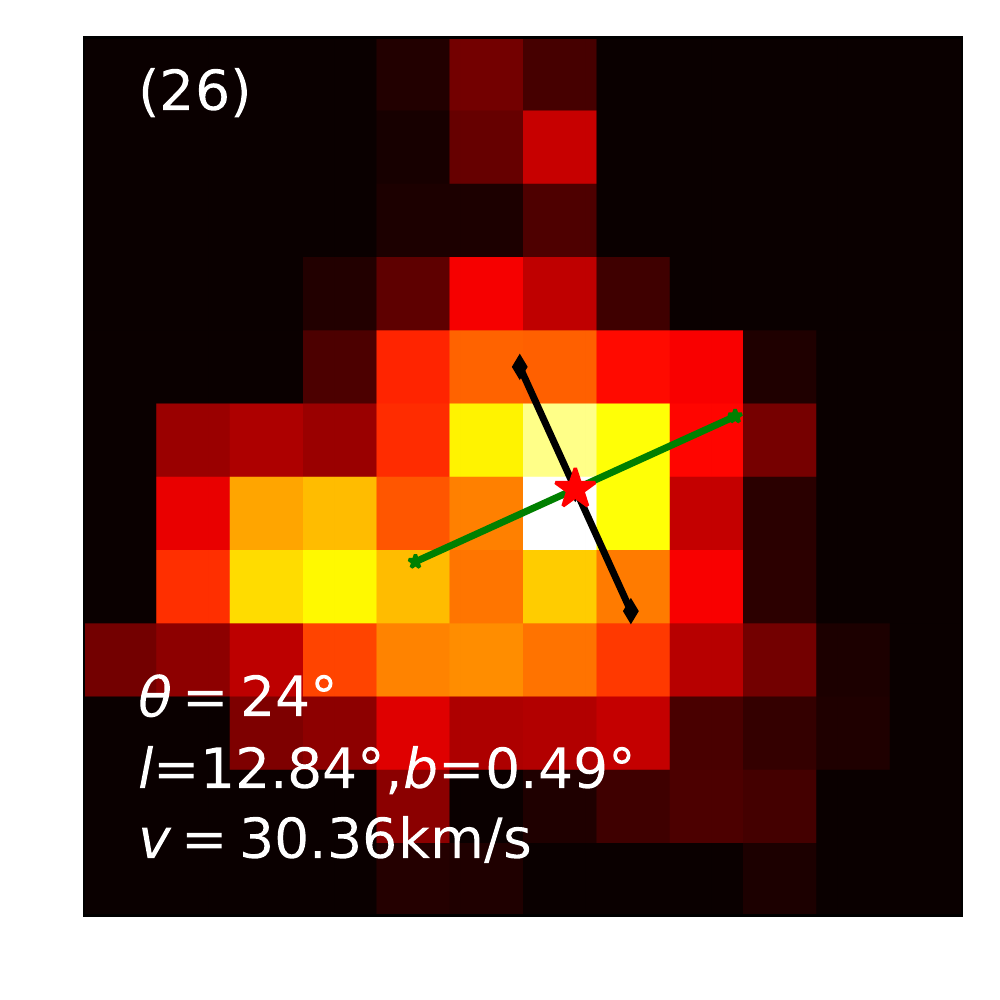}}
\end{minipage}\begin{minipage}[t]{0.24\textwidth}
		\centering
		\centerline{\includegraphics[width=1.9in]{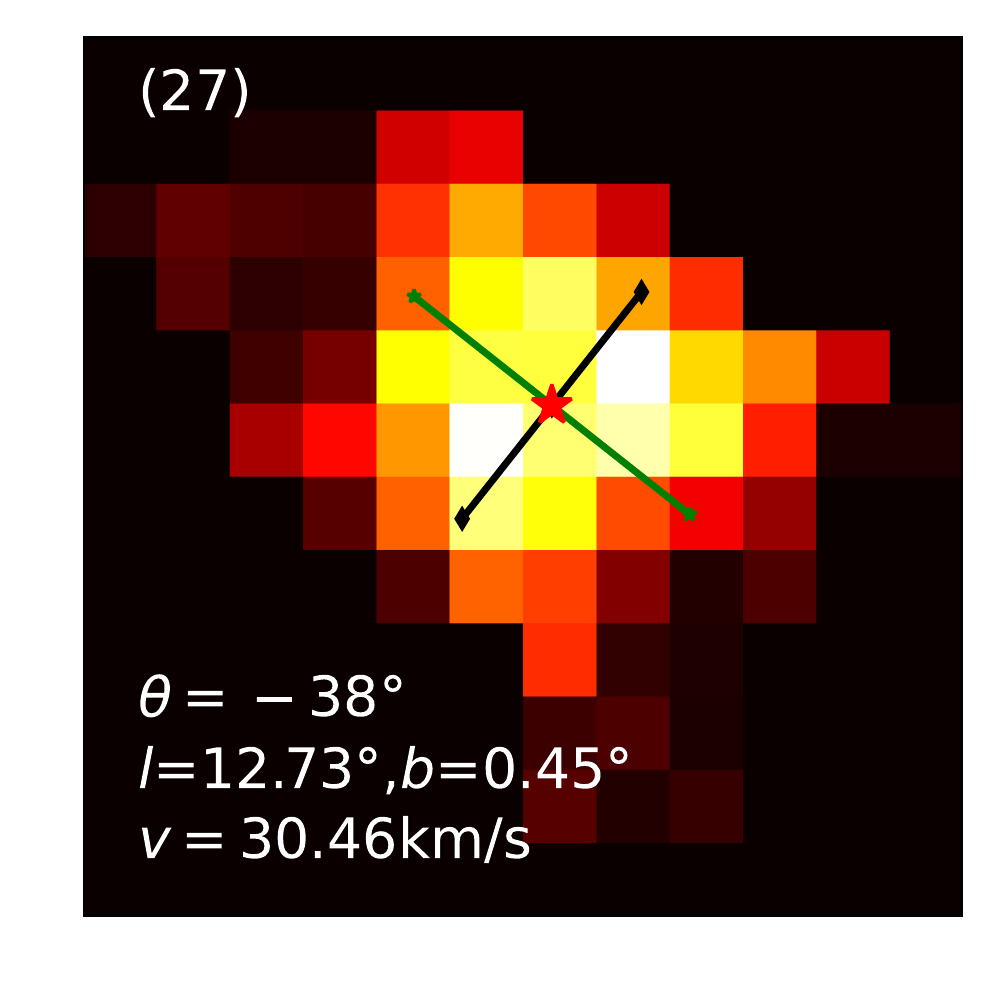}}
\end{minipage}\begin{minipage}[t]{0.24\textwidth}
		\centering
		\centerline{\includegraphics[width=1.9in]{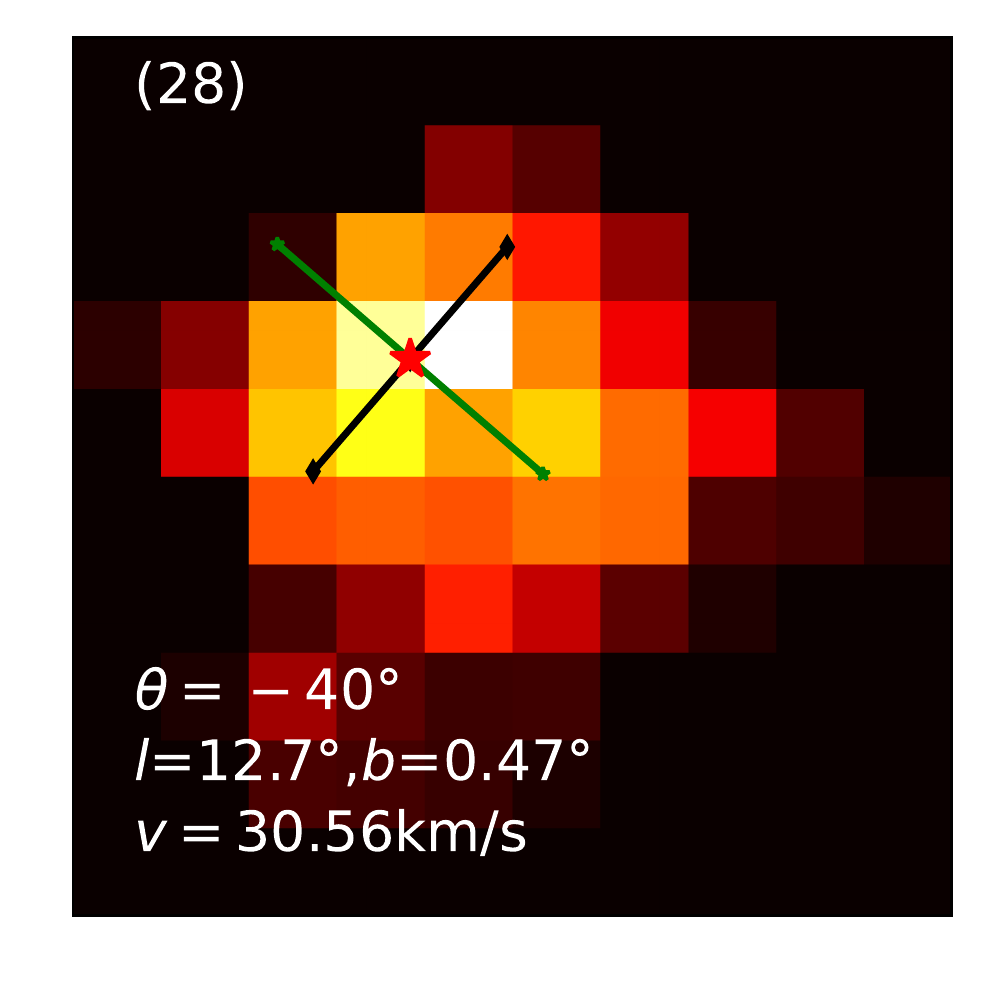}}
\end{minipage}

	\begin{minipage}[t]{0.24\textwidth}
		\centering
		\centerline{\includegraphics[width=1.9in]{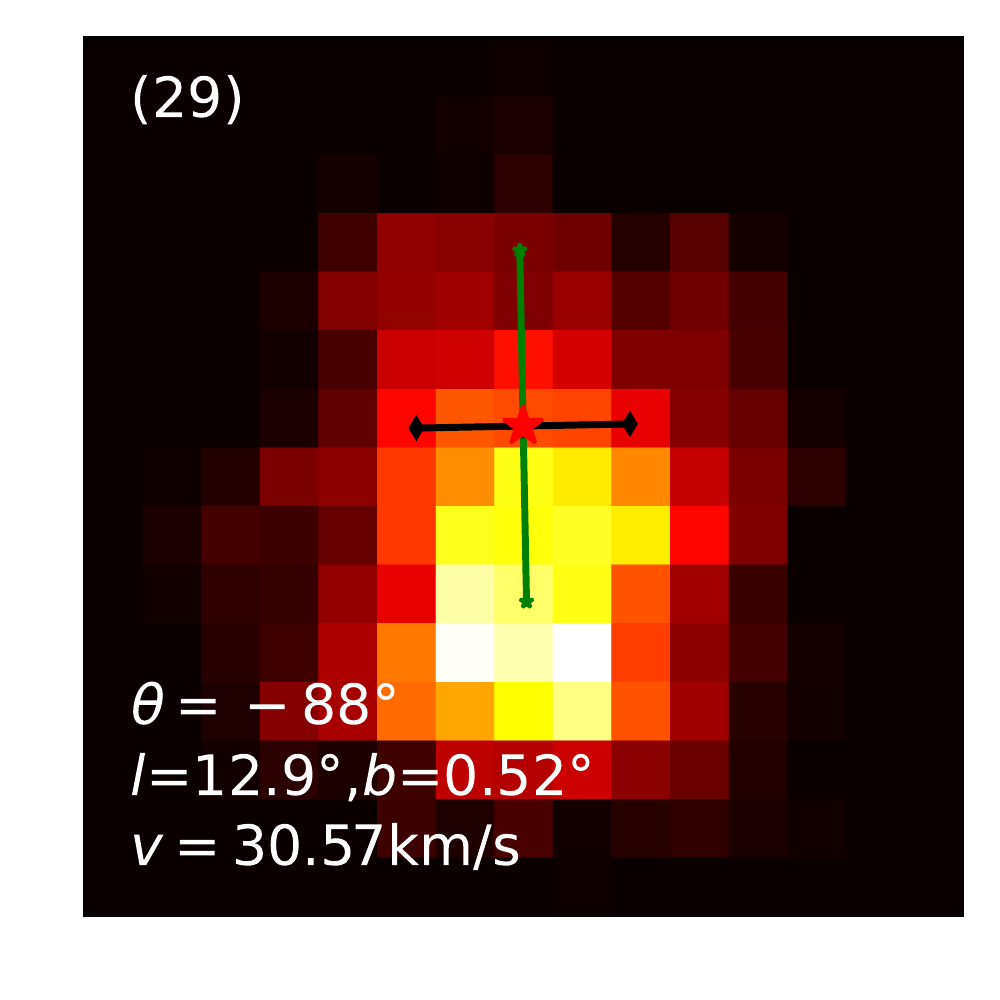}}
\end{minipage}\begin{minipage}[t]{0.24\textwidth}
		\centering
		\centerline{\includegraphics[width=1.9in]{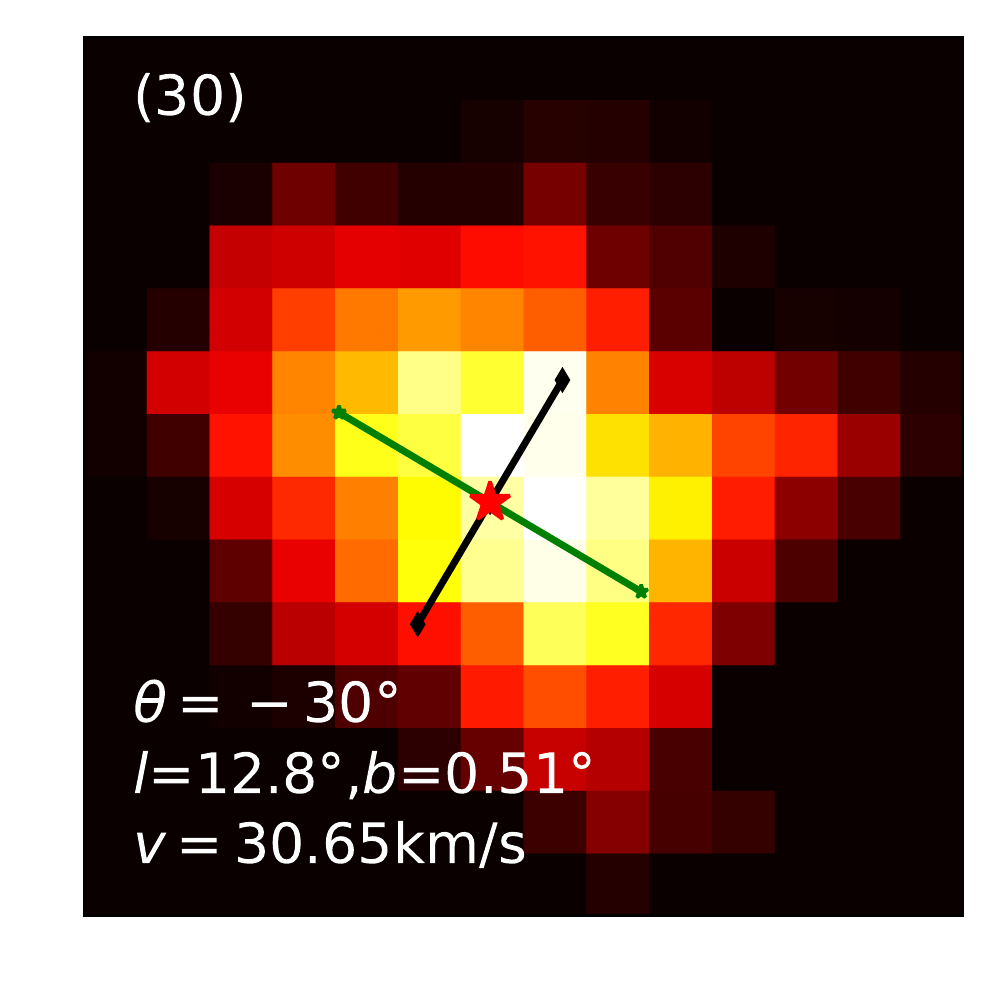}}
\end{minipage}\begin{minipage}[t]{0.24\textwidth}
		\centering
		\centerline{\includegraphics[width=1.9in]{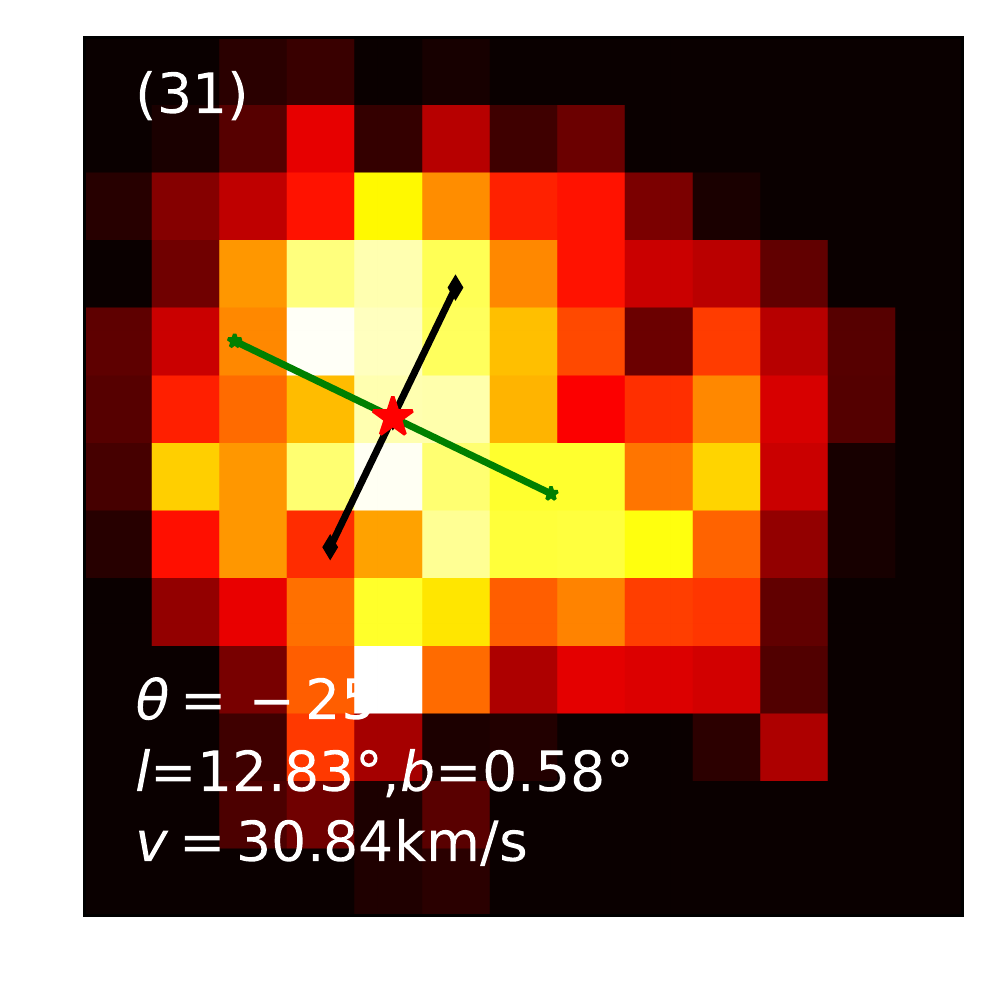}}
\end{minipage}\begin{minipage}[t]{0.24\textwidth}
		\centering
		\centerline{\includegraphics[width=1.9in]{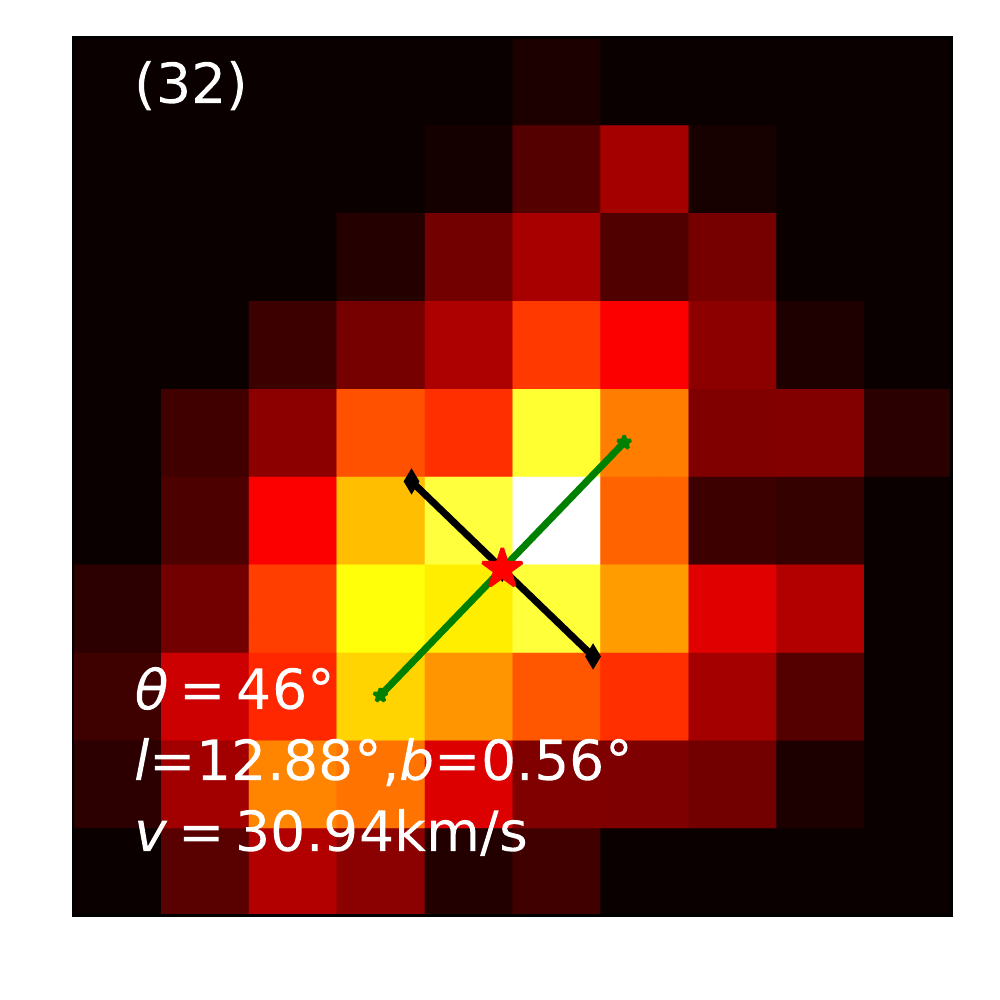}}
\end{minipage}

	\begin{minipage}[t]{0.24\textwidth}
		\centering
		\centerline{\includegraphics[width=1.9in]{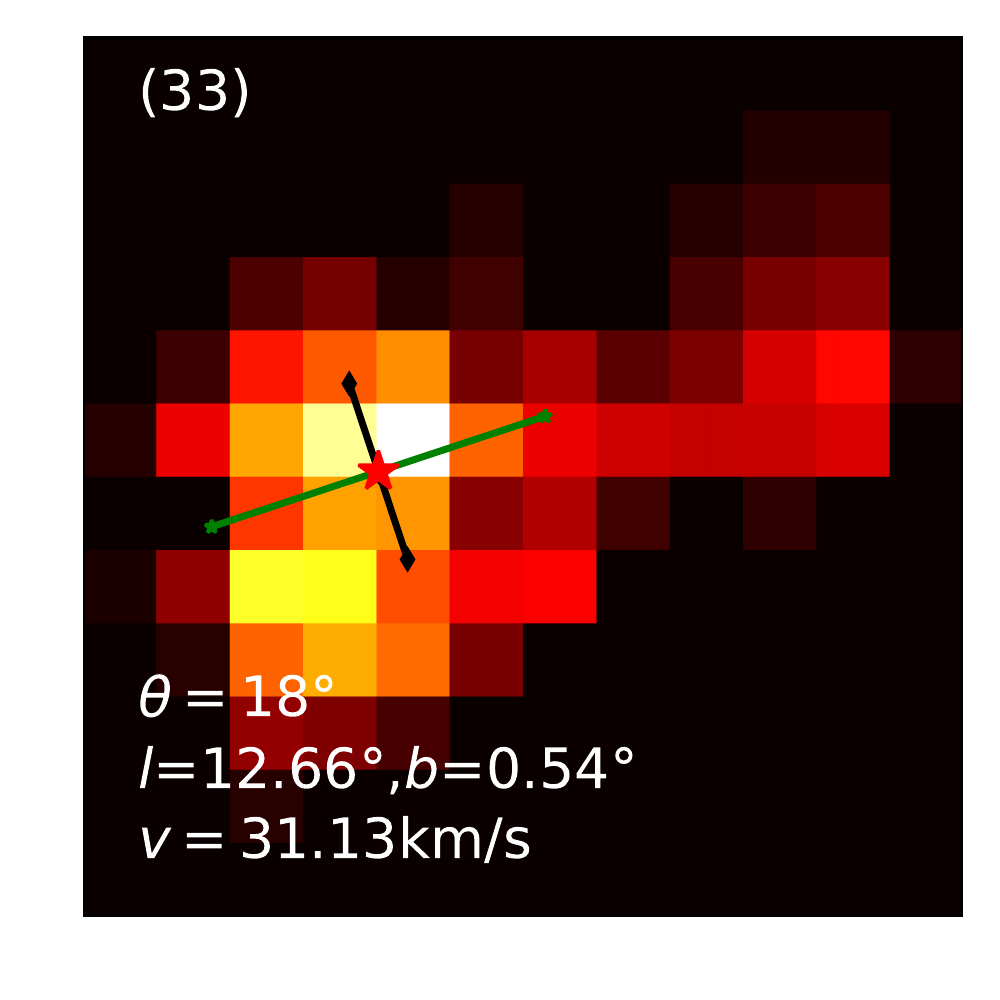}}
\end{minipage}\begin{minipage}[t]{0.24\textwidth}
		\centering
		\centerline{\includegraphics[width=1.9in]{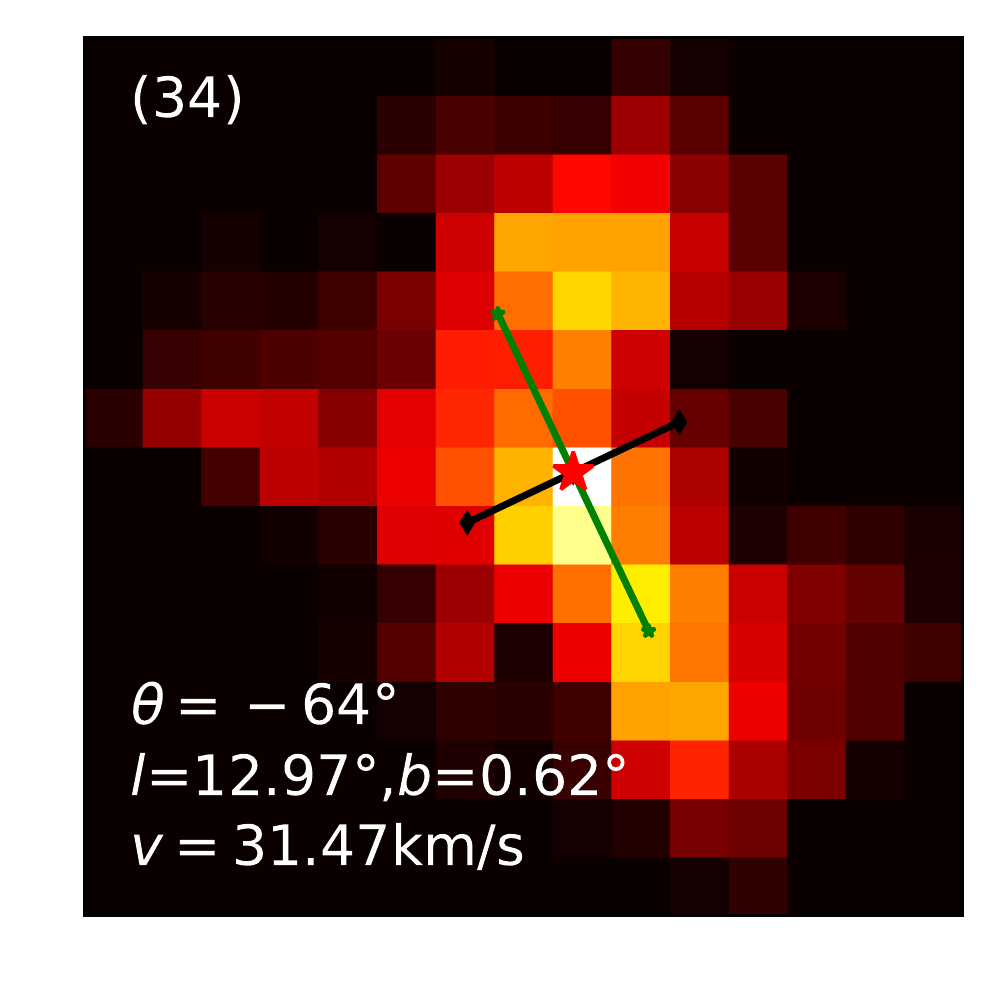}}
\end{minipage}\begin{minipage}[t]{0.24\textwidth}
		\centering
		\centerline{\includegraphics[width=1.9in]{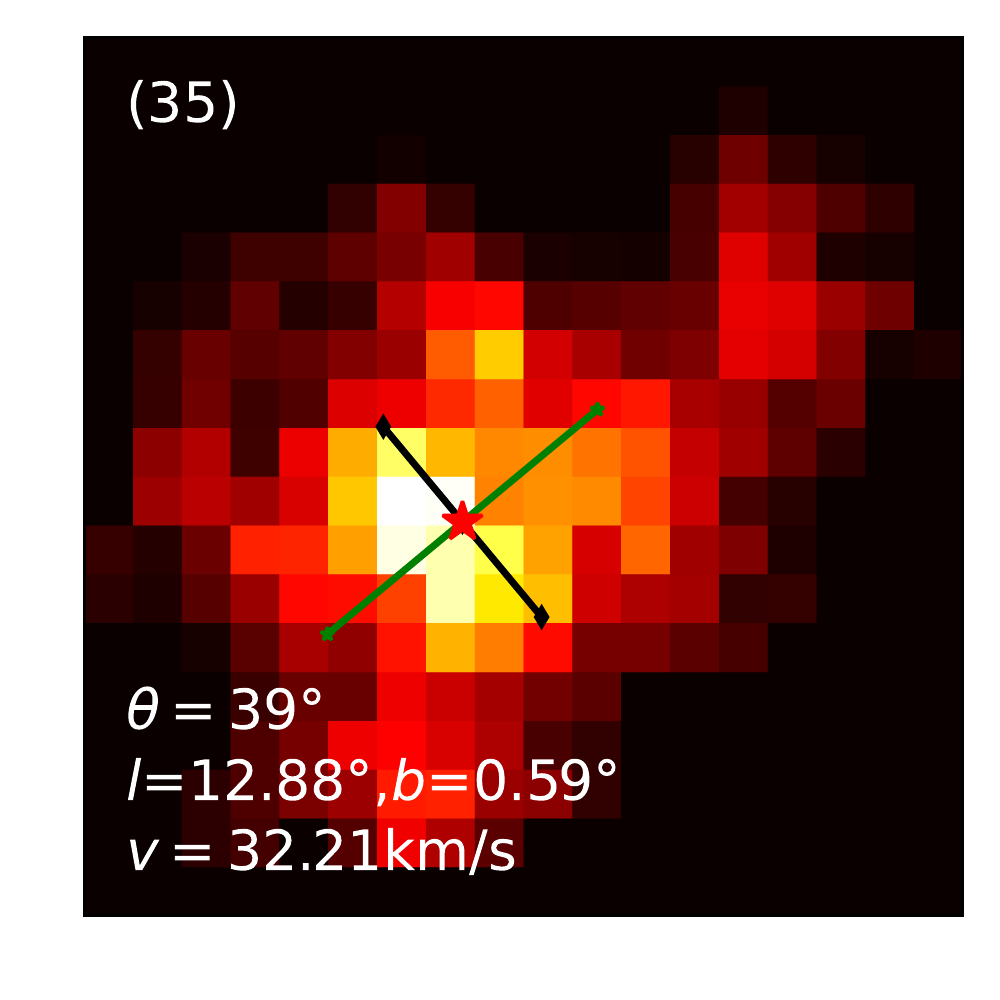}}
\end{minipage}\begin{minipage}[t]{0.24\textwidth}
		\centering
		\centerline{\includegraphics[width=1.9in]{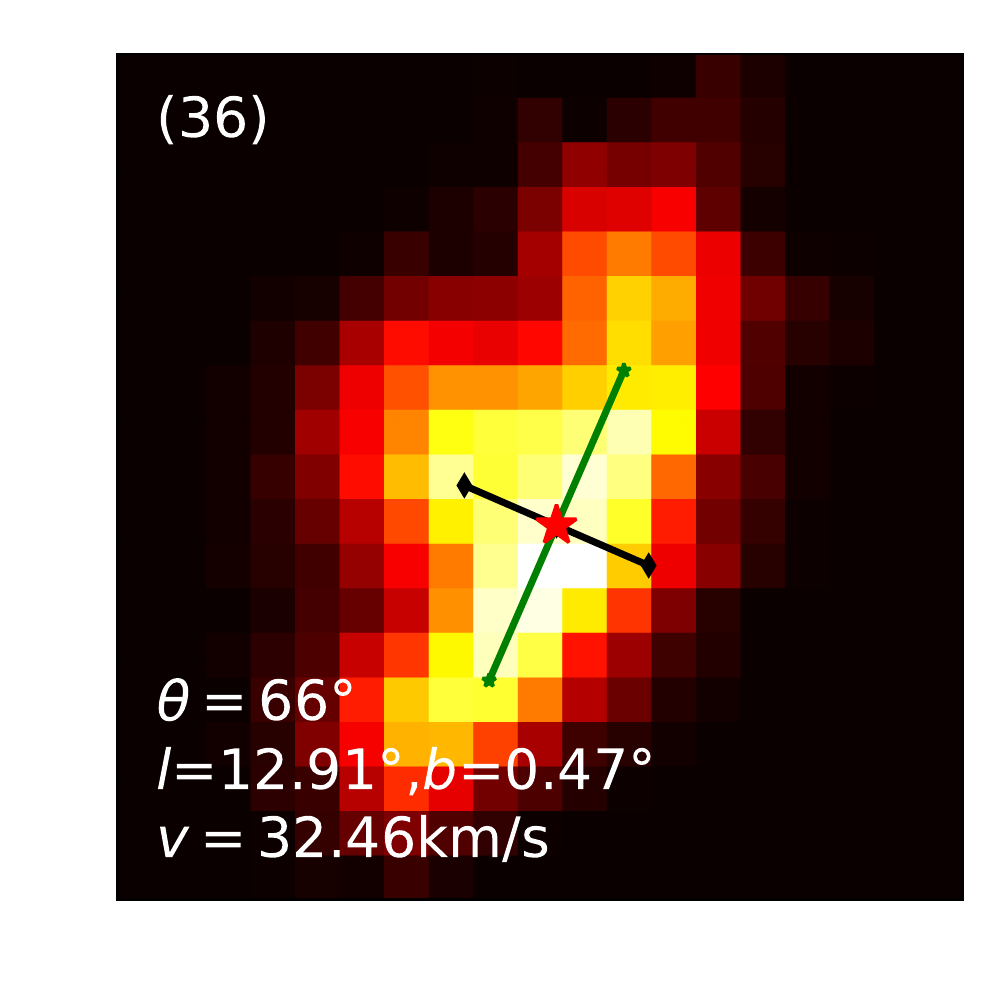}}
\end{minipage}\caption{Velocity-integrated intensity images of clumps in the high overlap area as shown in Figure \ref{Fig_C18O_0}. Each panel is presented in the same manner as in Figure \ref{Fig_C18O_1}.}
	\label{Fig_C18O_3}
\end{figure*}

Partial clumps in the high overlap area as described in Section \ref{sec:3.3.3} are shown in Figures \ref{Fig_C18O_2} and \ref{Fig_C18O_3}. They are used to visually evaluate the usability of FacetClumps in detecting molecular clumps. 

\end{document}